\documentclass[preprint,onecolumn,12pt]{elsarticle}

\usepackage{amsmath}
\usepackage{amssymb}
\usepackage{mathtools}
\usepackage{tikz}
\usepackage[a4paper]{geometry}
\usepackage{lipsum}
\usetikzlibrary{trees}

\newcommand{\pin}{\par\noindent}

\usepackage[hidelinks,colorlinks=true,linkcolor=blue, citecolor=blue]{hyperref}
\newtheorem{theorem}{Theorem} 
\biboptions{numbers,sort&compress}
\journal{Nuclear Physics B}

\begin{document}

\begin{frontmatter}



\title{Holographic unitary renormalization group for correlated\\ electrons~-~I: a tensor network approach}


\author{Anirban Mukherjee}
\ead{am14rs016@iiserkol.ac.in}
\author{Siddhartha Lal \corref{1}}
\cortext[1]{Corresponding author}
\ead{slal@iiserkol.ac.in}

\address{Department of Physical Sciences, Indian Institute of Science Education and Research-Kolkata, W.B. 741246, India \\
This article is registered under preprint number: arXiv:2004.06897v1}

\begin{abstract}
We present a unified framework for the renormalisation of the Hamiltonian and eigenbasis of a system of correlated electrons, unveiling thereby the interplay between electronic correlations and many-particle entanglement. For this, we extend substantially the unitary renormalization group (URG) scheme introduced in Refs.\cite{anirbanmotti,anirbanmott2,mukherjee2020}. We recast the RG as a discrete flow of the Hamiltonian tensor network, i.e., the collection of various $2n$-point scattering vertex tensors comprising the Hamiltonian. The renormalisation progresses via unitary transformations that block diagonalizes the Hamiltonian iteratively via the disentanglement of single-particle eigenstates. This procedure incorporates naturally the role of quantum fluctuations. The RG flow equations possess a non-trivial structure, displaying a feedback mechanism through frequency-dependent dynamical self-energies and correlation energies. The interplay between various UV energy scales enables the coupled RG equations to flow towards a stable fixed point in the IR. The effective Hamiltonian at the IR fixed point generically has a reduced parameter space, as well as number of degrees of freedom, compared to the microscopic Hamiltonian. Importantly, the vertex tensor network is observed to govern the RG flow of the tensor network that denotes the coefficients of the many-particle eigenstates. The RG evolution of various many-particle entanglement features of the eigenbasis are, in turn, quantified through the coefficient tensor network. In this way, we show that the URG framework provides a microscopic understanding of holographic renormalisation: the RG flow of the vertex tensor network generates a eigenstate coefficient tensor network possessing a many-particle entanglement metric. We find that the eigenstate tensor network accommodates sign factors arising from fermion exchanges, and that the IR fixed point reached generically involves a trivialisation of the fermion sign factor. Several results are presented for the emergence of composite excitations in the neighbourhood of a gapless Fermi surface, as well as for the condensation phenomenon involving the gapping of the Fermi surface. 
\end{abstract}

\begin{keyword}
unitary renormalization group, holographic entanglement renormalization, tensor networks, correlated electrons, fermionic criticality, arXiv:2004.06897
\end{keyword}

\end{frontmatter}
\tableofcontents
%\linenumbers

\section{Introduction}
\pin
The renormalization group (RG) is a formalism that enables the description of complex microscopic models in terms of simpler effective theories at infrared (IR) energy scales, btained via the integrating out of high-energy (UV) degrees of freedom~\cite{Wilson1974,Wilson1975,kadanoff1976,M.Fisher1998}. This aids in studying critical phenomena through the identification of universality classes~\cite{kadanoff1967static, wegner1994}. An advanced version of the Wilsonian RG program, known as functional RG (FRG), deals with problems involving electronic correlations via the RG flow of the Grassmanian many-body action~\cite{Wetterich1993,metzner2012functional,
salmhofer2001fermionic,kopietz2001exact}. Here, the exact Wetterich equations~\cite{Wetterich1993} incorporate all orders of quantum fluctuations by accounting for the entire hierarchy of $2n$-point vertex RG flow equations~\cite{kugler2018multiloop,tagliavini2019}. This formalism has been successful in capturing a wide variety of strongly correlated phases of electronic quantum matter, e.g., the pseudogap, strange metal, d-wave superconductivity etc.~ \cite{katanin2004,rohe2005, giering2012,hille2020pg,wang2009, vilardi2019}.
\pin
In recent times, many-particle quantum entanglement~\cite{li2008entanglement,chen2010local,yao2010entanglement,
thomale2010nonlocal} has emerged as an important feature for the study of quantum many-body systems with strong correlations, e.g., quantum spin liquids, fractional quantum hall phases, high Tc superconductors etc.~\cite{balents2010spin,
castelnovo2012spin,banerjee2016proximate}. In such systems, a marked (and sometimes dramatic) change in the nature of many-particle entanglement is observed as a signature of quantum criticality~\cite{osterloh2002scaling,osborne2002entanglement,
vidal2003entanglement,calabrese2004,orus2008universal,
laflorencie2016quantum}. In order to characterize the nature of many-particle entanglement in quantum many body system in the IR, as well as near criticality, an entanglement renormalization group (ERG) based on tensor network (TN) states has emerged as an indispensable tool~\cite{vidal2007,vidal2008,evenbly2009,orus2014practical,evenbly2011}. 
For instance, TN states such as matrix product states~(MPS) (developed initially in DMRG~\cite{white1992density}) has been shown as being highly accurate for studying the ground state properties of 1D gapped phases. In $2D$, tree tensor network states~(TTN)\cite{tagliacozzo2009simulation,murg2010simulating} and projected entangled pair states~(PEPS) are useful for studying gapped phases~(see references in \cite{orus2014practical}). The multiscale entanglement renormalization group ansatz (MERA) is yet another tensor network RG program in which each layer of RG transformations is organized as a stacking of a layers of tensor products of two-local unitary operations (i.e., perform entanglement renormalization) and a layer composed of isometries that remove the disentangled qubits (i.e., carry out the process of coarse-graining). MERA has been  used for studying both quantum criticality~ \cite{giovannetti2008,pfeifer2009entanglement,
swingle2016renormalization}, as well as gapped topological quantum liquids~\cite{aguado2008,swingle2016renormalizationTop,wen2016,
gu2016,gerster2017}. 
\pin
We present here an unitary RG (URG) program for electronic states that generates RG flows, on the one hand, the entire hierarchy of $2n$-point vertex tensors comprising the Hamiltonian, and on the other hand, the entire set of many-body wavefunction coefficient tensors that govern the renormalization of the eigenbasis. Indeed, the renormalization of the wavefunction coefficient tensors comprise the entanglement RG flow. In this way, URG provides a unified framework by which we obtain both the vertex and entanglement RG flows. Further, we show that the the vertex RG flows feed into the entanglement renormalization. As a result, stable fixed points are reached simultaneously for both. 
URG is carried out via a sequence of unitary disentanglement operations on a graph, each of whose nodes corresponds to one electronic state. Each unitary operation on the graph disentangles an electronic state from the rest (the coupled subspace), leading simultaneously to block diagonalisation of the Hamiltonian in the occupation number basis. This involves the removal of off-diagonal terms with respect to a given electronic state $|N\rangle$, making good quantum numbers of the occupation numbers ($1$ and $0$) of that state.
\pin
The URG formalism was introduced in Ref.\cite{anirbanmotti,anirbanmott2,pal2019}, and applied to (a) the repulsive 2D fermionic Hubbard model on a square lattice at half-filling and with hole doping, as well as (b) the Kagome XXZ antiferromagnet at finite magnetic field. In Ref.\cite{anirbanmotti,anirbanmott2}, a comparative study of the URG and FRG programs was offered using the two-particle vertex RG flow equations obtained from the two approaches. In URG, the RG flow equations for the various $2n$-point vertices have several important features. First, While the FRG scheme is exact in principle, in practice it involves a truncation in the loop expansion~\cite{metzner2012,tagliavini2019}. In contrast, the URG equations are non-perturbative, with contributions from all loops resummed into closed-form analytic expressions. 
Second, the URG equations possess a non-trivial denominator containing the renormalized correlation energies and self-energies, i.e., the number diagonal pieces of the renormalized many-body Hamiltonian. 
Further, the denominator of the vertex URG flow equations have an explicit dependence on an energy scale ($\omega$) for quantum fluctuations that arises from the renormalization of the remnant off-diagonal terms in the coupled subspace. Thus, the retardation effects observed in FRG~\cite{salmhofer2004} are manifest in the URG scheme as well. 
\pin
Although the URG flows do not involve loop truncation approximations, 
the complete heirarchy of $2n$-point vertex URG equations are challenging to solve as they are coupled nonlinearly. As a result, in Refs.\cite{anirbanmotti,anirbanmott2,pal2019}, we truncated the heirarchy of the $2n$-point vertex URG flows at six-point vertices. In the present work, we aim to improve on this substantially be presenting a unified treatment of the entire hierarchy of $2n$-point vertices. Finally, the nonperturbative nature of the URG equations yields stable fixed points that are governed by the quantum fluctuation energy scale $\omega$. It is also noteworthy that, upon reaching a stable fixed point, we can construct an effective Hamiltonian from the final values of the $2n$-point vertices. In this way, we have obtained effective Hamiltonians and eigenstates at the stable fixed points for 2D Hubbard model (both at $1/2$-filling, as well as upon hole-doping away from it) in Refs.\cite{anirbanmotti,anirbanmott2} and the Kagome XXZ antiferromagnet at finite field in Ref.\cite{pal2019}.
\pin
The URG has also been validated quantitatively in Refs.\cite{anirbanmotti,anirbanmott2} for the 2D Hubbard model with high accuracy by benchmarking two quantities obtained from the URG against other numerical methods~\cite{leblanc2015solutions}: the ground state energy per particle (within an errorbar of $10^{-4}t$), and the doublon fraction. Recently, in Ref.\cite{mukherjee2020}, we have also implemented the URG as a reverse renormalization group flow, i.e., by starting from the many- body eigenstates of the effective Hamiltonian for the Mott insulating state at IR stable fixed point of the 2D Hubbard model at $1/2$-filling, we have reconstructed the eigenstates of the parent model in the subspace associated with the most relevant scattering diagrams. Ref.\cite{mukherjee2020} shows that the reverse URG procedure generates an entanglement holographic mapping (EHM) network~\cite{lee2016,qi2013}  along the RG flow direction. This is a generalization of MERA that involves only unitary transformations. 
\pin
In the present work, we aim at extending the URG framework in several important ways. First, we show that the unitary operation for a given RG step is determined by the Hamiltonian obtained from the previous RG step. As a result, the action of the unitary operation on the many-body eigenstates naturally involve RG flows of the wavefunction coefficient tensor that incorporate contributions from the RG flow of all
$2n$-point vertices. As a result, the bulk of the EHM generated along the URG direction is composed of various 2-point, 4-point, 6-point and all higher order correlators. This is in contrast with the EHM formulation of Ref.\cite{lee2016}, where the bulk is composed of two- point correlators. Furthermore, we show that the quantum fluctuation scales ($\omega$) themselves undergo a non-trivial renormalization in the bulk of the EHM. The resulting interplay between the RG dynamics of quantum fluctuations and that of the Hamiltonian shows that the bulk of the EHM manifestly possesses non-trivial quantum as well as RG dynamics~\cite{lee2010,kskim2017,kskim2019}. We also offer here comparisons between URG and other entanglement RG methods. URG is carried out on generally on a graph, such that the notion of a physical distance is not essential for its implementation. This is a crucial departure from the implementation of MERA and EHM networks that depend upon a real-space geometry~\cite{evenbly2009}. Note that URG should also be contrasted with the continuous unitary transformation~(CUT) based RG schemes\cite{glazekWilson1993,glazekWilson1994,wegner1994} that successively band diagonalize the Hamiltonian over an infinite number of steps; URG involves a discrete set of unitary rotations that block diagonalise the Hamiltonian in a finite number of steps. Instead, the URG method is related to the strong-disorder RG approaches of Dasgupta et al.~\cite{ma1979sk}, Fisher~\cite{fisher1992random}, Rademaker et al.~\cite{rademaker2016explicit} and You et al.~\cite{you2016entanglement}. The philosophy of URG is similar to the 
entanglement based CUT (E-CUT) RG of Ref.\cite{sahin2017entanglement} in that both attempt to bridge the Hamiltonian RG with entanglement (tensor network) RG.
\pin
The rest of the work is organized as follows. In Sec. \ref{RG_flow_strategy}, we provide analytical derivation of the disentangling unitary transformation, and the block-diagonal Hamiltonian that results from the  
iterative application of a sequence of such unitary transformations.
Specifically, the electronic states are disentangled in the order of the single-particle energy, from higher to lower.
In this way, the entire heirarchy of $2n$-point scattering vertex tensors evolve from UV to IR via a series of disentanglement transformations. This allows us to interpret the Hamiltonian RG  as a vertex tensor network RG, where the network is formed from the $2n$-point scattering vertices. Sec.\ref{eig_base_renorm} is devoted to incorporating the effects of Hamiltonian block diagonalization on its eigenbasis. This is carried out by via applying the unitary transformations to perform disentanglement of UV degrees of freedom, and noting the subsequent entanglement renormalization of the remnant IR degrees of freedom. In this way, the Hamiltonian (vertex) tensor network is shown to govern the eigenstate tensor network (itself an EHM network~\cite{mukherjee2020}). We demonstrate that the eigenstate tensor network accommodates fermion exchange sign factors arising from the vertex renormalisations, and that the IR fixed point reached generically involves a trivialisation of the fermion sign factor.  Additionally, we show that the Hilbert space geometry quantified by a many-particle entanglement metric also undergoes a RG flow. 
Sections \ref{fate_single_p_exc} and \ref{bound_state_form} are devoted to demonstrating the usage of the URG for a general model of interacting electrons with translation invariance. Specifically, we show the existence of log-divergences in one-particle and two-particle self-energies that result in the breakdown of the Landau quasiparticle picture and the gapping of the Fermi surface respectively. We demonstrate that various sum rules are obeyed by the URG method, and reach very broad conclusions for the emergence of novel states of fermionic quantum matter. We conclude in Sec.\ref{conclusions}. The details of various calculations are provided in several appendices.
\section{Summary of main results}
We summarise here the main results of the paper for the benefit of the reader. 
\begin{itemize}
\item In Sec.\ref{RG_flow_strategy}, we provide the derivation of the unitary disentanglement operator~\cite{anirbanmotti} and the form of the rotated Hamiltonian (eq.\eqref{rot_ham}). The rotated Hamiltonian is found to commute with the number operator ($\hat{n}_{j}$) associated with the disentangled $|j\rangle$, and generates an integral of motion. Following this, we compare and contrast the URG with other unitary transformation-based RG methods (e.g., continuous unitary transformation (CUT) RG, spectrum bifurcation RG (SBRG), strong disorder RG etc.). For instance, in CUT RG, the off-diagonal matrix elements connecting energy states with highest energy differences are eliminated in a perturbative fashion via an infinitesmal Schrieffer-Wolff transformation, such that a succession of such transformations makes the Hamiltonian increasingly \emph{band-diagonal}. Owing to the perturbative nature of the transformation, the disentanglement between electronic degrees of freedom is partial. On the other hand, every unitary of the URG disentangles perfectly the highest energy electronic qubit from the rest degrees of freedom at a given RG step, thereby \emph{block-diagonalizing} the Hamiltonian in Fock space (eq.\eqref{rot_ham}). 
\item In Sec.\ref{RG_flow_strategy}, we provide a detailed description of the number-diagonal and off-diagonal parts of the disentangling unitary operator. The RG evolution of the Hamiltonian's spectrum is tracked as a function of the quantum fluctuation scale $\omega$ that originates from non-commutativity between off-diagonal and diagonal parts of the Hamiltonian. 
\item In Sec.\ref{RG_flow_strategy}, we present the scattering vertex tensor network representation of the complete Hamiltonian in eqs.\eqref{D-X decomposition}. From this, we obtain the entire heirarchy of the $2n$-point vertex RG flow equations (eq.\eqref{RG_flow_heirarchy}). The non-perturbative nature of the heirarchy of RG equations is seen from the fact that they are closed-form expressions that include all orders of loops in the couplings. Finally, in Fig.\ref{EHMnetwork}, we represent the renormalization as a Hamiltonian vertex tensor network whose construction is equivalent to that of an exact holographic mapping (EHM) network.
\item In Sec.\ref{eig_base_renorm}, we study the renormalization of the eigenbasis generated by the unitary transformations of the URG method. In this way, we obtain the entire family of RG equations for the many- body coefficients to comprise the eigenstates (eq.\eqref{coefficient renormalization}). These RG equations also account for the electron-exchange signatures generated by the $2n$-point scattering processes. \item In a generic Hamiltonian comprising of electronic dispersion and attractive four-fermionic interactions, we apply the URG to demonstrate the RG flow of the effective Hamiltonian towards that of the reduced BCS model. Alongside, we show that the eigenbasis renormalizes towards an eigen-subspace for the reduced BCS model composed of Anderson BCS pseudospins~\cite{anderson1958random}, thereby mitigating the Fermion sign problem in the IR effective theory.
\item Further, in Sec.\ref{eig_base_renorm}, we also show the renormalisation flow of the Fubini-Study metric of the many-body Hilbert space, and use this to classify flows to both gapless as well as gapped fixed point theories.
\item Finally, starting from a generic model of interacting fermions, we depict in Sec.\ref{Fermi_surface} and Sec.\ref{bound_state_form} the emergence of two-particle one-hole composite and pseudospin degrees of freedom for gapless and gapped phases respectively.
\end{itemize}
\section{Hamiltonian RG flow} \label{RG_flow_strategy}
\begin{figure}
\centering
\includegraphics[width=0.7\textwidth]{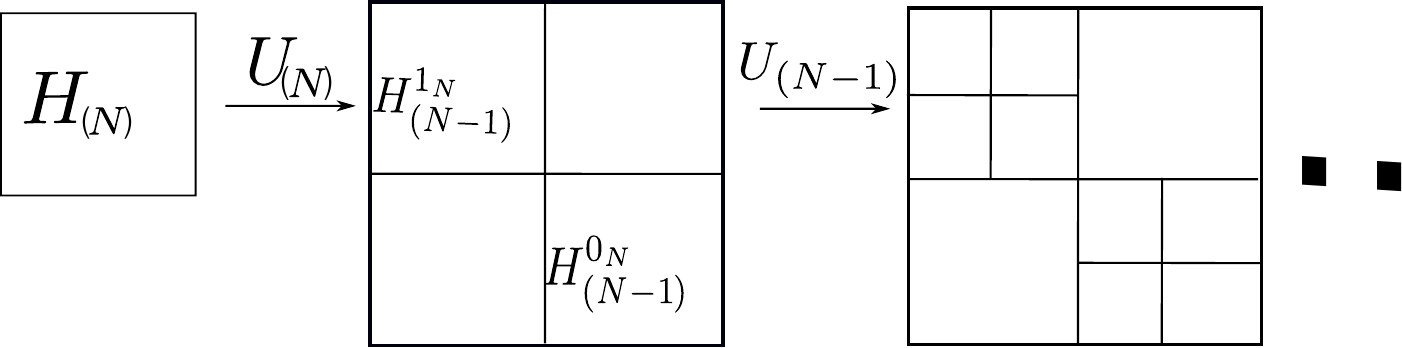} 
\caption{Representation of the Hamiltonian as a block matrix in the occupation number basis of single-electron states, and where the unitary rotation removes the off-diagonal blocks.}\label{Hamiltonian_flow_diagram}
\end{figure}

The renormalization group program will be set up in this section in order to describe flow of effective Hamiltonians and their associated eigenspaces across a range of energyscales at zero temperature. This range in energyscales arises from the \textit{quantum fluctuations} associated with the non-commutativity between off-diagonal and diagonal parts of the Hamiltonian in the occupation-number representation of the single-electron states. The renormalization group flow involves resolving these quantum fluctuations with respect to a single electronic state at every RG step. Below, we will first develop the Hamiltonian RG program in section. 
\par\noindent\\
{\bf \textit{Renormalization Group as Fermion Occupation Number Block Diagonalisation}}\label{RG}
\pin
Single electron states constituting an electronic system with N degrees of freedom (d.o.f) can be assigned indices ranging from $1$ to $N$. The index $j\in (1,\ldots,N)$ 
refers to a collection of attributes that label the electron creation ($c^{\dagger}_{j}$), annhilation ($c_{j}$) and occupation number ($\hat{n}_{j}=c^{\dagger}_{j}c_{j}$) operators in the second-quantized representation. For example, $j\equiv \lbrace n,\mathbf{r},\mathbf{\sigma}\rbrace$ refers to band index ($n$), position vector ($\mathbf{r}$) and spin ($\sigma$). The electron creation and annihilation operators satisfy the usual on-site commutation and anti-commutation relation dictated by the Pauli exclusion principle
\begin{eqnarray}
\lbrace c_{j},c^{\dagger}_{j}\rbrace =1~,~[c^{\dagger}_{j},c_{j}] =2\hat{n}_{j}-1.\label{2nd_quantized}
\end{eqnarray}
The numerical ordering of the indices $1$ to $N$ describe the single-particle energy eigenvalues sorted in ascending order. 
(For degenerate energy values the states are labelled via a specified random choice.)
The Hamiltonian H governing the dynamics of this system will contain two kinds of terms: (i) scattering terms that are off-diagonal in occupation number basis, i.e., causing fluctuations in the occupancy of a electronic state, and (ii) self/correlation energy terms that are diagonal in occupation number basis, i.e., causing a shift in energy associated with a given electronic occupancy configuration.
Such a partitioning of the Hamiltonian matrix was formalised in the context of quantum mechanical perturbation theory by Lowdin \cite{lowdin1951note,lowdin1962studies,lowdin1982partitioning}, and independently by Feshbach \cite{Feshbach1958, Feshbach1962}. Via this technique, the Hamiltonian($H$) is represented as a block matrix, in the occupation number basis of the state N: $\lbrace |0_{N}\rangle,|1_{N}\rangle\rbrace$,
\begin{eqnarray}
H &=&\hat{n}_{N}H\hat{n}_{N}+(1-\hat{n}_{N})H(1-\hat{n}_{N})+(1-\hat{n}_{N})H\hat{n}_{N}+\hat{n}_{N}H(1-\hat{n}_{N})\label{partitioned_H}\\
&=&\begin{bmatrix}Tr_{N}(H\hat{n}_{N}) & Tr(Hc_{N})\\
Tr(c^{\dagger}_{N}H) & Tr_{N}(H(1-\hat{n}_{N}))
\end{bmatrix}~.\label{blockmatrix}
\end{eqnarray} 
The occupation number operators $\hat{n}_{N}=c^{\dagger}_{N}c_{N}$ and $1-\hat{n}_{N}=c_{N}c^{\dagger}_{N}$ represent electron and hole subspace projections in the second quantized notation respectively. 
We define the unitary transformation $U_{(N)}$ as that which decouples state $N$ from all others, i.e, it block diagonalises $H$ by removing the off-diagonal quantum fluctuation blocks, resulting in the new Hamiltonian $H_{N-1}=U_{(N)}HU_{(N)}^{\dagger}$ (see fig(\ref{Hamiltonian_flow_diagram}b)). $U_{(N)}$ is determined by the \textit{decoupling equation},
\begin{eqnarray}
\hat{n}_{N}\hat{U}_{(N)}HU^{\dagger}_{(N)}(1-\hat{n}_{N})=0~.\label{decoupling condition}
\end{eqnarray}
In this way, the label $(N)$ (representing the decoupling of state $N$) is the first step of the RG transformations. Below we recollect the steps from Ref.\cite{anirbanmotti} for deriving $U_{(N)}$ for a general fermionic Hamiltonian $H$. 
\pin
 The fermionic Hamiltonian $H_{(N)}=H$ can, very generally, be decomposed as 
\begin{eqnarray}
H_{(N)}=H^{D}_{(N)}+H^{X,N}_{(N)}+H_{(N)}^{X,\bar{N}},
\end{eqnarray}    
where the number diagonal part of the Hamiltonian ($H^{D}_{(N)}$) is associated with n-particle self/correlation energies, and the term $H_{(N)}^{X,\bar{N}}$ represents coupling only among the other degrees of freedom $\lbrace 1,\ldots,N-1\rbrace$. These comprise the diagonal blocks in the block matrix representation of $H$ eq.\eqref{blockmatrix}. $H_{(N)}^{X,N}=c^{\dagger}_{N}Tr(H_{(N)}c_{N})+h.c.$  represents the off-diagonal blocks in $H$ eq.\eqref{blockmatrix} that are responsible for quantum fluctuations in the occupation number of state $N$. We are searching for a rotated many-body basis of states $|\Psi\rangle$'s in which the old Hamiltonian $H_{(N)}$ attains a block diagonal form $H_{(N-1)}=H^{D}_{(N-1)}+H_{(N-1)}^{X,\bar{N}}$
\begin{eqnarray}
(H^{D}_{(N)}+H^{X,N}_{(N)}+H_{(N)}^{X,\bar{N}})|\Psi\rangle = (H^{D}_{(N-1)}+H_{(N-1)}^{X,\bar{N}})|\Psi\rangle~.~~~~\label{blockDiagcond}
\end{eqnarray}
To proceed further in solving this equation, we write $|\Psi\rangle$ in the occupation number basis of states $0_{N}$ and $1_{N}$
\begin{eqnarray}
|\Psi\rangle = a_{1}|\Psi_{1},1_{N}\rangle + a_{0}|\Psi_{0},0_{N}\rangle~,\label{superposition}
\end{eqnarray}
where the pair of states $|\Psi_{1}\rangle$ and $|\Psi_{0}\rangle$  belong to the remnant $2^{N-1}$ dimensional Hilbert space of $1,..,N-1$ single electron degrees of freedom. The $2$ dimensional Hilbert space of the electron $N$ is spanned by $|1_{N}\rangle$ and $|0_{N}\rangle$. 
Replacing eq.\eqref{superposition} in eq.\eqref{blockDiagcond}, we obtain a  set of simultaneous equations
\begin{eqnarray}
a_{1_{N}}(\hat{\omega}_{(N)}-Tr_{N}(\hat{n}_{N}H^{D}_{(N)})\hat{n}_{N})|\Psi_{1},1_{N}\rangle &=& a_{0_{N}}c^{\dagger}_{N}Tr_{N}(H_{(N)}c_{N})|\Psi_{0},0_{N}\rangle~,\nonumber\\
a_{0_{N}}(\hat{\omega}_{(N)}-Tr((1-\hat{n}_{N})H^{D}_{(N)})(1-\hat{n}_{N}))|\Psi_{0},0_{N}\rangle &=& a_{1_{N}}Tr_{N}(c^{\dagger}_{N}H_{(N)})c_{N}|\Psi_{1},1_{N}\rangle~,~~~\label{transition-eqns}
\end{eqnarray}
where 
\begin{equation}
\hat{\omega}_{(N)}=H^{D}_{(N-1)}+H^{X,\bar{N}}_{(N-1)}-H^{X,\bar{N}}_{(N)}~.\label{QFoperator}
\end{equation}
In reaching the above simultaneous equations, we have used Appendix~\ref{block matrix} to obtain the occupation number representations of the diagonal/off-diagonal parts of $H$.
From equation set \ref{transition-eqns}, we deduce the following equations
\begin{eqnarray}
\eta^{\dagger}_{(N)}\eta_{(N)}|\Psi_{1},1_{N}\rangle &=& |\Psi_{1},1_{N}\rangle\implies\eta^{\dagger}_{(N)}\eta_{(N)}=\hat{n}_{N}~,\label{electron-op}\\
\eta_{(N)}\eta^{\dagger}_{(N)}|\Psi_{0},0_{N}\rangle &=& |\Psi_{0},0_{N}\rangle\implies \eta_{(N)}\eta^{\dagger}_{(N)}=1-\hat{n}_{N},~~~~~\label{hole-op}
\end{eqnarray}
where $\eta^{\dagger}_{(N)}$ and $\eta_{(N)}$ are defined as
\begin{eqnarray}
\eta^{\dagger}_{(N)}&=&\frac{1}{\hat{\omega}_{(N)}-Tr_{N}(H^{D}_{(N)}\hat{n}_{N})\hat{n}_{N}}c^{\dagger}_{N}Tr_{N}(Hc_{N}),\label{eh-transition-1} \\
\eta_{(N)}&=&\frac{1}{\hat{\omega}_{(N)}-Tr_{N}(H^{D}_{(N)}(1-\hat{n}_{N}))(1-\hat{n}_{N})}Tr_{N}(c^{\dagger}_{N}H)c_{N}.~~~~~\label{eh-transition-2}
\end{eqnarray}
Finally, the above equations enable us to relate $|\Psi\rangle$ and $|\Psi_{1},1_{N}\rangle$ via a similarity transformation as follows
\begin{eqnarray}
|\Psi\rangle  = a_{1}(1+\eta_{(N)})|\Psi_{1},1_{N}\rangle =a_{1}\exp(\eta_{(N)}))|\Psi_{1},1_{N}\rangle~.~~~~
\end{eqnarray} 
The similarity transformation $\exp(\eta_{(N)}))$ can be used to construct a unitary operator $U_{(N)}$ \cite{shavitt1980quasidegenerate,suzuki1982construction},
\begin{flalign}
U_{(N)}=\exp \frac{\pi}{4}(\eta^{\dagger}_{(N)}-\eta_{(N)})=\frac{1}{\sqrt{2}}(1+\eta^{\dagger}_{(N)}-\eta_{(N)})~.\label{Unitary_op}
\end{flalign}
The property of a unitary transformation $U_{(N)}U^{\dagger}_{(N)}=U^{\dagger}_{(N)}U_{(N)}=I$ can be immediately checked from the  
anti-commutation relation $\lbrace\eta_{(N)},\eta^{\dagger}_{(N)}\rbrace =1$. Via applying the unitary operator $U_{(N)}$ on $H$, we will obtain the form of the rotated Hamiltonian in the next section. 
\newpage
\par\noindent\\{\bf \textit{Derivation for the rotated Hamiltonian $U_{N}HU^{\dagger}_{N}$}}\label{derive-rot-H}
\pin 
We note that the rotated Hamiltonian should be purely diagonal in the occupation-number basis states $1_{N}$ and $0_{N}$. In order to verify this, we decompose the rotated Hamiltonian into diagonal and off-diagonal components
\begin{eqnarray}
U_{N}HU_{N}^{\dagger} &=& H_{1}+H_{2},\nonumber\\
H_{1}&=&\frac{1}{2}\bigg[H+[\eta^{\dagger}_{N}-\eta_{N},H]+\eta_{N}H\eta^{\dagger}_{N}+\eta^{\dagger}_{N}H\eta_{N}\bigg]~,\nonumber\\
H_{2}&=&\frac{1}{2}\bigg[H^{X}_{N}-\eta^{\dagger}_{N}Tr_{N}(c^{\dagger}_{N}H)c_{N}\eta^{\dagger}_{N}-\eta_{N}c^{\dagger}_{N}Tr_{N}(Hc_{N})\eta_{N}\bigg]~,
\end{eqnarray}
where the off-diagonal component $H_{2}$ must vanish. To show that, we first set up the preliminaries(using eq.\eqref{eh-transition-1} and eq.\eqref{eh-transition-2})
\begin{eqnarray}
\eta^{\dagger}_{N}\eta_{N}=\hat{n}_{N}&\implies & \hat{\omega}-Tr_{N}(H^{D}\hat{n}_{N})\hat{n}_{N}=c^{\dagger}_{N}Tr_{N}(Hc_{N})\eta_{N}~,\nonumber\\
&\implies &\eta_{N}c^{\dagger}_{N}Tr_{N}(Hc_{N})
\eta_{N} = Tr_{N}(c^{\dagger}_{N}H)c_{N}~.\label{vanishing-offdiag}
\end{eqnarray}
The definition of $H^{X}_{N}=c^{\dagger}_{N}Tr_{N}(Hc_{N})+h.c.$, along with eq.\eqref{vanishing-offdiag}, then implies that $H_{2}=0$. In the other component, $H_{1}$, we first  unravel the terms $\eta_{N}H\eta^{\dagger}_{N}$ and $\eta^{\dagger}_{N}H\eta_{N}$. Using eq.\eqref{quantum fluctuation scale}, eq.\eqref{eh-transition-1} and eq.\eqref{eh-transition-2}, we obtain
\begin{eqnarray}
&&\hspace*{-1cm}\frac{1}{\hat{H}'-Tr_{N}(H\hat{n}_{N})\hat{n}_{N}}c^{\dagger}_{N}Tr_{N}(Hc_{N}) = c^{\dagger}_{N}Tr_{N}(Hc_{N})\frac{1}{\hat{H}'-Tr_{N}(H(1-\hat{n}_{N}))(1-\hat{n}_{N})}~,~\nonumber\\
&&\hspace*{-1cm}\implies Tr_{N}(H\hat{n}_{N})\hat{n}_{N}c^{\dagger}_{N}Tr_{N}(Hc_{N})=c^{\dagger}_{N}Tr_{N}(Hc_{N})Tr_{N}(H(1-\hat{n}_{N}))(1-\hat{n}_{N})~.~~~~~~~\label{electron-hole-transition}
\end{eqnarray}
The above relation 
then allows us to simplify $\eta_{N}H\eta^{\dagger}_{N}$ and $\eta^{\dagger}_{N}H\eta_{N}$ as follows
\begin{eqnarray}
\eta_{N}H\eta^{\dagger}_{N} &=& Tr_{N}(H(1-\hat{n}_{N}))(1-\hat{n}_{N})~,\nonumber\\
\eta^{\dagger}_{N}H\eta_{N} &=& Tr_{N}(H\hat{n}_{N})\hat{n}_{N}~.\label{Piece 1 of H1}
\end{eqnarray}
Next, we deduce $[\eta^{\dagger}_{N}-\eta_{N},H]$, i.e., the renormalization of the Hamiltonian using the relations obtained above
\begin{eqnarray}
[\eta^{\dagger}_{N}-\eta_{N},H]=2\tau_{N}\lbrace c^{\dagger}_{N}Tr_{N}(Hc_{N}),\eta_{N}\rbrace~.\label{ren_H}
\end{eqnarray} 
Finally, by combining the result $H_{2}=0$ together with eqs.\ref{Piece 1 of H1} and \ref{ren_H}, we obtain the form of the rotated $H$
\begin{eqnarray}
U_{N}HU^{\dagger}_{N}&=& \frac{1}{2}Tr_{N}(H)+\tau_{N}Tr_{N}(H\tau_{N})+\tau_{N}\lbrace c^{\dagger}_{N}Tr_{N}(Hc_{N}),\eta_{N}\rbrace~.\label{rot_ham}
\end{eqnarray}
One can easily check that the rotated Hamiltonian $[U_{N}HU^{\dagger}_{N},\hat{\tau}_{N}]=0$, i.e., $\tau_{N}$ is an integral of motion. Turning to the quantum fluctuation operator $\omega$ (eq.\eqref{quantum fluctuation scale}), we note that its eigenvalues represent energy scales for the fluctuations in the occupation number of state $|N\rangle$. 
\par\noindent
We will now put our unitary disentangling tranformation in context with the unitary transformations used in various other RG methods, including continuous unitary transformation (CUT) RG~\cite{glazekWilson1993,glazekWilson1994,wegner1994,savitz2017stable}
strong disorder RG~\cite{rademaker2016explicit,monthus2016flow}                                                                                                                                                                        and spectrum bifurcation RG~\cite{you2016entanglement}. We recall that CUT RG schemes aim, via the iterative application of unitary transformations, to remove off-diagonal entries coupling various energy configurations using a variety of choices for the RG flow generator. The goal is, in this way, to make the Hamiltonian matrix more band-diagonal. Nevertheless, this implementation of the RG in terms of unitary transformations eventually becomes perturbative in nature, as at any given RG step, the rotated Hamiltonian cannot be computed exactly owing to the appearance of an infinite series expansion in the couplings. Instead, an effective Hamiltonian is obtained perturbatively through a truncation of the coupling expansion. This is also true of the recently developed entanglement-CUT RG scheme~\cite{sahin2017entanglement}, where the RG flow of the entanglement content between operators is studied using tensor networks. 
Similarly, in various recent strong disorder RG schemes~\cite{rademaker2016explicit,monthus2016flow}, the generator of transformations is chosen such that certain terms in the Hamiltonian can be dropped. As with the CUT RG, this leads to only the partial disentanglement of a single electronic degree of freedom at any given RG step. Finally, in the spectrum bifurcation RG scheme~\cite{you2016entanglement}, the Hamiltonian is made progressively block diagonal at each RG step via the iterative application of local unitary rotations along with coarse-graining transformations that are perturbative in nature.
\par\noindent
This should be contrasted with the non-local nature of the unitary operations employed in our RG scheme (eq.\eqref{Unitary_op}), that implement non-perturbative coarse-graining transformations through the precise disentanglement of one electronic state at every step.
Further, unlike the RG schemes discussed above, we obtain close-form analytic expressions for the rotated Hamiltonian at every step of the RG transformations.
Finally, our Hamiltonian RG flow evolves across multiple quantum-fluctuation scales, the eigenvalues ($\omega$) of $\hat{\omega}$ eq.\eqref{QFoperator}. This helps obtain effective theories for various subparts of the many-body spectrum. 
\par\noindent
This brings us to an important outcome of our RG transformation scheme $H\to U_{N}HU^{\dagger}_{N}$~:-\emph{if along the RG flow, one of the energy eigenvalues of $\hat{\omega}$ operator matches with an eigenvalue of the diagonal operator $H^{D}$, we obtain a stable fixed point of the RG transformations that is signalled via the vanishing of the off-diagonal blocks in the occupation basis of the electronic state being disentangled at that step.} 
This can be seen by starting from equation eq.\eqref{eh-transition-2}, with $\eta_{N}$ acting on any one of  the eigenstates of the $\hat{\omega}$ operator (say $|\Phi_{1},1_{\mathbf{k}\sigma}\rangle$) with eigenvalue $\omega$  
\begin{eqnarray}
(\omega-Tr_{N}(H^{D}\hat{n}_{N})|\Phi_{1},1_{N}\rangle &=& c^{\dagger}_{N}Tr(Hc_{N})\eta_{N}|\Phi_{1},1_{N}\rangle\nonumber\\
\textrm{Det}(\omega-Tr_{N}(H^{D}\hat{n}_{N}))=0 &\implies & H^{X}_{N}|\Phi_{1},1_{N}\rangle =0~.
\label{quantum_fluc_switch_off}
\end{eqnarray}
This shows that if one of the eigenenergies of $H^{D}$ becomes equal to fluctuation energy scale $\omega$, a stable fixed point is reached due to a vanishing off-diagonal block~\cite{glazek2004}. In what follows we initially discuss various features of this unitary transformation, rotated Hamiltonian and using it propose a Hamiltonian renormalization group scheme. Eventually from there we arrive at a heirarchy of n-point scattering vertex flow equations.
\par\noindent
Note that with the removal of the off-diagonal blocks, the Hamiltonian $H_{(N-1)}$ commutes with $\hat{n}_{N}$, i.e. $[H_{(N-1)},\hat{n}_{N}]=0$, generating a good quantum number ($\hat{n}_{N}$).
 Further, the new Hamiltonian blocks $H^{1_{N}}_{(N-1)}$ and $H^{0_{N}}_{(N-1)}$ Fig.\ref{Hamiltonian_flow_diagram}individually have dimensions ($2^{N-1}\times 2^{N-1}$) halved compared to that of $H$ ($2^{N}\times 2^{N}$). 
We note that the idea behind using unitary block diagonalization transformations to decouple partitions has been discussed in the past mainly in the context of nuclear physics, quantum chemistry\cite{vanVleck1929,Okubo1954,bloch1958-1,bloch1958-2,
bloch1958determination,coester1958,coester1960short,
klein1974degenerate,brandow1967linked,brandow1979formal,
suzuki1983degenerate}. In this work we provide an concrete form for the unitary ansatz that satisfies the decoupling equation eq.\eqref{decoupling condition}.
\par\noindent
For the next RG step, $H_{(N-1)}$ is written in a block representation with respect to the next d.o.f (say, the electronic state $N-1$) and the entire procedure is repeated iteratively. This is represented by the flow diagram Fig.(\ref{Hamiltonian_flow_diagram}a). The iterated block diagonalisation leads to a RG recursion {\it flow} relation for the Hamiltonian
\begin{eqnarray}
H_{(j-1)}=U_{(j)}H_{(j)}U_{(j)}^{\dagger}~,~\forall ~j\leq l\leq N~~~\label{RG flow}
\end{eqnarray}
where $H_{(N)}=H$ is the bare Hamiltonian. As seen earlier, given that $[H_{(j-1)},\hat{n}_{l}]=0$, the eigenvalues of the $\hat{n}_{l}$ operators are associated with a set of integrals of motion that are generated under the RG flow. Another aspect of the unitary RG transformation is that it preserves the Hilbert space of $N$ fermionic states by preserving the canonical anticommutation relation
 \begin{eqnarray}
 \lbrace c^{\dagger(j)}_{i},c^{(j)}_{k}\rbrace =\delta_{ik}~,~  \lbrace c^{(j)}_{i},c^{(j)}_{k}\rbrace = 0~,~\label{canoc-commute}
 \end{eqnarray}
where $c^{\dagger(j)}_{i} = U_{(j)}c^{\dagger(j+1)}_{i}U^{\dagger}_{(j)}$ is the rotated creation operator. Similarly, the unitary operation leads to rotation of the annhilation operator $c^{(j)}_{i}$ and the occupation number operator $\hat{n}^{(j)}_{i}$. These rotated operators form the Pauli group $P^{(j)}$, composed of the direct product of all possible $4^{N}$ combinations of N matrices, where for every label $j$ there are $4$ matrices: $I_{i}$, $\tau^{(j)}_{i}=\hat{n}^{(j)}_{i}-\frac{1}{2}$, $\tau_{x,i}^{(j)}=\frac{1}{2}(c^{\dagger(j)}_{i}+c_{i}^{(j)})$ and $\tau_{y,i}^{(j)}=\frac{i}{2}(c^{\dagger(j)}_{i}-c_{i}^{(j)})$~:
\begin{eqnarray}
P^{(j)} &=& \lbrace I,\lbrace \tau^{(j)}_{x,1},\tau^{(j)}_{y,1},\tau^{(j)}_{1},...,\tau^{(j)}_{x,N},\tau^{(j)}_{y,N},\tau^{(j)}_{N}\rbrace,\nonumber\\
&&\lbrace \tau^{(j)}_{x,1}\tau^{(j)}_{x,2},\underbrace{...}_{\binom{N}{2} \text{elements}},\tau^{(j)}_{x,N-1}\tau^{(j)}_{x,N}\rbrace,\lbrace ...xy...\rbrace,\nonumber\\
&&\lbrace ...xz...\rbrace,\lbrace ...yy...\rbrace, \lbrace ...yz...\rbrace,\lbrace ...zz...\rbrace ,...\rbrace ~. \label{pauligroup}
\end{eqnarray}
The set of operators denoted $\lbrace...xy...\rbrace$ is the collection of product of the $\tau^{(j)}_{x,l}$ and $\tau^{(j)}_{y,m}$ Pauli matrices for a pair of electronic states; the other $\lbrace ...\rbrace$ set of operators are defined similarly.
\pin
The rotated operators composing the Pauli group at step $j$($P^{(j)}$) are emergent from the Pauli group of the earlier step ($P^{(j+1)}$): $P^{(j)}=U_{(j+1)}P^{(j+1)}U_{(j+1)}^{\dagger}$,  where the $U_{(j)}$ non-local unitary operations are generalized Clifford group transformations. We recall that a restricted class of Clifford group rotations~\cite{gottesman1998heisenberg}, employed in the spectrum bifurcation RG (SBRG) of Ref. \cite{you2016entanglement}, rotates one Pauli operator to another while preserving the Pauli group $P$. Together with the Schrieffer-Wolff transformations (see Ref.\cite{bravyi2011schrieffer} a review), such restricted Clifford transformations are observed to remove the entanglement content of the Hilbert space only approximately. On the other hand, the general Clifford group transformations ($U_{(j)}$) presented above cause non-local rotations in the many-body space of tensor product of Pauli operators, leading to the exact removal of entanglement between a given state $j$ and the rest of the coupled states. This leads to the emergence of conserved quantities, together with the morphing of entanglement content within the coupled space.
As the states have been labelled in ascending order of energy, the iterative decoupling of states and the concomitant renormalization of the effective Hamiltonians represents a flow from the utraviolet (UV) to the infrared (IR).
We recall that the strategy of repeatedly decoupling one fermion state at every step was independently proposed by Choi~\cite{choi1969treatment} and Wilson~\cite{Wilson1970} for treating degeneracies within Rayleigh-Schroedinger perturbation theory and in the formulation of a Hamiltonian RG method applied to the meson-nucleon interaction problem respectively.
\begin{figure}
\centering
\includegraphics[width=0.55\textwidth]{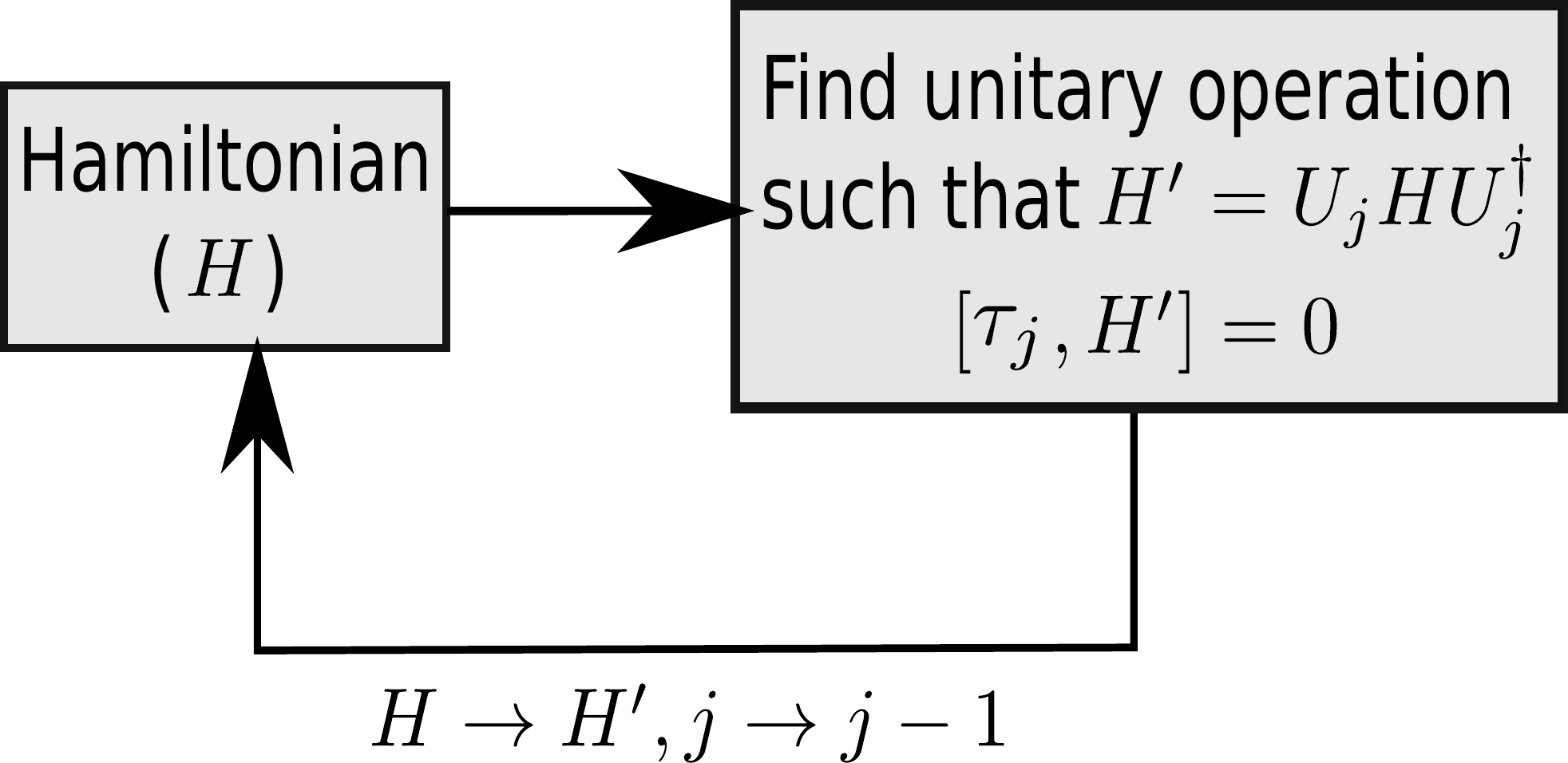}
\caption{Schematic representation of the Hamiltonian RG flow. The feedback loop $j\to j-1$ depicts the iterative RG process. Each RG step involves the creation of one integral of motion $\tau_{j}$ that commutes with the rotated Hamiltonian.}\label{HamRGFlow}
\end{figure}
\par\noindent
The form of the unitary operator for an arbitrary RG step $j$ is given similar to eq.\eqref{Unitary_op}
\begin{eqnarray}
U_{(j)}=\frac{1}{\sqrt{2}}[1+\eta_{(j)}-\eta^{\dagger}_{(j)}],~\lbrace\eta^{\dagger}_{(j)},\eta_{(j)}\rbrace =1,\label{Unitary operator}~~~~
\end{eqnarray}
Here, $\eta_{(j)}$ and $\eta^{\dagger}_{(j)}$ are the electron-hole (e-h) transition operators associated with the $j$th RG step, they possess the usual anti-commutation relation as well as satisfy the following relations
\begin{eqnarray}
(1-\hat{n}_{j})\eta_{(j)}\hat{n}_{j}&=&\eta_{(j)}~,~\hat{n}_{j}\eta_{(j)}(1-\hat{n}_{j})=0\nonumber\\
\eta_{(j)}^{2}&=&0~,~\left[\eta^{\dagger}_{(j)},\eta_{(j)}\right]=2\hat{n}_{j}-1~.\label{eta-operator-rel}
\end{eqnarray}
It can be easily seen that the operators $\eta_{(N)}$ and $\eta^{\dagger}_{(N)}$ defined in eqs.(\ref{eh-transition-1}, \ref{eh-transition-2}) satisfies the above properties.
Using the above relations we obtain an simplified form for the Hamiltonian flow equation eq.\eqref{RG flow},
\begin{eqnarray}
\hat{H}_{(j-1)} &=&\frac{1}{2}Tr_{j}(\hat{H}_{(j)})+Tr_{j}(\hat{H}_{(j)}\tau_{j})+\tau_{j}\lbrace c^{\dagger}_{j}Tr_{j}(H_{(j)}c_{j}),\eta_{(j)}\rbrace~.~~~~\label{Hamiltonian_renormalization}
\end{eqnarray}
Here, the off-diagonal term $c^{\dagger}_{j}Tr_{j}(H_{(j)}c_{j})$ is a compact representation of the multi-particle scattering terms via which the state labeled $j$ is coupled to the remaining $j-1$ d.o.f. Importantly, as shown in Appendix~\ref{block matrix}, this compact representation of the off-diagonal block respects the signatures arising from fermion exchanges.  
It is evident from the expression eq.\eqref{Hamiltonian_renormalization} that the renormalization of the Hamiltonian blocks marked by the two values of the occupation number of the $j$th state ($\hat{H}^{1_{j}}_{(j-1)}$ and $\hat{H}^{0_{j}}_{(j-1)}$) take place in an opposite manner. Further, the approach makes manifest the renormalization of the Hamiltonian as an outcome of the mixing between UV and IR d.o.fs
~\cite{minwalla2000noncommutative}. 
\par\noindent
It is easily seen from Fig.(\ref{Hamiltonian_flow_diagram}) that the Hamiltonian spectrum of $H_{(j)}$ splits into two branches via the unitary operation $U_{(j)}$. This branching is a many-body analog of the avoided level-crossing mechanism. 
At the RG step $j$, there are $2^{j}$ decoupled blocks marked by the occupation numbers of decoupled states and
every individual block has $2^{N-j}$ coupled many-body configurations. We note that a similar approach to an iterative decoupling of states has been investigated in the context of spectrum bifurcation RG (SBRG)~\cite{you2016entanglement}. 
In contrast to our method, however, the effective Hamiltonian generated by the SBRG method involves a perturbative treatment of inter-particle interactions.
\\
\par\noindent{\bf \textit{Form of the Unitary operator $U_{(j)}$}} 
\pin 
At this point, we compare the form of the Unitary operator $U_{j}$ with those that have appeared in the literature. First, $U_{(j)}$ can be cast in the familiar Van Vleck form~\cite{vanVleck1929} 
\begin{equation}
U_{(j)}=\exp[\frac{\pi}{4}(\eta^{\dagger}_{(j)}-\eta_{(j)})]~,
\label{formofunitaryoperator}
\end{equation}
providing a geometric representation of a rotation angle (whose value here is $\pi/4$) between the old and new configuration subspaces~\cite{kvaal2008geometry}. This $\pi/4$ unitary gate has the property: $(U_{(N)})^{4}=-I$. In this form, it can be compared to the unitary exponential wave operator investigated in the context of coupled cluster theory~\cite{westhaus1973cluster,mukherjee1975correlation,
kutzelnigg1977pair,pal1982certain}. 
We further note that the form eq.\eqref{Unitary operator} is a simplified version of the canonical transformation advocated first by Shavitt and Redmond~\cite{shavitt1980quasidegenerate} (see also~\cite{westhaus1981connections,suzuki1982construction}).  Unlike the wave operator of coupled cluster theory, however, it is clear from the relations given earlier in eqs.(\ref{eta-operator-rel}) that the unitary operator in eq.\eqref{Unitary operator} does not involve a infinite series expansion.  
\par\noindent
Recently, in the context of the phenomenon of many-body localisation in systems with both strong correlations as well as disorder (see, e.g., \cite{pal2010many}), versions of the strong disorder RG~\cite{ma1979sk,dasgupta1980low,aoki1982decimation,fisher1992random,fisher1994random,
 fisher1995critical,fisher1999phase} have been proposed. These involve a RG flow arising from the diagonalization of Hamiltonians via the iterative application of unitary operators~\cite{rademaker2016explicit,rademaker2017many,monthus2018strong,monthus2018many,
 monthus2017many}. 
A similar renormalisation scheme for Hamiltonians had also been proposed earlier by White~\cite{white2002numerical} in the context of the quantum chemistry of molecular clusters. These schemes rely essentially on Jacobi's method for the iterative diagonalisation of Hermitian matrices~\cite{jacobi1846ueber}. This involves the application of a unitary displacement operator which removes the largest off-diagonal element in the Hamiltonian matrix: 
$D=\exp(i\lambda(X^{\dagger}-X))$, where $\lambda$ is the displacement angle and the operator $X$ satisfies the properties: $X^{2}=0$, $XX^{\dagger}X=X$. A comparison with our unitary operator reveals that the $\eta_{(j)}$ transition operator given earlier satisfies the same properties as $X$, as well as certain others (see eqs.\ref{eta-operator-rel})). This leads to an important difference in terms of the results obtained from the two approaches. The strong-disorder style RG schemes leads to an effective Hamiltonian at low energies that is an expansion in terms of products of the emergent local integrals of motion ($\tau_{i}$) that are generated, i.e., $H_{eff} = \sum_{i}\epsilon_{i}\tau_{i} + \sum_{ij}V_{ij}\tau_{i}\tau_{j} + \sum_{ijk}V_{ijk}\tau_{i}\tau_{j}\tau_{k}+\ldots$. On the other hand, we will see below that the effective Hamiltonian reached from our RG scheme involves only a small number of such terms, as the rest can be shown to be irrelevant under the RG transformations. Another important difference is that in our RG, the iterative decoupling procedure can be inhibited if a stable fixed point theory (involving residual occupation number quantum fluctuations due to the remaining coupled states) is attained. This leads to breakdown of the adiabatic continuity between the states at higher and lower energies, signalling instead the emergent condensation of composite degrees of freedom at low energies. This, as we shall see later, is a non-perturbative outcome of the UV-IR mixing. In this case, the local integrals of motion will not form a complete set. 
\par\noindent
It is also worth noting the efforts of Wilson~\cite{Wilson1970}, who attempted the partitioning of the Hamiltonian into low and high energy subspaces. To decouple the blocks at a particular step of the RG flow, Wilson proposed an operator $R$ that mixes states between the two subspaces. The effective Hamiltonian then obtained posseses a Bloch-Brandow form~\cite{bloch1958-1,bloch1958-2,brandow1967linked}
\begin{eqnarray}
H_{eff}&=& \frac{1+R-R^{\dagger}}{\sqrt{1+R^{\dagger}R+RR^{\dagger}}}H\frac{1+R^{\dagger}-R}{\sqrt{1+R^{\dagger}R+RR^{\dagger}}}~,
\end{eqnarray}
which can be seen as the action of a unitary transformation
$U_{W}=\exp(\text{arctanh}(R-R^{\dagger}))$ on $H$~\cite{vanVleck1929,Okubo1954,shavitt1980quasidegenerate,westhaus1973cluster,
suzuki1982construction,kvaal2008geometry}~:~$H_{eff}=U_{W}HU_{W}^{\dagger}$.
\par
The $\eta_{(N)}$ transition operator obtained by us is analogous to Wilson's $R$ operator, with the difference that, in our case, the partitioning (or decoupling procedure) is carried out in the fermion occupation number (i.e., Fock) basis  
of the state $N$. The relations eq.\eqref{eta-operator-rel} then allow for a simplification of both the unitary operator as well as the effective Hamiltonian at every step of the RG procedure. In Appendix \ref{URGconnectionCUT}, we will also elaborate on how to recast our RG scheme in terms of a sequence of continuous unitary transformations~\cite{glazekWilson1993,glazekWilson1994,wegner1994}. 
\begin{figure}
\centering
\includegraphics[scale=0.45]{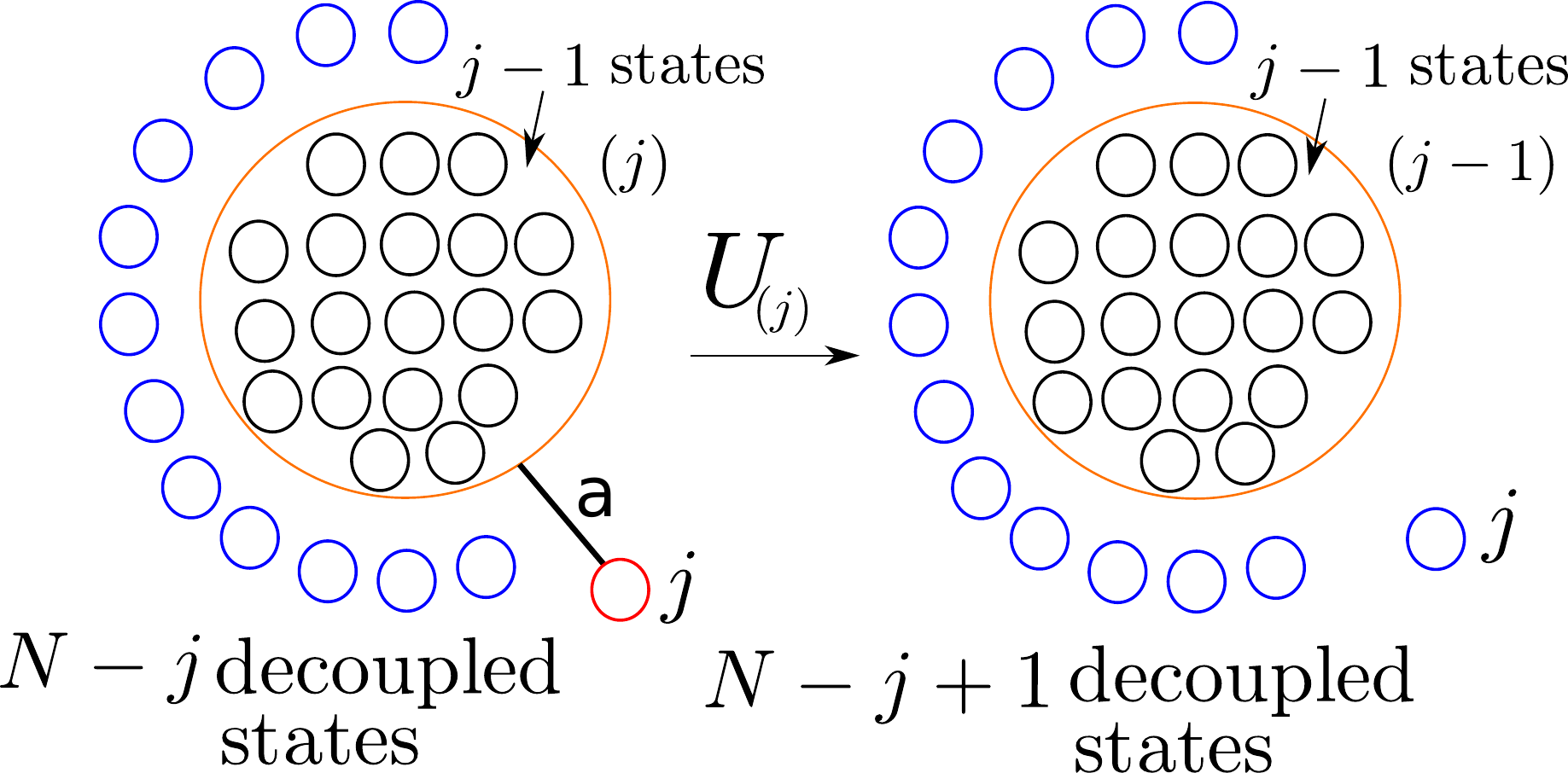}
\caption{Disentangling the state labeled $j$ via the unitary operation $U_{(j)}$. The black circles represent the remnant coupled single-particle states, red circles the states to be decoupled, and the blue circles represent the states that are already decoupled. The black line marked `a' represents the entanglement between the states $\lbrace 0, j-1\rbrace$ and the state $j$ that is to be removed in the RG step $j$.}\label{decoupling states}
\end{figure}
\par\noindent\\{\bf \textit{Form of the e-h transition operator} $\eta_{(j)}$}\\ 
As derived in eq.\eqref{eh-transition-1} and eq.\eqref{eh-transition-2} the solution to the decoupling equation eq.\eqref{decoupling condition} for the $j$th RG step determines the form of transition operator $\eta_{(j)}$ 
\begin{eqnarray}
\eta_{(j)}  = Tr_{j}(c^{\dagger}_{j}H_{(j)})c_{j}\frac{1}{\hat{\omega}_{(j)} -Tr_{j}(H^{D}_{(j)}\hat{n}_{j})}~.\label{e-h transition operator}
\end{eqnarray}  
Here, the operator $\hat{\omega}_{(j)}$ is associated with the $2^{j-1}$ energy eigenvalues that represent the \textit{quantum fluctuation} energyscales of the $\lbrace 1\ldots j-1 \rbrace$ coupled single-particle states having $2^{j-1}$ occupation number 
configurations (see fig(\ref{decoupling states})). There are also $2^{N-j+1}$ number diagonal configurations associated with the subspace of states (labelled $\lbrace j+1,\ldots , N\rbrace$) that have been decoupled under the RG and are no longer affected by quantum fluctuations. The  
complete configuration space comprises, thereby, of $2^{N-j+1}2^{j-1}=2^{N}$ many body states.
\par
\begin{figure}[h!]
\centering
\includegraphics[width=0.5\textwidth]{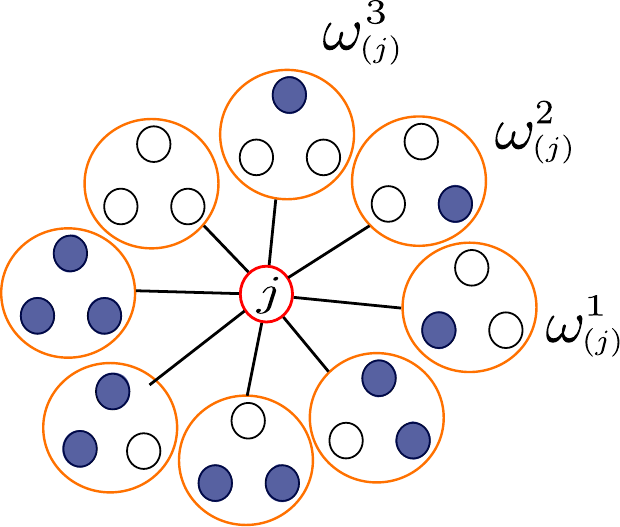} 
\caption{An example describing the quantum fluctuations originate from the coupling of 3 electronic states with the state $j$ via off-diagonal Hamiltonian blocks. The fluctuation operator $\hat{\omega}$ then describes the energy scales $\omega^{1}_{(3)},...,\omega^{8}_{(3)}$ for occupation number fluctuations generated about the $2^{3}=8$ number diagonal configurations of the coupled states, with blue/white filled circles representing occupied/unoccupied states respectively.}\label{multireference_energy_scales}
\end{figure}

\noindent 
The quantum fluctuations originating from scattering processes, $Tr_{j}(c^{\dagger}_{j}H_{(j)})c_{j}$, occupy off-diagonal terms in the Hamiltonian matrix represented in the occupation number basis of single-particle states. The term $Tr_{j}(c^{\dagger}_{j}H_{(j)})c_{j}$ involves various 1-particle, 2-particle and higher correlated scattering events of the state $j$ with the degrees of freedom $\lbrace 1,\ldots, j-1\rbrace$. Such an expansion of the off-diagonal part of the Hamiltonian  
is similar to the cluster expansion employed in the coupled cluster methods~ \cite{coester1958,coester1960short}. For example, in the case of the generic four-Fermi interaction model for a system of $N$ spinless fermions 
\begin{eqnarray}
H^{4}= \sum_{ij}h_{ij}[c^{\dagger}_{i}c_{j}+h.c.] + \sum_{ijkl}V_{ijkl}c^{\dagger}_{i}c^{\dagger}_{j}c_{k}c_{l}~,\label{fourspinlessfermimodel}
\end{eqnarray} 
the partitioning in the occupation number representation of the state labelled $N$ (i.e, the first RG step) gives rise to an off-diagonal block represented as the sum of 1 and 2-particle scattering vertices 
\begin{eqnarray}
{\textrm Tr}_{N}(c^{\dagger}_{N}H^{4}_{(N)})c_{N}=\lbrace\sum_{i}h_{iN}c^{\dagger}_{i}+\sum_{ijk}V_{ijkN}c^{\dagger}_{i}c^{\dagger}_{j}c_{k}\rbrace c_{N}~.\label{expression_of_transition}
\end{eqnarray}
In Appendix \ref{block matrix}, we show that this compact representation of the off-diagonal block using partial trace operations with respect to a given single-particle state respects signatures arising from fermion exchanges. The operator $Tr_{N}(H^{D}_{(j)}\hat{n}_{j})$ appearing in the transition operator eq.\eqref{e-h transition operator} refers to all the $n$-particle self energies in the number occupation subspace ($I\otimes \hat{n}_{j}$) of state $j$, and appear along the diagonal of the Hamiltonian. Again, for the case of four Fermi Hamiltonian ($H^{4}$) 
\begin{equation}
Tr_{N}(H^{4,D}_{(N)}\hat{n}_{j})=\sum_{i}h_{i}n_{i}+\sum_{ij}V_{ij}n_{i}n_{j}~,
\end{equation}
i.e., it contains the one-particle self-energy and two-particle correlation energies. 
\par\noindent\\
{\bf \textit{Origin of the quantum fluctuation scales $\hat{\omega}_{(j)}$}}\label{QF scales}\\
The Hamiltonian can, very generally, be decomposed into three parts 
\begin{equation}
H_{(j)}=H^{D}_{(j)}+H^{X,j}_{(j)}+H_{(j)}^{X,\bar{j}}~,
\end{equation}
where the number diagonal part of the Hamiltonian ($H^{D}_{(j)}$) is associated with n-particle self energies,  
while the term 
$H_{(j)}^{X,j}$ represents quantum fluctuations in the occupation number of state $j$, i.e., $n$-particle scattering induced coupling between the state $j$ and the other $\lbrace 1,\ldots, j-1\rbrace$ states (shown via the black line labelled by `a' in Fig.(\ref{decoupling states})). Finally, the term $H_{(j)}^{X,\bar{j}}$ represents coupling among the other degrees of freedom $\lbrace 1,\ldots,j-1\rbrace$ (shown via the orange circle in Fig.(\ref{decoupling states})). 
Following eq.\eqref{superposition} and eq.\eqref{transition-eqns} an explicit form for the quantum fluctuation operator $\hat{\omega}_{j}$ can be derived 
\begin{eqnarray}
\hat{\omega}_{(j)}=H_{(j-1)}^{D}+\Delta H_{(j)}^{X,\bar{j}}~.\label{quantum fluctuation scale}
\end{eqnarray}
The two components of $\hat{\omega}_{j}$ encode the renormalisation of different aspects of the remaining coupled degrees of freedom $\lbrace 1, j-1\rbrace$: the first ($\Delta H_{(j)}^{X,\bar{j}}= H_{(j-1)}^{X,\bar{j}}-
H_{(j)}^{X,\bar{j}}$) corresponds to the renormalization of various scattering vertices, 
and the second ($H_{(j-1)}^{D}$) to the renormalization of various $n$-particle self energies. 
\pin
The operator $\hat{\omega}_{(j)}$ has $2^{j-1}$ eigenvalues ($\omega_{(j)}^{i}$) corresponding to the $2^{j-1}$ occupation number configurations for the remaining $j-1$ coupled single-electron states, and these are determined as the RG proceeds. Fig.(\ref{multireference_energy_scales}) illustrates the $8$ configurations of three electronic states labelled by the associated quantum fluctuation energyscales $\omega_{(j)}^{i}$ of a 4th state's (labelled $j$). In this way, the renormalisation procedure outlined above is a multireference method, i.e., the RG steps resolve the multiple energyscales for quantum fluctuations in an iterative fashion.
\par\noindent
From the form of the operator $\hat{\omega}_{(j)}$, we can also obtain its RG flow equation
\begin{eqnarray}
\Delta\hat{\omega}_{(j)}=\Delta H^{D}_{(j-1)}+\Delta H^{X,\bar{j}}_{(j-1)}-\Delta H^{X,\bar{j}}_{(j)}.
\end{eqnarray}
This flow equation naturally encodes the interplay of the RG dynamics of the Hamiltonian (seen from the second order discrete derivative of the off-diagonal component's $H^{X,\bar{j}}$) with the RG dynamics of the quantum fluctuations ($\omega$). In this way, we realise the bulk of the EHM as manifestly possessing non-trivial quantum as well as RG dynamics~\cite{lee2010,kskim2017,kskim2019}.
\par\noindent
Another important outcome of the decoupling procedure can now be understood. The vanishing of the off-diagonal scattering processes via, say, 
\begin{equation}
H^{X,j,0\to 1}_{(j*)}|\Psi_{(j),0_{j}}\rangle=0
\label{quantum_fluc_switch_off1}
\end{equation}
at RG step $j$ and fluctuation scale $\omega_{(j)}^{i}=E^{i}_{(j),1_{j}}$  
leads to $E^{i}_{(j),1_{j}}$ becoming the exact eigenvalue of $H^{D}_{(j),1_{j}}$ containing all $n$-particle self energies of the state $j$. No further decoupling of states can take place at this energyscale, signalling that a fixed point of the RG transformations has been reached. 
Indeed, this is the many-body analog of the fixed point condition obtained from the non-perturbative similarity RG flow by Glazek and Wilson~\cite{glazek2004} for the case of a simple discrete Hamiltonian. 
\\
\par\noindent
{\bf \textit{Spectral decomposition of $\hat{\omega}_{(j)}$ and $\eta_{(j)}$}}\label{spectral_decomposition}\\ 
The operator $\hat{\omega}_{(j)}$ can be given a spectral representation (eq.\eqref{QFoperator}),
  \begin{eqnarray}
 \hat{\omega}_{(j)}=\sum_{i=1}^{2^{j-1}}\omega_{(j)}^{i}\hat{O}_{(j)}(\omega_{(j)}^{i})~,
 \label{quantum fluctuation eigenvalue}
 \end{eqnarray}  
where $\hat{O}_{(j)}(\omega_{(j)}^{i})$  is a projection operator that projects onto one among the $2^{j-1}$ eigenstates of the operator~$\hat{\omega}_{(j)}$. This eigenstates represent many body entangled configurations of the $\lbrace 1,j-1\rbrace$ single-particle states.  With this, a spectral representation can also be found for the e-h transition operator,
\begin{eqnarray}
\eta^{\dagger}_{(j)} &=& \sum_{i}\eta^{\dagger}_{(j)}(\omega_{(j)}^{i})\hat{O}_{(j)}(\omega_{(j)}^{i})~,~\nonumber\\
\eta^{\dagger}_{(j)}(\omega_{(j)}^{i}) &=& \frac{1}{\omega^{i}_{(j)} -Tr_{j}(H^{D}_{(j)}\hat{n}_{j})}c^{\dagger}_{j}Tr_{j}(H_{(j)}c_{j})~
\label{e-h transition operator1}
\end{eqnarray} 
Thus, $\eta^{\dagger}_{(j)}(\omega_{(j)}^{i})$ represents the collection of all scattering processes that takes the electronic state $j$ between the unoccupied and occupied configuration, in turn causing quantum fluctuations involving many-body configurations of the coupled $\lbrace 0,j-1\rbrace$ single-particle states (Fig.\ref{multireference_energy_scales}). The term $(\omega^{i}_{(j)} -Tr_{j}(H^{D}_{(j)}\hat{n}_{j}))^{-1}$ represent the operator Green's function for the diagonal part of the renormalised Hamiltonian ($H^{D}_{(j)}$). As $H^{D}_{(j)}$ is in general a sum of the products of number diagonal operators ($n_{i}-1/2$) with lengths lying between $1$ and $N$, we will see below that this Green's function plays a key role in determining the heirarchy of RG equations for members of the $n$-particle interaction vertices (eq.\eqref{RG_flow_heirarchy}).
\par\noindent
We can also define 
a wave-operator $\Omega_{(j)}=\exp(-\eta^{\dagger}_{(j)})$, such that   
$|\Psi^{1_{j},i}_{(j)}\rangle = \Omega_{(j)}|\Psi_{(j),i}\rangle$~.  
This wave-operator is an equivalent non-unitary transformation that can be employed in solving the decoupling equations and obtaining the effective Hamiltonian~\cite{suzuki1983degenerate}. 
A similar wave-operator appears in the generalized Bloch equations and multireference coupled-cluster methods~\cite{lindgren1987connectivity,mukherjee1989use,mavsik1998multireference} for solving the decoupling equations between two subspaces of many-body configurations.
\par\noindent
The spectral decomposition of $\hat{\omega}$ can be used to recast the decoupling equations eq.\eqref{transition-eqns} in the form employed in multireference Brillouin-Wigner perturbation theory~\cite{mavsik1998multireference,hubavc2010brillouin}
\begin{eqnarray}
|\Psi_{(j),1}(\omega_{(j)}^{i}),1_{j}\rangle &=&\eta^{\dagger}_{(j)}(\omega_{(j)}^{i})|\Psi_{(j),0},0_{j}\rangle~,~\nonumber\\
|\Psi_{(j),0}(\omega_{(j)}^{i}),0_{j}\rangle &=&\eta_{(j)}(\omega_{(j)}^{i})|\Psi_{(j),1},1_{j}\rangle~.
\end{eqnarray}
The multireference nature of the approach allows for the treatment of dynamical fluctuations associated with the $2^{j-1}$ occupation number configurations of the $\lbrace 1,j\rbrace$ coupled states, i.e.,  
it enables the exploration of ground state as well as excited states of the Hamiltonian spectrum ~\cite{wenzel1992basis,wenzel1996multireference,mavsik1997multireference}. 
An important distinction can now be made. In the standard Brillouin-Wigner method, the denominator of the frequency resolved e-h transition operator, $\eta^{\dagger}_{(j)}(\omega_{(j)}^{i})$ contains a many-body Hamiltonian with off-diagonal terms (of the kind indicated by $H_{X}$ in our formalism). This leads to a Dyson series-like expansion of the propagator, whose truncation makes the approach perturbative. On the other hand, we have recast the denominator in $\eta^{\dagger}_{(j)}(\omega_{(j)}^{i})$ such that it contains only the number diagonal many-body operator $H^{D}_{(j)}$ at every RG step. This resolves the problem associated with the inversion of a large many-body matrix, allowing the closed form derivation of flow equations along multiple many-body configuration channels without the need for truncation of any series expansion. 
\par\noindent 
We use the spectral representations of $\hat{\omega}_{(j)}$ and $\eta^{\dagger}_{(j)}$ to rewrite the Hamiltonian flow equation eq.\eqref{Hamiltonian_renormalization} generated via URG
\begin{eqnarray}
\Delta\hat{H}_{(j)} &=&\sum_{i=1}^{2^{j-1}}\Delta \hat{H}_{(j)}(\omega_{(j)}^{i})\hat{O}_{(j)}(\omega_{(j)}^{i})~,~\nonumber\\
\Delta \hat{H}_{(j)}(\omega_{(j)}^{i}) &=&\left(\hat{n}_{j}-\frac{1}{2}\right)\lbrace c^{\dagger}_{j}Tr_{j}(H_{(j)}c_{j}),\eta_{(j)}(\omega_{(j)}^{i})\rbrace~.
\label{specrepofHamRG}
\end{eqnarray}
The dependence of $\omega_{(j)}^{i}$ in $\Delta \hat{H}_{(j)}(\omega_{(j)}^{i})$ reflects retardation effects in the effective Hamiltonian at multiple energyscales, and date back to the early work of Breit~\cite{breit1929effect}. Frequency dependent effects have recently also been taken account of within the functional RG approach to strongly correlated condensed matter systems~\cite{uebelacker2012self}, as well as in QCD in the form of the dynamical renormalisation group formalism of Refs.\cite{boyanovsky1999dynamical,boyanovsky2001non,boyanovsky2003dynamical}. 
\par
\begin{figure}
\centering
\includegraphics[width=0.5\textwidth]{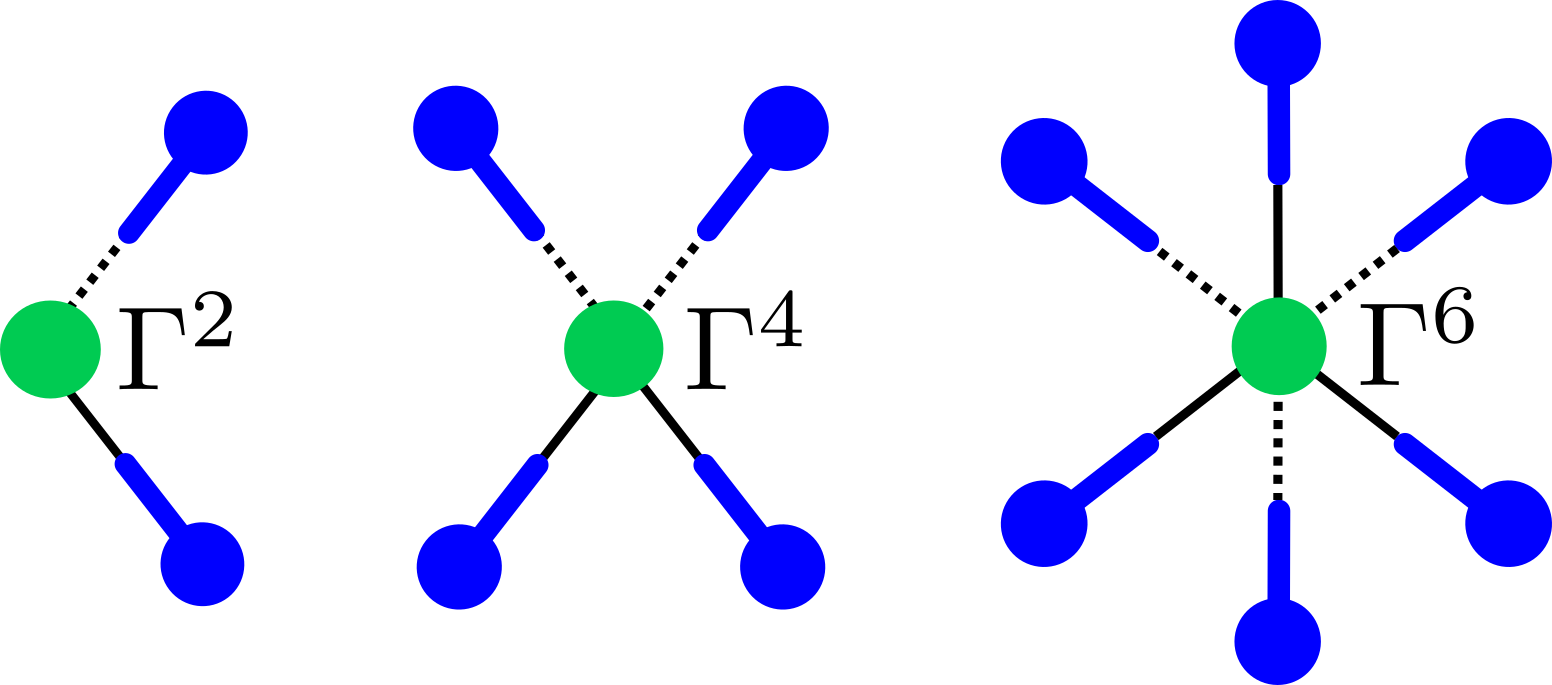}
\caption{The 2-point $\Gamma^{2}$ , 4-point $\Gamma^{4}$ and 6-point $\Gamma^{6}$ scattering vertex tensors. The blue legs represent electronic states, the dashed/solid line represents the outgoing/incoming electrons.}\label{vertextensors}
\end{figure}

\noindent
A multireference formalism leading to effective Hamiltonians at various energy scales deserves to be contrasted with the single-reference Wilsonian approach to RG: there, we typically obtain only the effective Hamiltonian at low energies via the application of projection operators.
In our formalism, the lowest energy state for the effective Hamiltonian $H_{(j)}(\omega_{(j)}^{i})$ 
can obtained from the asymptotic imaginary time ($\tau$) evolution operation $\exp(-\tau H_{(j)}(\omega_{(j),i}))$ on any arbitrary state $|\Psi(\omega_{(j),i})\rangle$ belonging to its Hilbert space~\cite{fradkin2013field,lindgren2006many}
\begin{eqnarray}
|\Psi^{0}_{(j)}(\omega_{(j)}^{i})\rangle = \lim_{\tau\to \infty}\exp(-\tau H_{(j)}(\omega_{(j)}^{i}))|\Psi(\omega_{(j)}^{i})\rangle.~~~\label{lowest_State}
\end{eqnarray} 
As a fixed point is reached for the Hamiltonian RG flow (i.e., $H_{(j)}(\omega_{(j)}^{i})\rightarrow H_{(j)}(\omega_{(j^{*})}^{i})$, via the condition eq.\eqref{quantum_fluc_switch_off}), the lowest energy state for that spectrum (eq.\eqref{lowest_State}) $|\Psi^{0}_{(j)}(\omega_{(j)}^{i})\rangle$ is also obtained. In this way, we can explore the spectrum around any fluctuation scale $\omega_{(j)}^{i}$. 
\par\noindent\\ \label{cluster_decomposition}
{\bf \textit{Cluster expansion and hierarchy of RG flow equations}}\\
The Hamiltonian operator's diagonal $(H^{D}_{(j)})$ and off-diagonal $(H^{X}_{(j)})$ parts at a given RG step $(j)$ can be written as a closed-form cluster decomposition
\begin{eqnarray}
H_{(j)}&=& H^{D}_{(j)}+H^{X}_{(j)}.\label{D-X decomposition}\\
H^{D}_{(j)}&=& \sum_{i=1}^{2^{j-1}}\sum_{n=1}^{N}\sum_{\alpha,\alpha'}\lbrace\tilde{c}^{\dagger}_{\alpha}\frac{\Gamma^{2n}_{\alpha\alpha'}(\omega^{i})}{2^{n}}\tilde{c}_{\alpha'}\hat{O}(\omega^{i})\rbrace_{(j)}~,\nonumber\\
H^{X}_{(j)} &=& \sum_{i=1}^{2^{j-1}}\sum_{n=1}^{a^{max}_{j}}\sum_{\alpha,\beta}\lbrace\tilde{c}^{\dagger}_{\alpha}\Gamma^{2n}_{\alpha\beta}(\omega^{i})\tilde{c}_{\beta}\hat{O}(\omega^{i})\rbrace_{(j)}~.\label{cluster decomposition}
\end{eqnarray}
We now clarify the various terms and notations appearing in these equations. The index $i$ is associated with the $2^{j-1}$ configurations of the coupled $j-1$ single-particle states. The index $n$ labels the various $n$-particle (or $2n$-point) scattering process, and runs from $1$ to $N$ for the contributions to $H^{D}_{(j)}$ for a system of $N$ particles. On the other hand, $n$ runs from $1$ to an upper limit $a_{j}^{max}$ (indicating the highest off-diagonal $n$-particle vertex possible at the RG step $j$, as discussed in detail in Appendix \ref{n-particle vertex highest}) for the contributions to $H^{X}_{(j)}$. The indices $\alpha, \alpha'$ and $\beta$ are defined as follows: $\alpha:=\{(l,\mu)\}$ is an ordered set of $n$ pairs of indices $l$ and $\mu$, where the index $0\leq l \leq j$ (the subspace of coupled single-particle states) and $\mu=0,1$ refers to either unoccupied ($0$) or occupied ($1$) state. Thus, the operator $\tilde{c}^{\dagger}_{\alpha}$ refers to a string of $n$ fermionic creation and annihilation operators for the $0,\ldots,n$ states in product. The index $\alpha':=\{(l,\bar{\mu})\}$, where $0\leq l \leq j$ once again but $\bar{\mu}$ is the complement of $\mu$, i.e., $\bar{\mu}=1,0$ refers to unoccupied ($1$) and occupied ($0$). Thus, by construction, the operator $\tilde{c}_{\alpha'}$ also refers to a string of $n$ fermionic creation and annihilation operators for the $0,\ldots,j$ states in product, and which are the complement to the string given by $\tilde{c}^{\dagger}_{\alpha}$. In this way, the product $\tilde{c}^{\dagger}_{\alpha}\tilde{c}_{\alpha'}$ defines a product of $n$ number operators. Thus, the vertex function associated with this term, $\Gamma^{2n}_{\alpha\alpha'}(\omega^{i})$, denotes the magnitude of the diagonal 
$n$-particle correlation energy $H^{D}_{(j)}$. The presence of the projection operator $\hat{O}$ ensures a sum over each member of $2^{j-1}$ many-body configurations. The closed form representation of $H$ eq.\eqref{cluster decomposition} can be interpreted as a tensor network formed from the $2n$-point vertex tensors. As shown in Fig.\ref{vertextensors}, the node of each such vertex tensor $\Gamma^{2n,(j)}_{\alpha\alpha'}(\omega^{i})$ represents the scattering process, while the blue legs of the tensors represents the electronic states. The black dashed/solid edges represents the outgoing/incoming electronic states respectively. The number $2n$ is the total number of incoming and outgoing lines.
\par\noindent
The sum over the indices $\alpha$ and $\alpha'$ is an anti-symmetrised summation over the indices $\mu$ and $\bar{\mu}$, as can be seen from the following. Under interchange of $\alpha$ and $\alpha'$, 
$\Gamma^{2n}_{\alpha\alpha'}$ satisfies the relation
\begin{equation}
\Gamma^{2n}_{\alpha\alpha'} = (-1)^{\sum_{i=1}^{n}(\bar{\mu}_{i}-\mu_{i})}\Gamma^{2n}_{\alpha'\alpha}~.
\end{equation}
This allows the recasting of $H^{D}_{(j)}$ as a sum over various $n$-particle vertex terms
\begin{equation}
\tilde{c}^{\dagger}_{\alpha}\frac{\Gamma^{2n}_{\alpha\alpha'}(\omega^{i})}{2^{n}}\tilde{c}_{\alpha'} = \Gamma^{2n}_{\alpha\alpha'} \prod_{s=1}^{n}\tau_{l_{s}}~,
\end{equation}
where $ \tau_{l_{s}}= (\hat{n}_{l_{s}}-\frac{1}{2})$ is the occupation number operator defined in an electron-hole symmetric fashion, and $0\leq l_{s}\leq j$. We illustrate this representation in an example of a 2-particle correlation energy
 \begin{eqnarray}
 &&\hspace*{-0.5cm}\Gamma^{4}_{\alpha\alpha'}\tau_{l_{1}}\tau_{l_{2}} = \frac{1}{4}\Gamma^{4}_{\alpha\alpha'}\bigg[\hat{n}_{l_{1}}\hat{n}_{l_{2}}-\hat{n}_{l_{1}}(1-\hat{n}_{l_{2}})-(1-\hat{n}_{l_{1}})\hat{n}_{l_{2}}+(1-\hat{n}_{l_{1}})(1-\hat{n}_{l_{2}})\bigg].~~~
 \end{eqnarray}
\par\noindent
In the same way, the index $\beta:=\{(l',\mu')\}$ refers to an ordered set of $n$ pairs of indices $l'$ and $\mu'$, where the index $0\leq l' \leq j$ (the subspace of coupled single-particle states) and $\mu'=0,1$ refers to either unoccupied ($0$) or occupied ($1$) state,
\emph{but} with $l'$ and $\mu'$ being distinct from $l$ and $\mu$. Then, the product $\tilde{c}^{\dagger}_{\alpha}\tilde{c}_{\beta}$ appearing in $H^{X}_{(j)}$ defines a product of fermionic annihilation and creation operators that transfer electrons between the sets of states given by $\{(l,\mu)\}$ and $\{(l',\mu')\}$. 
Naturally, the vertex function associated with this term, $\Gamma^{2n}_{\alpha\beta}(\omega^{i})$, denotes the magnitude of the off-diagonal element for $n$-particle (or $2n$-point) scattering.  
\par
\begin{figure}
\centering
\includegraphics[scale=0.5]{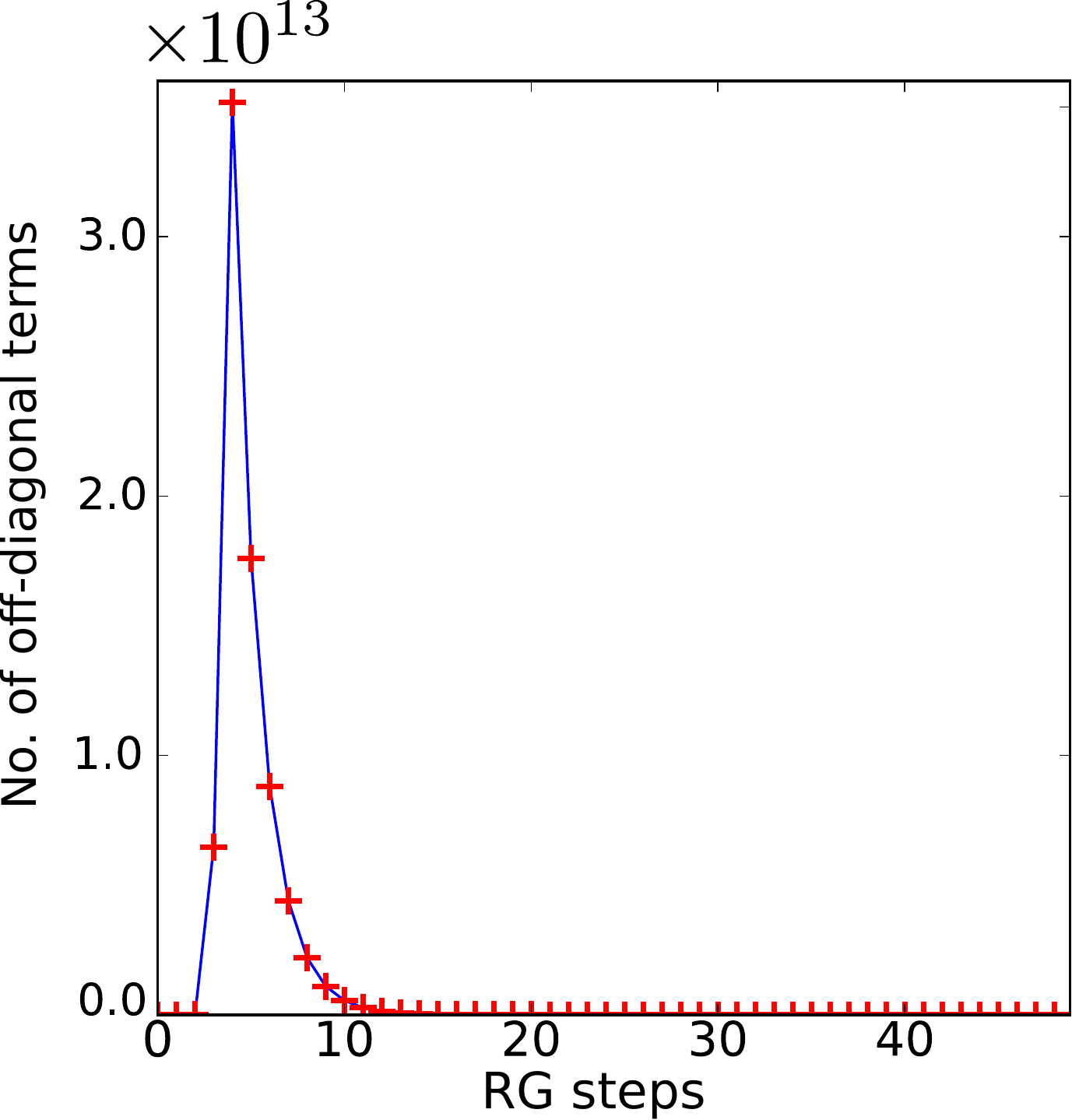} 
\caption{Variation of the total number of off-diagonal terms $K_{j}$ (red pluses, eq.\eqref{tot. no. off-diag} defined in Appendix \ref{n-particle vertex highest}) with RG step $j$. The analysis has been carried out for $N=50$ electronic states and $a_{0}=4$. The dimensionality is given by dim($\mathcal{C}^{50}=2^{50}$).} \label{off_diag_tot}
\end{figure}

\noindent
In the off-diagonal $n$-particle terms within $H^{X}_{(j)}$, the ordering of any given creation and annihilation operator string can be seen as a permutation of the normal ordered form arranged in an ascending sequence of the entries of the labels $\alpha$ and $\beta$. The sign of this permutation can be accommodated within the $n$-particle vertex 
\begin{equation}
\Gamma^{2n}_{\alpha\beta} = sgn(\mathcal{P}_{\alpha\beta})\Gamma^{2n}_{(\alpha_{0}\beta_{0})}~,~
\end{equation}
where $(\alpha_{0}\beta_{0})$ represents a two-level sorting of the indices $\alpha$ and $\beta$. $\mathcal{P}_{\alpha\beta}$ represents the permutation operation on the sorted list. The first involves a sorting of the labels $\mu$ and $\mu'$ in a descending fashion: 
this creates a normal ordered string, i.e., $c^{\dagger}$ operators followed by $c$ operators. The second sorting involves a further arrangement of the labels $l$ and $l'$ in ascending order.  
The sign of the permutation generated in this process, $sgn(\mathcal{P}_{\alpha\beta})$, nullifies the fermion sign of the unsorted list. 
It can also be seen that the largest off-diagonal operator string has a length of $a_{j}$. We have shown in Appendix (\ref{n-particle vertex highest}) that $a_{j}$ has a non-monotonic behaviour with the RG step $j$: it increases initially, peaks and then falls till the fixed point is reached. A similar behaviour is displayed in Fig.(\ref{off_diag_tot}) for the total number of off-diagonal terms with RG step $j$ (eq.\eqref{tot. no. off-diag}). Further, the off-diagonal parts of the Hamiltonian can be seen to describe both number conserving as well as non-conserving terms. For example, the $n=1$ vertices with $\alpha=(l_{1},1), \beta=(l_{2},0)$ and $\alpha'=(l_{1},1), \beta'=(l_{2},1)$ correspond to the following 
number conserving and non-conserving terms respectively
\begin{equation}
\tilde{c}^{\dagger}_{\alpha}\Gamma^{2}_{\alpha\beta}\tilde{c}_{\beta} =\Gamma^{2}_{\alpha\beta}c^{\dagger}_{l_{1}}c_{l_{2}}~,~
\tilde{c}^{\dagger}_{\alpha'}\Gamma^{2}_{\alpha'\beta'}\tilde{c}_{\beta'} =\Gamma^{2}_{\alpha'\beta'}c^{\dagger}_{l_{1}}c^{\dagger}_{l_{2}}.
\end{equation}
\par\noindent
In order to obtain the vertex RG flow equations from the renormalized Hamiltonian, we decompose it into a sum of strings comprised of number diagonal and off-diagonal operators. This decomposition is carried out as follows. First, we write one spectral component of the rotated Hamiltonian $H_{(j)}$ using eq.\eqref{cluster decomposition}, i.e., $H_{(j)}(\omega^{i}_{(j)})$ as a \emph{cluster expansion} of 2-point, 4-point, 6-point and higher order vertices
\begin{eqnarray}
H_{(j)}(\omega^{i}_{(j)}) = H^{2}_{(j)}(\omega^{i}_{(j)}) + H^{4}_{(j)}(\omega^{i}_{(j)}) + H^{6}_{(j)}(\omega^{i}_{(j)}) + \ldots~.~~~
\end{eqnarray}
The term $H^{2}_{(j)}(\omega^{i})$ can, very generally, be decomposed into diagonal and off-diagonal parts, $H^{2}_{(j)}=H^{2,D}_{(j)}+H^{2,X}_{(j)}$, and each of the two parts renormalized via contributions from all $2n$-point vertices. For instance, the contribution to $H^{2,D}_{(j)}$ is given by
\begin{eqnarray}
&&\hspace*{-2cm}
H^{2,D}_{(j)}= \bigg(\sum_{l}\Gamma^{2,(j)}_{\alpha\alpha'}(\omega^{i}_{(j)})\sigma_{l} +2\sum_{\alpha,\gamma}\lbrace\Gamma^{2}_{\alpha\gamma}G^{2}_{\gamma\gamma'}\Gamma^{2}_{\gamma\alpha'}\rbrace^{(j)}(\omega^{i}_{(j)})\tau_{j}\sigma_{l}+8\sum_{\alpha,\gamma}\lbrace\Gamma^{4}_{\alpha\gamma}G^{6}_{\gamma\gamma'}\Gamma^{4}_{\gamma\alpha'}\rbrace^{(j)}(\omega^{i}_{(j)})\tau_{j}\sigma_{j_{1}}\sigma_{j_{2}}\sigma_{l}\nonumber\\
&&\hspace*{-1cm}+32\sum_{\alpha,\gamma}\lbrace\Gamma^{6}_{\alpha\gamma}G^{10}_{\gamma\gamma'}\Gamma^{6}_{\gamma\alpha'}\rbrace^{(j)}(\omega^{i}_{(j)})\tau_{j}\sigma_{j_{1}}\sigma_{j_{2}}\sigma_{j_{3}}\sigma_{j_{4}}\sigma_{l}+\ldots\bigg)\times \bigg(1+\sum_{k_{1}=j+1}^{N}\tau_{k_{1}}+\sum_{\substack{k_{1}\neq k_{2}\\ \in \lbrace j+1,N\rbrace}}^{N}\tau_{k_{1}}\tau_{k_{2}}+..\bigg)\nonumber\\
&=&\sum_{l}\Gamma^{2,(j-1)}_{\alpha\alpha'}
\tau_{l}\left(\sum_{\substack{i=1, \\ \lbrace k_{1},\ldots ,k_{i}\rbrace}}^{N-j+1}\prod_{l=1}^{i}\tau_{k_{l}}\right)~,
\label{various_terms}
\end{eqnarray}
where
\begin{eqnarray}
\Gamma^{2,(j-1)}_{\alpha\alpha'}(\omega^{i}_{(j)})&=& \Gamma^{2,(j)}_{\alpha\alpha'}(\omega^{i}_{(j)})+\sum_{\alpha,\gamma}\lbrace\Gamma^{p_{1}}_{\alpha\gamma}G^{2p_{1}-2}_{\gamma\gamma'}\Gamma^{p_{1}}_{\gamma'\alpha'}\rbrace^{(j)}(\omega^{i}_{(j)})~,~~~~~~\label{2-point-D}
\end{eqnarray}
$G^{2p_{1}-2}_{\gamma\gamma'}$ is the Green's function containing the correlation energies of $p_{1}-1$ particle labelled $j_{1},\ldots , j_{p_{1}-1}$
\begin{eqnarray}
\hspace*{-0.7cm}
G^{2p_{1}-2}_{\gamma\gamma'}=\frac{2^{p_{1}-1}\prod_{s=1}^{p_{1}-1}\sigma_{j_{s}}\tau_{j}}{\omega^{i}_{(j)}-\sum\limits_{\alpha,l=1}^{p_{1}-1}(\Gamma^{2l,(j)}_{\alpha\alpha'}(\omega^{i}_{(j)})+\Gamma^{2l+2,(j)}_{\alpha\alpha'}(\omega^{i}_{(j)})\tau_{j})\prod^{l}_{s=1}\sigma_{n_{s}}}~,~~~
\end{eqnarray}
and the operators $\tau_{k_{i}}=n_{k_{i}}-\frac{1}{2}$ in eq.\eqref{various_terms} represent decoupled degrees of freedom that commute with the Hamiltonian $H_{(j)}$.  Note, however, that the operators $\sigma_{n_{s}}=n_{n_{s}}-\frac{1}{2}$ do not commute with $H_{(j)}$ as they belong to the coupled space, and the labels $n_{1},...,n_{l}\in j_{1},...,j_{j}$ lie within $1,...,j$.
\pin
Similarly, the contribution to $H^{2,X}_{(j)}$ is given by
\begin{eqnarray}
\hspace*{-0.5cm}&&
H^{2,X}_{(j)}= \bigg(\sum_{l}\Gamma^{2,(j)}_{\alpha\beta}(\omega^{i}_{(j)})c^{\dagger}_{l}c_{l'}+2\sum_{\alpha,\gamma}\lbrace\Gamma^{2}_{\alpha\gamma}G^{2}_{\gamma\gamma'}\Gamma^{2}_{\gamma\beta}\rbrace^{(j)}(\omega^{i}_{(j)})\tau_{j}c^{\dagger}_{l}c_{l'}\nonumber\\
\hspace*{-0.5cm}&+&8\sum_{\alpha,\gamma}\lbrace\Gamma^{4}_{\alpha\gamma}G^{6}_{\gamma\gamma'}\Gamma^{4}_{\gamma\beta}\rbrace^{(j)}(\omega^{i}_{(j)})\tau_{j}\sigma_{j_{1}}\sigma_{j_{2}}c^{\dagger}_{l}c_{l'}+2\sum_{\alpha,\gamma}\lbrace\Gamma^{6}_{\alpha\gamma}G^{10}_{\gamma\gamma'}\Gamma^{6}_{\gamma'\beta}\rbrace^{(j)}(\omega^{i}_{(j)})\tau_{j}\sigma_{j_{1}}\sigma_{j_{2}}\sigma_{j_{3}}\sigma_{j_{4}}c^{\dagger}_{l}c_{l'}+..\bigg) \nonumber\\
\hspace*{-0.5cm}&\times &\bigg(1+\sum_{k_{1}=j+1}^{N}\tau_{k_{1}}+\sum_{\substack{k_{1}\neq k_{2}\\ \in \lbrace j+1,N\rbrace}}^{N}\tau_{k_{1}}\tau_{k_{2}}+\ldots\bigg)\nonumber\\
&=&\sum_{l,l'}\Gamma^{2,(j-1)}_{\alpha\beta}
c^{\dagger}_{l}c_{l'}\left(\sum_{\substack{i=1, \\ \lbrace k_{1},\ldots ,k_{i}\rbrace}}^{N-j+1}\prod_{l=1}^{i}\tau_{k_{l}}\right)~,\label{2-point-X}
\end{eqnarray}
where
\begin{eqnarray}
\Gamma^{2,(j-1)}_{\alpha\beta}(\omega^{i}_{(j)})&=& \Gamma^{2,(j)}_{\alpha\beta}(\omega^{i}_{(j)})+2^{p_{1}-1}\sum_{\alpha,\gamma}\lbrace\Gamma^{p_{1}}_{\alpha\gamma}G^{p_{1}+p_{2}-2}_{\gamma\gamma'}\Gamma^{p_{2}}_{\gamma'\beta}\rbrace^{(j)}(\omega^{i}_{(j)})~.~~~~~~
\end{eqnarray} 
Similarly, the renormalisations of $H^{4}_{(j)}$ and $H^{6}_{(j)}$ are given by
\begin{eqnarray}
H^{4}_{(j)} &=& \left[\sum_{l_{1}.l_{2}}\Gamma^{4,(j-1)}_{\alpha\alpha'}\sigma_{l_{1}}\sigma_{l_{2}}+\sum_{\substack{l_{1},l_{2},\\ l_{3},l_{4}}}\Gamma^{4,(j-1)}_{\alpha\beta}c^{\dagger}_{l_{1}}c^{\dagger}_{l_{2}}c_{l_{3}}c_{l_{4}}\right]\left(\sum_{\substack{i=1, \\ \lbrace k_{1},\ldots ,k_{i}\rbrace}}^{N-j+1}\prod_{l=1}^{i}\tau_{k_{l}}\right),\label{cont-H4}\\
H^{6}_{(j)} &=& \bigg(\sum_{l_{1}.l_{2}}\Gamma^{6,(j-1)}_{\alpha\alpha'}\sigma_{l_{1}}\sigma_{l_{2}}\sigma_{l_{3}}+\sum_{\substack{l_{1},l_{2},l_{3}\\ l_{4},l_{5},l_{6}}}\Gamma^{6,(j-1)}_{\alpha\beta}c^{\dagger}_{l_{1}}c^{\dagger}_{l_{2}}c^{\dagger}_{l_{3}}c_{l_{4}}c_{l_{5}}c_{l_{6}}\bigg)\left(\sum_{\substack{i=1, \\ \lbrace k_{1},\ldots ,k_{i}\rbrace}}^{N-j+1}\prod_{l=1}^{i}\tau_{k_{l}}\right)~,~~~~~\label{cont-H6}
\end{eqnarray}
where $\Gamma^{4,(j-1)}_{\alpha\beta}$ and $\Gamma^{6,(j-1)}_{\alpha\beta}$ are given by
\begin{eqnarray}
\Gamma^{4,(j-1)}_{\alpha\beta}(\omega^{i}_{(j)})&=& \Gamma^{4,(j)}_{\alpha\beta}(\omega^{i}_{(j)})+2^{p_{1}-1}\sum_{\alpha,\gamma}\lbrace\Gamma^{p_{1}}_{\alpha\gamma}G^{p_{1}+p_{2}-4}_{\gamma\gamma'}\Gamma^{p_{2}}_{\gamma'\beta}\rbrace^{(j)}(\omega^{i}_{(j)})~,\label{4-point-vertex-RG}\\
\Gamma^{6,(j-1)}_{\alpha\beta}(\omega^{i}_{(j)})&=& \Gamma^{6,(j)}_{\alpha\beta}(\omega^{i}_{(j)})+2^{p_{1}-1}\sum_{\alpha,\gamma}\lbrace\Gamma^{p_{1}}_{\alpha\gamma}G^{p_{1}+p_{2}-6}_{\gamma\gamma'}\Gamma^{p_{2}}_{\gamma'\beta}\rbrace^{(j)}(\omega^{i}_{(j)})\label{6-point-vertex-RG}~.~~~
\end{eqnarray}
\begin{figure}
\centering
\includegraphics[width=0.7\textwidth]{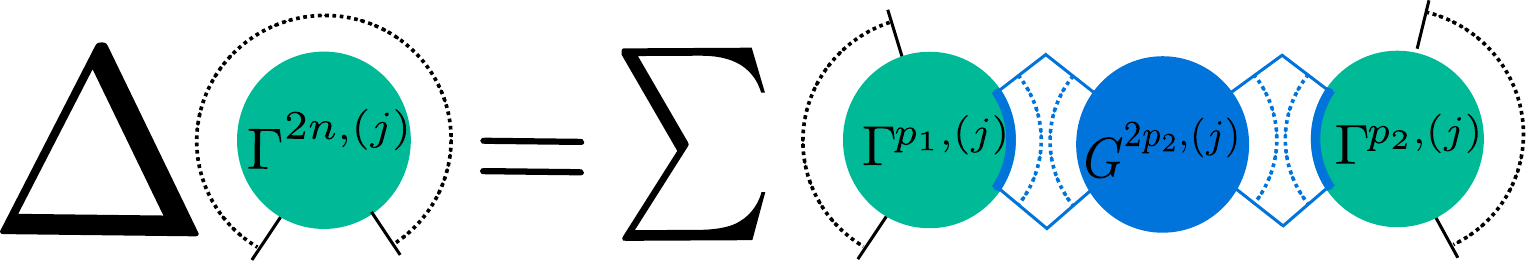} 
\caption{The RG evolution of the $2n$-point vertices $\Gamma^{2n,(j)}$. The green disk with multiple legs in black/blue (shown via a span of dots) on the left hand side of the equation represents a $2n$-point vertex($\Gamma^{2n,(j)}$), while on the right, it represents a $p_{1}$-point ($\Gamma^{p_{1},(j)}$) and a $p_{3}$-point vertex ($\Gamma^{p_{3},(j)}$) respectively. The blue disk (sandwiched between the two green disks) represents a $2p_{2}$-point Green's function ($G^{2p_{2},(j)}$) in which the blue legs represent the states involved in the scattering process. The blue legs of the Green's function $G^{2p_{2},(j)}$ are shown to be contracted with the legs of the green disks (also marked in blue), so as to generate the tensor legs of the $2n$-point vertex. The index $p_{1}$ and $p_{3}$ are summed over, as discussed in the main text.}
\label{RG_flow_diagram}
\end{figure}
\pin
We now present one of the important results of our work. Using the method of induction, we generalize the expressions for the 2-, 4- and 6-point vertex RG flow eqns. eq.\eqref{2-point-D}, eq.\eqref{2-point-X}, eq.\eqref{4-point-vertex-RG} and eq.\eqref{6-point-vertex-RG} in order to recast the Hamiltonian flow equation in terms of an entire hierarchy of $2n$-point vertex RG equations $\Gamma^{2n}_{\alpha\beta}$
(see Fig.(\ref{RG_flow_diagram}))
\begin{eqnarray}
\Delta \Gamma^{2n,(j)}_{\alpha\beta}(\omega^{i})&=&\sum_{p_{1},p_{3}}^{2a_{j}^{max}}\sum_{\gamma}\{\Gamma^{p_{1}}_{\alpha\gamma}G^{2p_{2}}_{\gamma\gamma'}\Gamma^{p_{3}}_{\gamma'\beta}\}^{(j)}(\omega^{i})~,
\label{RG_flow_heirarchy}
\end{eqnarray}
where $\alpha:=\{(l,\mu)\}$ is an ordered set of $p_{1}-p_{2}$ pairs of indices, and $l$ and $\mu$ are defined precisely as earlier. The indices $\gamma:=\{(l',\mu')\}$ and $\gamma':=\{(l',\bar{\mu}')\}$ are ordered sets of $p_{2}$ pairs of indices, where $l'$ and $\mu'$ are defined similarly to $l$ and $\mu$, and $\bar{\mu}'$ is the complement of $\mu'$. The index $\beta:=\{(l'',\mu'')\}$ is also an ordered set of $p_{3}-p_{2}$ pairs of indices, where $l''$ and $\mu''$ are defined similarly to $l$ and $\mu$. The indices $p_{1}$ and $p_{3}$ take value from the set of even positive integers lying in $[n+1,2a_{j}^{max}]$ and $[2,2a_{j}^{max}]$ respectively while $p_{2}$ takes values among the set of all positive integers lying in $[1,2a_{j}^{max}-n]$, such that $p_{1}+p_{3}-2p_{2}=2n$.
\par\noindent
The Green's function $G^{2p_{2},(j)}_{\gamma\gamma'}(\omega_{(j)}^{i})$
contains all correlation energies ($\Gamma^{2k,(j)}_{\delta\delta'}$) for $1\leq k\leq p_{2}$ particles
\begin{eqnarray}
 G^{2p_{2},(j)}_{\gamma\gamma'} &=& \left(\omega_{(j)}^{i}-\sum_{k=1}^{p_{2}}\sum_{\delta\delta'}\Gamma^{2k,(j)}_{\delta\delta'} \prod_{s=1}^{k}(\hat{n}_{l_{s}}-\frac{1}{2})\right)^{-1}2^{p_{2}}\prod_{s=1}^{p_{2}}(\hat{n}_{l_{s}}-\frac{1}{2})~,
 \label{n-particle_green_fn}
 \end{eqnarray}
where the indices $\delta:=\{(l,\mu)\}$ and $\delta':=\{(l,\bar{\mu})\}$ are ordered sets of $k$ pairs of indices, $l$ and $\mu$ are defined as earlier, and $\bar{\mu}$ is the complement of $\mu$. The index $l_{s}$ denotes the entries of $l$ that appear within $\delta$. As shown in the diagrammatic representation of Fig.(\ref{RG_flow_diagram}), various $2p_{2}$-point Green's functions connect the $p_{1}$- and $p_{3}$-point interaction vertices in renormalising the $2n$-point vertex.
The appearance of the frequency dependent correlation energies within the Green's function 
leads to two non-perturbative features of the RG transformations. First, the interplay of the multireference quantum fluctuation scale $\omega_{(j)}^{i}$ 
and the correlation energies in the vertex RG flows eq.\eqref{RG_flow_heirarchy} enables the distillation of the relevant vertices from the irrelevant ones. Second, following the discussion of eq.\eqref{quantum_fluc_switch_off}, the fixed points of the RG equations are given by the poles of eq.\eqref{n-particle_green_fn} and allows the fixed point effective Hamiltonians to be derived. In subsequent sections, we will demonstrate the fixed point effective Hamiltonians that arise from the RG treatment of various microscopic models. 
\par\noindent
It should be noted that the feedback from correlation energies and the hierarchical nature of RG flow equations are also features of the FRG scheme~\cite{uebelacker2012self}. For instance, the recent multiloop functional RG scheme~\cite{kugler2018multiloop,tagliavini2019} contains a systematic way of dealing with various $2n$-point vertices,  
but the heirarchy of flow equations typically requires a truncation. The unitary RG formulation for the Hamiltonian presented here does not, however, need any truncation. Instead, its non-perturbative nature overcomes the limitations of both a coupling expansion~\cite{you2016entanglement} as well as a cluster expansion~\cite{rademaker2016explicit} prevalent in other Hamiltonian RG formulations.
\pin\\
{\bf\textit{Relation between quantum ($\omega$) and thermal ($T$) fluctuation scales}}\\
In the RG formalism described so far the nature of the renormalized Hamiltonian at an RG step $j$ is intrinsically associated with an emergent quantum fluctuation scale $\omega_{(j)}$ eq.\eqref{quantum fluctuation eigenvalue}. Importantly the fluctuation scale  determines, whether the low energy spectrum is gappedor gapless at the IR fixed point. This energy scale $\omega_{(j)}$ was shown in \cite{anirbanmotti} to be equivalent to a thermal energy scale $k_{B}T$  upto which a given fixed point theory $H^{*}(\omega)$ and its low energy Hilbert space persist. This relation is derived in Ref.\cite{anirbanmotti,anirbanmott2} and has the form
\begin{eqnarray}
T = \frac{1}{k_{B}\pi^{2}}\mathcal{P}\int^{\infty}_{-\infty}d\omega'\frac{\Sigma^{*}(\omega)}{\omega-\omega'}.\label{Thermal scale}
\end{eqnarray}
The above relation shows that the finite lifetime $\tau$ of the single-particle states can be viewed as an effective temperature scale arising out of the unitary disentanglement: it is the highest temperature
upto which the one-particle excitations can survive. Beyond it, the one-particle excitations are replaced by 2e-1h composite excitations. 
We will see in later sections that the RG transformations lead generically to two scenarios: the first involves the iterative block diagonalisation procedure reaching a fixed point with a gapless Fermi surface, and the second the reaching of a fixed point involving the gapping of the Fermi surface via the formation and condensation of bound states. The temperature scale we have just obtained has a meaning for both scenarios. For the first, it indicates the lifetime of the gapless excitations in the neighbourhood of the Fermi surface. On the other hand, it indicates the regime of validity of the emergent condensate at finite temperatures for the second scenario. 
\begin{figure}
\centering
\includegraphics[width=0.55\textwidth]{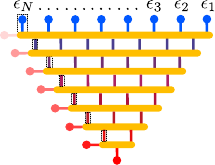}\caption{Hamiltonian (or vertex) tensor network representation of URG. The blue legs represents the holographic boundary made of the physical fermionic qubits coupled via the bare $2n$-point vertex tensors. The yellow blocks represent nonlocal unitary disentanglers that iteratively map the boundary qubits to the bulk decoupled qubits (on left of each yellow block) with energies varying from high (light red) to low (deep red). Color variation of the input legs into each subsequent unitary operator (yellow block) depicts the variation in the nonlocal structure of the vertex tensor network within the remnant coupled states as the RG flows from UV to IR.}
\label{EHMnetwork}
\end{figure}
\pin\\
{\bf \textit{Tensor network representation of the unitary RG}}\\
Similar to the case of SBRG~\cite{you2016entanglement},
the unitary transformation based RG shown in eq.\eqref{RG flow} preserves the Hilbert space (eq.\eqref{canoc-commute}) and the spectrum. Thus, by following Ref.\cite{qi2013exact,lee2016}, we demonstrate that this RG has a holographic interpretation in the form of an emergent vertex tensor network (see Fig.\ref{EHMnetwork}). In another work Ref.\cite{mukherjee2020} we have shown that indeed the entanglement renormalization group flow generated by the unitary transformations $U_{(j)}$ also describes an EHM network. In the next section below we will formulate the many body coefficient (i.e. a tensor) RG flow generated by URG. 
\par\noindent
In the RG step $j$, the unitary transformation $U_{(j)}$ (yellow block in Fig.\ref{EHMnetwork}) causes the disentanglement of exactly one electronic state (blue leg enclosed by a dotted rectangle) from the rest of the electronic states (blue legs). In this process, $U_{(j)}$ leads to the renormalization of the Hamiltonian and its associated eigenbasis. This causes the blue physical electronic states (i.e., the occupied/unoccupied basis of fermionic qubits) from the left of the {\it holographic boundary} to be mapped onto the red {\it emergent bulk physical qubits}. The initially entangled boundary qubits (blue legs) are arranged in descending order of the single-electron energy contribution $\epsilon_{i}$ from left to right, i.e., $\epsilon_{N}\geq \ldots\geq\epsilon_{1}$. The Hilbert space of the bulk qubits is spanned by Pauli gates $\tau_{x,i}^{(j)}$,$\tau_{y,i}^{(j)}$ and $\tau^{(j)}_{i}$).
\par\noindent 
The colour variation of the disentangled qubits from light (high energy) to deep red (low energy) in proceeding between the various layers of Fig.\ref{EHMnetwork} represents the RG flow from UV (high energy physical qubits) to IR (low energy physical qubits). This also reflects the fact that, due to the nonlocal nature of the unitary transformations, the nature of entanglement within the remnant coupled states morphs along with disentanglement of the boundary qubits. As discussed in an  earlier section, this results from the fact that the unitary gates we have presented here belong to a generalised notion of the Clifford gates discussed in Ref.\cite{gottesman1998heisenberg}: at every RG step, the Pauli group morphs according to eq.\eqref{pauligroup}. 
In this way, the tensor network structure shown in Fig.\ref{EHMnetwork} represents the RG flow of the entire set of 2n-point vertex tensors (eq.\eqref{RG_flow_heirarchy})
\section{Geometry of Eigenbasis Renormalization}\label{eig_base_renorm}
\begin{figure}[h!]
\centering
\includegraphics[width=1.0\textwidth]{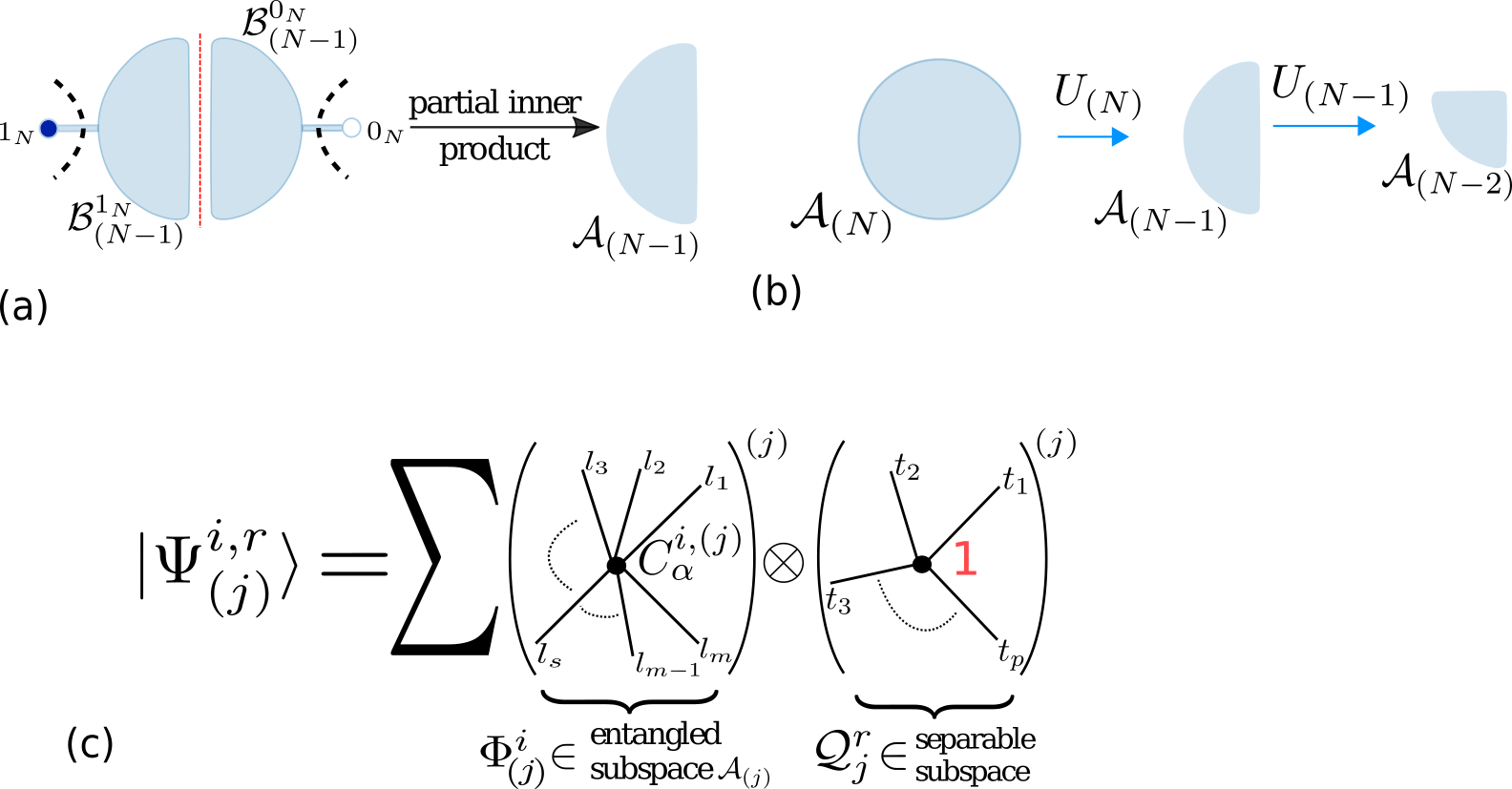}
\caption{(a) The entangled subspace $\mathcal{A}_{N-1}$ resulting from the partial inner product of the eigenstates in $\mathcal{B}_{(N-1)}$ with the occupation states $1_{N},0_{N}$  of the disentangled electronic state $N$. (b) The shrinking \textit{entangled subspace} $\mathcal{A}_{(N)}\to \mathcal{A}_{(N-1)}\to\mathcal{A}_{(N-2)}$ upon the action of successive unitary operations. (c) The eigenstate of renormalized Hamiltonian $|\Psi^{i,r}_{(j)}\rangle$ is represented as a sum over tensor coefficients $C^{i,(j)}_{\alpha}$, where $\alpha:=(l_{1},..l_{m})$ are the legs of tensors labelling the occupied states. This state is represented as a direct product of $|\Phi^{i}_{(j)}\rangle\in  \mathcal{A}_{j}$ belonging to the space of entangled states, and $|\mathcal{Q}_{j}^{r}\rangle$ belonging to the separable subspace.}\label{subspace_decomposition}
\end{figure}

Having formulated the RG procedure for the Hamiltonian, we will now provide a geometric view of many-body eigenbasis renormalization in terms of the RG flow of Fubini-Study quantum distances~\cite{anandan1990geometry} between many body states. As discussed earlier, the bare (or starting) Hamiltonian $H_{(N)}$ of the RG flow has $N$ coupled electronic degrees of freedom, with and eigenbasis of $2^{N}$ eigenstates $\mathcal{B}_{(N)} = \lbrace |\Psi^{i}\rangle,i\in[1,2^{N}]\rbrace$ that satisfy the eigenvalue relation
\begin{eqnarray}
H_{(N)}|\Psi^{i}\rangle =E^{i}|\Psi^{i}\rangle~.
\end{eqnarray}
The eigenbasis $\mathcal{B}_{(N)}$ is renormalized via the same unitary rotation that block diagonalizes $H_{(N)}$~(eq.\eqref{decoupling condition})
\begin{equation}
H_{(j-1)}=U_{(j)}H_{(j)}U^{\dagger}_{(j)}~,~\mathcal{B}_{(j)}=U_{(j+1)}\mathcal{B}_{(j+1)}~.
\label{Eigenbasis_renormalization}
%|\Psi^{i}_{(j-1)}\rangle= U_{(j)}|\Psi^{i}_{(j)}\rangle~.
\end{equation}
This ensures the \textit{spectrum preserving} nature of RG flow
\begin{eqnarray}
H_{(j)}|\Psi^{i}_{(j)}\rangle \hspace*{-0.05cm} &=& \hspace*{-0.05cm}E^{i}|\Psi^{i}_{(j)}\rangle \hspace*{-0.05cm} \Rightarrow \hspace*{-0.05cm} H_{(j-1)}|\Psi^{i}_{(j-1)}\rangle \hspace*{-0.05cm} = \hspace*{-0.05cm} E^{i}|\Psi^{i}_{(j-1)}\rangle~.~~~~~~\label{spectrum_preservation}
\end{eqnarray}
We have already seen that the Hamiltonian is block diagonalized in the occupation number basis of $j$, i.e., $[H_{(j-1)},\hat{n}_{j}]=0$, such that the occupation eigenvalues of $\hat{n}_{j}$ ($1_{j}$ and $0_{j}$) are good quantum numbers that label the renormalized eigenstates. By writing the Hamiltonian in terms of diagonal ($H^{D}$) and off-doagonal ($H^{X}$) parts as before, the renormalized eigenstates will have a renormalized value of the \textit{quantum distance} measured with respect to the separable eigenstates ($\phi_{l}$) of $H^{D}$.  
This allows us to observe the \textit{geometry of eigenbasis renormalization}. In this way, by using Shimony's geometric measure of entanglement~\cite{shimony1995degree}, we will show that the quantum distance RG will guide the renormalization flow of many-body entanglement.   
\begin{figure}
\centering
\includegraphics[width=0.6\textwidth]{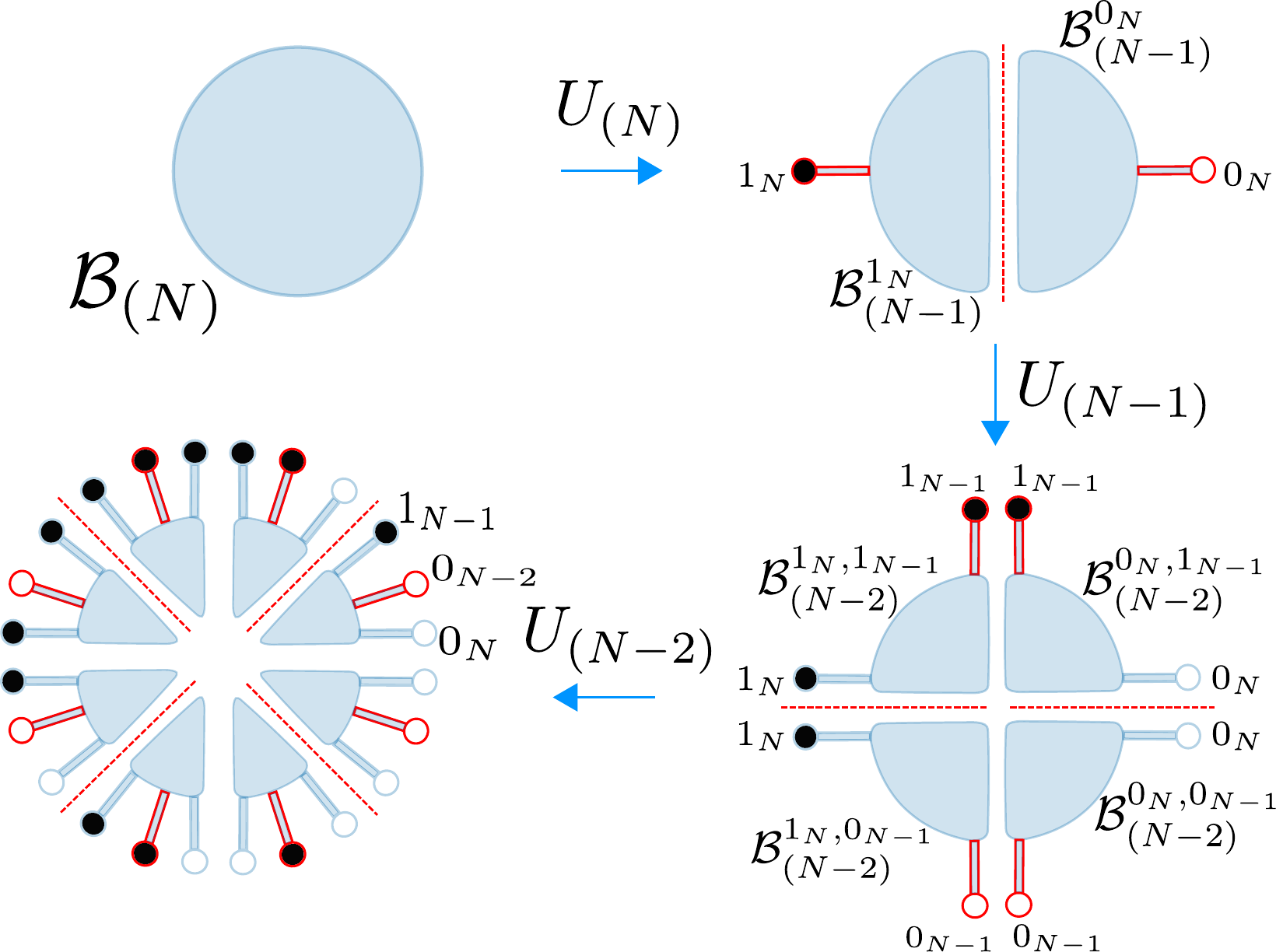}
\caption{The  RG evolution of the Hamiltonian's eigenbasis $\mathcal{B}_{(N)}$ (blue disk, top left) via the disentangling unitary operations. Upon the application of the first unitary operation $U_{(N)}$, $\mathcal{B}_{(N)}$ bifurcates into two blue semicircular disks representing the eigenbasis of the entangled electronic states $\mathcal{B}^{0_{N}}_{(N-1)}$, $\mathcal{B}^{1_{N}}_{(N-1)}$ labelled by the occupation of the disentangled state. The red handles (on the blue semicircular disks) with filled (black)/unfilled (white) circles represent the occupied ($1_{N}$)/unoccupied ($0_{N}$) configurations of the disentangled state. Upon subsequent application of the next unitary operation $U_{(N-1)}$, the blue disk further bifurcates into four quadrants, and so on.}
\label{basis_partitioning}
\end{figure}
\newpage
\pin\\
{\bf \textit{Eigenbasis renormalization scheme}}\\
The iterative block diagonalization of the Hamiltonian and concomitant renormalisation of the eigenbasis (eq.\eqref{Eigenbasis_renormalization}) partitions $\mathcal{B}_{(j)}$ into a direct sum of two halves (see Fig.\ref{basis_partitioning}) labelled by the occupation number 
of the decoupled state $j+1$ 
\begin{equation}
\mathcal{B}_{(j)} = \mathcal{B}^{0_{j+1}}_{(j)}\oplus\mathcal{B}^{1_{j+1}}_{(j)}~.
\end{equation}
As the RG procedure is iterative, every RG step halves the sub-bases obtained at the earlier step into two equal parts. As observed earlier, the subparts of the renormalized eigenbasis $\mathcal{B}_{(j)}$ can be denoted by the collection of occupation numbers of all the decoupled states ($\mathcal{Q}_{j}$~,~$\lbrace j+1,N\rbrace$), represented by blue and red handles with filled ($1_{l}$)/ unfilled ($0_{l}$) circles in Fig.(\ref{basis_partitioning})). These subspaces satisfy the following completeness relation: $\mathcal{B}_{(j)} = \bigoplus_{\mathcal{Q}_{j}} \mathcal{B}^{\mathcal{Q}_{j}}_{(j)}$, where $\mathcal{Q}_{j}$ labels the $2^{N-j}$ number diagonal configurations that complete the \textit{separable subspace}. The configuarations of the separable subspace is visualised in Fig.(\ref{basis_partitioning}) in terms of the filled/unfilled circles on the handles attached to any one light blue subpart. The state $|\mathcal{Q}_{j}\rangle$ which is a collection of the separable electronic state occupation numbers is composed of a string of 1's and 0's, and can be represented as a tensor. This tensorial representation is visualised in Fig. \ref{subspace_decomposition}(c) by treating the occupied electronic state configurations as legs of the object marked 1 in red. 
\par\noindent
The iterative decoupling of the eigenbasis into smaller sub-bases (Fig.\ref{basis_partitioning}) indicates that the renormalized eigenstates at RG step $j$ possess an interaction-driven many-body entanglement that is limited to within the subspace of coupled states labelled $\lbrace 1,j\rbrace$ (shown by light blue filled regions in Fig.(\ref{basis_partitioning})). On the other hand, the entanglement of the decoupled states $\lbrace j+1,N\rbrace$ is limited to that arising from the Pauli exclusion principle for fermions. This allows us to write a many-body eigenstate at RG step $(j)$, $|\Psi^{i,r}_{(j)}\rangle$, labelled by pair of indices $(i,r)$. Here, the index $r$ indicates the configurations 
belonging to the separable (or decoupled) subspace ($\mathcal{Q}_{j}^{r}$), and $i$ indicates the many-body configuration involving the states $\lbrace 1,j\rbrace$ that are still coupled. The configuration of coupled states is then described uniquely by the index $\alpha_{1}$ (defined similarly to $\alpha$ in (eq.\eqref{RG_flow_heirarchy}), $\alpha_{1}:=\{(l,\mu)\}$ is an ordered set of $m$ pairs of indices, $1\leq l\leq j$ and $\mu=1$ throughout. Thus, $\alpha_{1}$ denotes the set of coupled occupied single-particle states $\lbrace l_{1},\ldots, l_{m}\rbrace$, as shown in Fig. \ref{subspace_decomposition}c.
\par\noindent
Thus, very generally, we can write the eigenstates $|\Psi^{i,r}_{(j)}\rangle$ as as sum over all $\alpha_{1}$ configurations
 \vspace*{-0.1cm}
 \begin{eqnarray}
|\Psi^{i,r}_{(j)}\rangle = \sum_{\alpha_{1}}C^{i,(j)}_{\alpha_{1}}|\alpha_{1}\rangle |\mathcal{Q}^{r}_{j}\rangle~.\label{configuration_space_expansion}
\end{eqnarray}
The coefficient $C^{i,(j)}_{\alpha_{1}}$ 
is a tensor with $m$ legs 
representing the superposition weight of the configuration of occupied single-electron states. The wave function at RG step $(j)$ (Fig.\ref{subspace_decomposition}c) is therefore a summation of all such tensors chosen from among the remaining $j$ coupled states.  
The index $\alpha_{1}$ is arranged as $l_{1}<l_{2}<\ldots<l_{m}$, such that an even/odd permutation of this order due to electron exchanges will be compensated by a signature ($+1/-1$) in the coefficient $C^{i,(j)}_{\alpha_{1}}$
\begin{eqnarray}
C^{i,(j)}_{\mathcal{P}\alpha_{1}}=e^{i\pi n_{\mathcal{P}}}C^{i,(j)}_{\alpha_{1}}~,\label{Fermion_tensor_property}
\end{eqnarray}
where $n_{\mathcal{P}}$ is the no. of electron exchanges in the permutation. The subspace of coupled states $\mathcal{A}_{(j)}$ (Fig.(\ref{subspace_decomposition})(a)) 
can be removed from eigenbasis partitions $\mathcal{B}_{(j)}$ obtained at RG step $j$ by taking a \textit{partial inner product} of $|\Psi^{i,r}_{(j)}\rangle$ with the configurations of the decoupled states ($|\mathcal{Q}_{j}^{i}\rangle$) 
\begin{equation}
\mathcal{A}_{(j)} = \lbrace |\Phi^{i}_{(j)}\rangle :=\langle\mathcal{Q}_{j}^{r}|\Psi^{i,r}_{(j)}\rangle ,~r=[1,2^{N-j}],~i=[1,2^{j}]\rbrace~.
\end{equation}
This partial inner product procedure preserves orthogonality between the basis elements of $\mathcal{A}_{(j)}$ (Fig.(\ref{subspace_decomposition})(a)).
\par\noindent 
In this way, at every RG step, one single-electron state is decoupled from the coupled subspace $\mathcal{A}_{j}$ (Fig. \ref{subspace_decomposition}b) and added to the separable subspace $\mathcal{Q}_{j}$. This leads to the partitioning of the occupation number eigenbasis $\mathcal{B}$ (Fig.\ref{basis_partitioning}), such that there is no superposition between states belonging to different occupation number sub-bases of the decoupled single-electron state. This is the many-body disentangling procedure in our renormalization group formalism. It should be noted that similar disentanglement procedures are employed in the Tensor Network Renormalization and the Multiscale Entangled Renormalization Ansatz approaches~\cite{vidal2007entanglement,evenbly2015tensor,
evenbly2015tensor2}, albeit for removing short-ranged many-body entanglement. In contrast, the unitary decoupling operation eq.\eqref{decoupling condition} in our method removes every type of entanglement between a given electronic state $j$ and all others, and the unitary operations comprising the corresponding tensor network are thus non-local in nature. This will be presented in the section below.
\pin
An important feature of the vertex tensor network shown in Fig.\ref{EHMnetwork} and the EHM network (generated from entanglement RG of correlated electron systems)\cite{mukherjee2020} must now be discussed. Implicit to the construction of this tensor network is the fact that it has another holographic dimension $\omega_{i}$, i.e., an eigenvalue of the quantum fluctuation operator $\hat{\omega}|\Psi_{i}\rangle =\omega_{i}|\Psi_{i}\rangle$ (eq.\eqref{quantum fluctuation eigenvalue}) corresponding to the eigenstate $\Psi_{i}$. The $2^{N}$ eigenvalues of $\hat{\omega}$ correspond to the $2^{N}$ orthogonal directions encoding the entire many-body eigenbasis $\mathcal{B}$. Separate tensor networks are then generated by projecting the master tensor network shown in Fig.\ref{EHMnetwork} along each of these $2^{N}$ directions. The nature of the individual projected tensor network encodes the entanglement content of the many-body eigenstate $|\Psi_{i}\rangle$ it describes. For instance, a projected tensor network corresponding to a gapless eigenstate will possess equal numbers of boundary and emergent bulk qubits\cite{lee2016}. On the other hand, a projected tensor network corresponding to a gapped eigenstate will possess a lesser number of emergent bulk qubits than the boundary qubits and the remnant will form a emergent tensor network with finite entanglement~\cite{mukherjee2020}. In geometrising these projected tensor networks, one can employ quantum information theoretic measures such as mutual information in computing the information geodesics (shortest distance) between any pair of boundary qubits \cite{mukherjee2020,hyatt2017,swingle2012b,evenbly2011}.
\par\noindent\\
{\bf \textit{Disentangling single-electron occupation number states}}\\
The many-body eigenstate $|\Phi^{i}_{(j)}\rangle$ within the coupled subspace $\mathcal{A}_{j}$ is transformed at RG step $j$ via the unitary evolution equation 
\begin{equation}
|\Phi^{i}_{(j-1)}\rangle = U_{(j)}|\Phi^{i}_{(j)}\rangle.\label{unitary_evolution_eqn_state}
\end{equation}			 	 
Inspired by the exponential form of the unitary operator $U_{(j)}=\exp(\frac{\pi}{4}(\eta^{\dagger}_{(j)}-\eta_{(j)}))$, we write the state $|\Phi^{i}_{(j)}\rangle$ as 
\begin{eqnarray}
|\Phi^{i}_{(j)}\rangle =  \frac{1}{\sqrt{2}}\Omega_{(j)}|\Phi^{i,0_{j}}_{(j)}\rangle~,
\label{wave_operator_wvfn}
\end{eqnarray}
where $\Omega_{(j)}=\exp(-\eta^{\dagger}_{(j)})$ is the wave operator discussed below eq.\eqref{e-h transition operator1}, and $|\Phi^{i,0_{j}}_{(j)}\rangle$ denotes the many-body state with the decoupled electronic state $j$ being unoccupied. 
The normalization factor above can be determined from the e-h transition operator relations given in eq.\eqref{eta-operator-rel}.  
Further, the e-h transition operator relations $(\eta^{\dagger}_{(j)})^{2}=0$ and $\hat{n}_{j}\eta^{\dagger}_{(j)}(1-\hat{n}_{j})=\eta^{\dagger}_{(j)}$ allow a remarkable simplification of the wave operator: $\Omega_{(j)} = 1-\eta^{\dagger}_{(j)}$. 
Thus, we find that $\Omega_{(j)}$ contains the entanglement content of the state $j$ with the rest of the coupled states. 
\par\noindent
Further, using eq.\eqref{formofunitaryoperator} and the form of the wave operator obtained just above, eq.\eqref{unitary_evolution_eqn_state} simplifies to
 \begin{eqnarray}
 |\Phi^{i}_{(j-1)}\rangle = \frac{1}{\sqrt{2}}U_{(j)}\hat{\Omega}_{j}|\Phi^{i,0_{j}}_{(j)}\rangle = |\Phi^{i,0_{j}}_{(j)}\rangle ~.\label{disentanglement_step}
 \end{eqnarray}
This confirms the disentanglement procedure from the action of the unitary operator on the eigenstates, and is a complementary view of basis partitioning (eq.\eqref{Eigenbasis_renormalization}). 
Finally, we can use eq.\eqref{disentanglement_step} together with eq.\eqref{wave_operator_wvfn} to obtain the renormalisation of the many-body eigenstate $|\Phi^{i,0_{j}}_{(j)}\rangle$ in going from RG step $j$ to step $j-1$
\begin{equation}
|\Phi^{i,0_{j-1}}_{(j-1)}\rangle = \sqrt{2}\Omega^{-1}_{(j)}|\Phi^{i,0_{j}}_{(j)}\rangle~,\label{wave_operator_RG_eqn}
\end{equation}
where $\Omega^{-1}_{(j-1)}=e^{\eta^{\dagger}}=1+\eta^{\dagger}$~.
A renormalisation of the eigenstate $|\Phi^{i,1_{j}}_{(j)}\rangle$ can be obtained similarly
\begin{equation}
|\Phi^{i,1_{j-1}}_{(j-1)}\rangle = \sqrt{2}(\Omega^{\dagger}_{(j)})^{-1}|\Phi^{i,1_{j}}_{(j)}\rangle~.\label{wave_operator_RG_eqn2}
\end{equation}
\par\noindent\\
{\bf \textit{RG flow of entanglement within subspace $\mathcal{A}_{j}$}}\\
We will now quantify the RG flow of entanglement through measures like Shimony's geometric measure and entanglement entropy. The flow is manifested in the renormalization of the tensors comprising the configuration space expansion of the states $|\Phi^{i}_{(j)}\rangle$ belonging to the coupled (or \textit{entangled}) subspace $(\mathcal{A}_{j})$
\begin{eqnarray}
|\Phi^{i}_{(j)}\rangle = \sum_{\alpha_{1}} C^{i,(j)}_{\alpha_{1}}|\alpha_{1}\rangle + \sum_{\beta_{1}} C^{i,(j)}_{\beta_{1}}|\beta_{1}\rangle~,\label{superposition_two_sets}
\end{eqnarray}
where $\alpha_{1}:=\{(l,\mu)\}$ is an ordered set of $m$ pairs of indices, $1\leq l\leq j$, $\mu=1$ throughout, and one of the occupied single-particle states is the state $j$. Similarly, $\beta_{1}:=\{(l,\mu)\}$ is an ordered set of $m$ pairs of indices, $1\leq l\leq j-1$, $\mu=1$ throughout, i.e., the occupied states do not include the state $j$. The tensor coefficients $C^{i,(j)}_{\alpha_{1}/\beta_{1}}$ fulfil the normalization condition: $\langle\Phi^{i}_{(j)}|\Phi^{i}_{(j)}\rangle=\sum_{\alpha_{1},\beta_{1}}\left(|C^{i,(j)}_{\alpha_{1}}|^{2}+|C^{i,(j)}_{\beta_{1}}|^{2}\right)=1$. 
The renormalization of tensor coefficients $C^{i,(j)}_{\alpha_{1}/\beta_{1}}$ 
proceeds via the wave operator operating on the eigenstates. Using the decomposition of the wave operator $\Omega_{(j)}$ (in terms of eqs.\ref{cluster decomposition} and eq.\eqref{quantum fluctuation eigenvalue}) in eq.\eqref{wave_operator_RG_eqn2}, we obtain
we obtain a set of tensor flow equations for $\Delta C^{i,(j)}_{\beta_{1}} = C^{i,(j-1)}_{\beta_{1}} - C^{i,(j)}_{\beta_{1}}$
\begin{eqnarray}
\Delta C^{i,(j)}_{\beta_{1}}=(\sqrt{N^{(j)}}-1)C^{i,(j)}_{\beta_{1}}-\sqrt{N^{(j)}}\sum^{a_{j}^{max}}_{n=1}\sum_{\alpha_{1},\alpha_{1}',\beta_{1}'}\text{sgn}(\alpha_{1},\alpha_{1}',\beta_{1}')\lbrace\Gamma^{2n}_{\beta_{1}^{'}\alpha_{1}}
G^{2\bar{p}}_{\alpha_{1}\alpha_{1}^{'}}C^{i}_{\alpha_{1}^{'}}\rbrace^{(j)},~~~
\label{Tensor_flow_eqn}
\end{eqnarray}
where $N^{(j)}$ is the normalization coefficient for the RG step $j$ given by
\begin{eqnarray}
(N^{(j)})^{-1} &=&
\sum_{\beta_{1}}\frac{1}{2}\bigg[C^{i,(j)}_{\beta_{1}}-\sum^{a_{j}^{max}}_{n=1}\sum_{\bar{p}=1}^{2n}\sum_{\alpha_{1},\alpha_{1}',\beta_{1}'}\text{sgn}(\alpha_{1},\alpha_{1}',\beta_{1}')\lbrace\Gamma^{2n}_{\beta_{1}^{'}\alpha_{1}}
G^{2\bar{p}}_{\alpha_{1}\alpha_{1}^{'}}C^{i}_{\alpha_{1}^{'}}\rbrace^{(j)}\bigg]^{2}~,~~~
\label{normalization}
\end{eqnarray}
sgn$(\alpha_{1},\alpha_{1}',\beta_{1}')$ is a fermion sign function arising from electron exchanges due to $n$-particle scattering processes, and the various indices are described as follows. The index $\alpha_{1}^{'}:=\{(l,\mu)\}$ is an ordered set of $\bar{m}+\bar{p}$ pairs of indices, $1\leq l\leq j$, $\mu=1$ throughout, and one of the occupied single-particle states is the state $j$. On the other hand, $\alpha_{1}:=\{(l,\mu)\}$ is an ordered set of $\bar{p}$ pairs of indices, $1\leq l\leq j$, $\mu=0$ throughout. Clearly, the set of values of $l$ taken within $\alpha_{1}$ is contained within the set of values of $l$ taken within $\alpha_{1}^{'}$. In this way, the indices $\alpha_{1}\alpha_{1}^{'}$ in the Green's function $G^{2\bar{p}}_{\alpha_{1}\alpha_{1}^{'}}$ indicates the $\bar{p}$ single-electron states undergoing a scattering via the vertex $\Gamma^{2n}_{\beta_{1}^{'}\alpha_{1}}$, and the sum over the index $\alpha_{1}$ involves only these $\bar{p}$ single-electron states. In turn, the index $\beta_{1}^{'}$ within the vertex $\Gamma^{2n}_{\beta_{1}^{'}\alpha_{1}}$ is an ordered set 
$\beta_{1}^{'}:=\{(l,\mu)\}$ of $p=2n-\bar{p}$ pairs of indices, $1\leq l\leq j-1$, $\mu=1$ throughout, such that the occupied single-particle states do not include the state $j$. Finally, the index that labels the flow equation of the tensor coefficient, $\beta_{1}:=\{(l,\mu)\}$, is an ordered set of $\bar{m}+p$ pairs of indices, $1\leq l\leq j-1$, $\mu=1$ throughout, such that the occupied single-particle states do not include the state $j$. The index $\beta_{1}$ emerges from the convolution of the indices $\alpha_{1}, \alpha_{1}^{'}$ and $\beta_{1}^{'}$, i.e., it is defined as $\beta_{1}:=(\beta_{1}^{'}\cup\alpha_{1}^{'})-\bar{\alpha}_{1}$, i.e., a set formed from the union of the sets $\beta_{1}^{'}$ and $\alpha_{1}^{'}$ and from which a set $\bar{\alpha}_{1}$ has been removed, where $\bar{\alpha}_{1}:=\{(l,\bar{\mu})\}$ is an ordered set of $\bar{p}$ pairs of indices, $1\leq l\leq j$, $\bar{\mu}=1$ throughout. Note that the set of values of $l$ with $\alpha_{1}$ is the same as those within $\bar{\alpha}_{1}$. However $\bar{\mu}$  values in $\bar{\alpha}_{1}$ are different from $\mu$ in $\alpha_{1}$. 
\pin
We will now describe the fermion sign function $sgn(\alpha_{1},\alpha_{1}',\beta_{1})$.   
For this, we write the state $|\alpha_{1}'\rangle$ (that contributes to the coefficient tensor flow eq.\eqref{Tensor_flow_eqn}) in second quantized notation as
\begin{eqnarray}
|\alpha'_{1}\rangle = |l_{\bar{m}+\bar{p}}\ldots l_{1}\rangle = c^{\dagger}_{l_{\bar{m}+\bar{p}}}\ldots c^{\dagger}_{l_{1}}|0\rangle~,
\end{eqnarray}
where $l_{1}<l_{2}<\ldots<l_{\bar{m}+\bar{p}-1}<l_{\bar{m}+\bar{p}}$ belong to the ordered set $\alpha_{1}'$. Similarly, the set $\alpha_{1}=\lbrace (b_{\bar{p}},0),\ldots, (b_{1},0)\rbrace$ and $\beta'_{1}=\lbrace (a_{p},1),\ldots, (a_{1},1)\rbrace$ (see also below eq.\eqref{Tensor_flow_eqn}). As $\alpha_{1}\subset\alpha_{1}'$, the label $b_{i}$ in $\alpha_{1}$ corresponds to a label $l_{k}$ in the list $\beta^{'}_{1}$. This information will be useful below in counting the electron exchanges. Now, the phase collected from counting the electron exchanges in the $n$-particle scattering process is accounted for as follows
\begin{eqnarray}
&&\hspace*{-1.2cm}\Gamma^{2n,(j)}_{\beta'_{1}\alpha_{1}}c^{\dagger}_{a_{p}}\ldots c^{\dagger}_{a_{1}}c_{b_{\bar{p}}}\ldots c_{b_{1}}|\alpha'_{1}\rangle, \nonumber\\
&&\hspace*{-1.2cm}=\exp\left(i\pi\sum_{i=1}^{b_{1}}n_{i}\right)\Gamma^{2n,(j)}_{\beta'_{1}\alpha_{1}}c^{\dagger}_{a_{p}}\ldots c^{\dagger}_{a_{1}}c_{b_{\bar{p}}}\ldots c_{b_{2}}|l_{\bar{m}+\bar{p}}\ldots l_{k+1}l_{k-1}\ldots l_{1}\rangle,~\nonumber\\
&&\hspace*{-1.2cm}=\exp\left(i\pi\sum_{\substack{i=1,\\ i\neq b_{1}}}^{b_{2}}n_{i}\right)\exp\left(i\pi\sum_{i=1}^{b_{1}}n_{i}\right)\Gamma^{2n,(j)}_{\beta'_{1}\alpha_{1}}c^{\dagger}_{a_{p}}\ldots c^{\dagger}_{a_{1}}c_{b_{\bar{p}}}\ldots c_{b_{3}}|l_{\bar{m}+\bar{p}}\ldots l_{k'+1}l_{k'-1}\ldots l_{k+1}l_{k-1}\ldots l_{1}\rangle~.~~~~~~
\end{eqnarray}
In the expression for the phase $\exp\left(i\pi\sum_{i=1}^{b_{1}}n_{i}\right)$, the number $n_{i}=1$ if $i\in \alpha_{1}'$ and $n_{i}=0$ otherwise. The labels $b_{1}$ and $b_{2}$ in set $\alpha_{1}$ correspond to labels $k$ and $k''$ in set $\alpha_{1}'$. In this way, the state resulting from the operation of the entire string of annihilation operators comprising the 2n-point scattering vertex is given by
\begin{eqnarray}
\Gamma^{2n,(j)}_{\beta'_{1}\alpha_{1}}c^{\dagger}_{a_{p}}\ldots c^{\dagger}_{a_{1}}c_{b_{\bar{p}}}\ldots c_{b_{1}}|\alpha_{1}'\rangle =\prod_{k=1}^{\bar{p}}P_{k}\Gamma^{2n,(j)}_{\beta'_{1}\alpha_{1}}c^{\dagger}_{a_{p}}\ldots c^{\dagger}_{a_{1}}|\alpha_{1}''\rangle~,
\end{eqnarray} 
where index $\alpha_{1}''=\lbrace(l,\mu)\rbrace$ is an ordered set of $\bar{m}$ pairs of indices with $\mu=1$. The net phase comprising the operation of $\bar{p}$ annihilation operators is given by $\prod_{k=1}^{\bar{p}}P_{k}$, where
\begin{eqnarray}
P_{k}=\exp\left(i\pi\sum_{i\in \alpha_{1}'-\rho}^{b_{k}}n_{i}\right)~,~\rho=\lbrace b_{1},\ldots,b_{k-1}\rbrace ~.
\end{eqnarray}
In the above summation, the index $i$ is restricted to the set $\alpha_{1}'-\rho$, as electrons in set $\rho$ are annihilated. Finally, the net electron exchange phase generated by the string of electron creation and annhilation operators is 
\begin{eqnarray}
sgn(\alpha_{1},\alpha_{1}',\beta_{1}')=\prod_{k=1}^{p}Q_{k}\prod_{k=1}^{\bar{p}}P_{k}~,\label{fermion_exchange_sign}
\end{eqnarray}
where $Q_{k}$ is given by
\begin{eqnarray}
Q_{k} = \exp(i(k-1)\pi)\exp\left(i\pi\sum_{\substack{i=1,\\ i\notin \gamma\cup\alpha_{1}}}^{a_{k}}n_{i}\right)~,
\end{eqnarray}
and the index $\gamma=\lbrace a_{1},\ldots, a_{k}\rbrace$ is a set of labels for the states where electrons are created. In the expression for $Q_{k}$, the number $n_{i}=1$ if $i\in\alpha_{1}''$ and $n_{i}=0$ otherwise.
\pin
We have seen earlier that the action of the unitary operator $U_{(j)}$ on state $|\Phi^{i}_{(j)}\rangle$ (eq.\eqref{unitary_evolution_eqn_state}) led to a subspace rotation of the state onto one of the occupation number configurations of state $j$, such that the projection along the other occupation number configuration axis is precisely zero
\begin{eqnarray}
\hspace*{-1cm}
|\Phi^{i,0_{j}}_{(j)}\rangle &=& \frac{1}{\sqrt{2}}(1+\eta^{\dagger}_{(j)} - \eta_{(j)})|\Phi^{i}_{(j)}\rangle\nonumber\\ 
&=&\frac{1}{\sqrt{2}}(1+\eta^{\dagger}_{(j)})\sum_{\gamma'_{1}}C^{i,(j)}_{\gamma'_{1}}|\gamma'_{1}\rangle + \frac{1}{\sqrt{2}}(1-\eta_{(j)})\sum_{\rho'_{1}}C^{i,(j)}_{\rho'_{1}}|\rho'_{1}\rangle\nonumber\\
\Rightarrow && \eta^{\dagger}_{(j)}\sum_{\gamma'_{1}}C^{i,(j)}_{\gamma'_{1}}|\gamma'_{1}\rangle + \sum_{\rho'_{1}}C^{i,(j)}_{\rho'_{1}}|\rho'_{1}\rangle = 0~,
\end{eqnarray}
where $\rho'_{1}$ and $\gamma'_{1}$ are defined identically to the indices $\alpha'_{1}$ and $\beta'_{1}$ defined earlier respectively. The index $\rho'_{1}=\lbrace(l,\mu)\rbrace$ is, similar to $\alpha_{1}$, an ordered set of $\bar{p}$ elements with state $j$ occupied. This leads to a constraint on the value of tensor coefficient $C^{i,(j)}_{\alpha_{1}'}$ given by
\begin{eqnarray}
C^{i,(j)}_{\alpha_{1}'}&=& 
-\sum^{a_{j}^{max}}_{k=1}
\sum_{\gamma_{1},\gamma_{1}',\rho_{1}'}sgn(\gamma_{1},\gamma_{1}',\rho_{1}')\lbrace\Gamma^{2k}_{\rho_{1}^{'}\gamma_{1}}
G^{4k-2\bar{p}}_{\gamma_{1}\gamma_{1}^{'}}C^{i}_{\gamma_{1}^{'}}\rbrace^{(j)}~,
\label{tensor_eqn_constraint}
\end{eqnarray}
where the index $\gamma_{1}$ are defined in the same way as $\alpha_{1}$. The index  $\gamma_{1}:=\lbrace (l,\mu)\rbrace$ is an ordered set of $2k-\bar{p}$ indices with all $\mu =0$ and the state $j$ excluded. This comprises the ($2k=\bar{p}+2k-\bar{p}$)-point off-diagonal vertex $\Gamma^{2k,(j)}_{\rho_{1}'\gamma_{1}}$. The index $\gamma_{1}'=\lbrace (l,\mu)\rbrace$ is an ordered set of $\bar{m}+2k-\bar{p}$ indices with all $\mu=1$ and the state $j$ is excluded. Similar to $\alpha_{1}$ and $\alpha'_{1}
$ we observe $\gamma_{1}$ is a subset of $\gamma'_{1}$. So the indices $\gamma_{1}\gamma_{1}'$ in the Green's function $G^{4k-2\bar{p}}_{\gamma_{1}\gamma_{1}'}$ represents only the $2k-\bar{p}$ single electron states which get scattered by the vertex $\Gamma^{2k,(j)}_{\rho_{1}'\gamma_{1}}$. Similar to $\beta_{1}$, the index $\alpha_{1}'$ emerges from the convolution of the indices $\gamma_{1}, \gamma_{1}^{'}$ and $\alpha_{1}^{'}$:~$\alpha_{1}':=(\rho_{1}'\cup\gamma_{1}^{'})-\gamma_{1}$. In other words, $\alpha_{1}'$ corresponds to a set formed from the union of the sets $\rho_{1}^{'}$ and $\gamma_{1}^{'}$, and from which a set $\gamma_{1}$ has been removed. 
\pin
Using the constraint eq.\eqref{tensor_eqn_constraint} together with the $n$-particle vertex flow eq.\eqref{RG_flow_heirarchy}, the tensor flow eq.\eqref{Tensor_flow_eqn} can be written as
\begin{eqnarray}
 \Delta C^{i,(j)}_{\beta_{1}}&=& (\sqrt{N^{(j)}}-1)C^{i,(j)}_{\beta_{1}}+\sqrt{N^{(j)}}\sum^{a_{j}^{max}}_{\bar{k}=1}\sum_{\gamma_{1},\gamma_{1}',\beta_{1}'}sgn(\gamma_{1},\gamma_{1}',\beta_{1}')\lbrace\Delta\Gamma^{2\bar{k}}_{\beta_{1}^{'}\gamma_{1}}
G^{4\bar{k}-2p}_{\gamma_{1}\gamma_{1}^{'}}C^{i}_{\gamma_{1}^{'}}\rbrace^{(j)}~,
\label{coefficient renormalization}~~~~~~
\end{eqnarray}
where the RG flow for $2\bar{k}$-point vertex ($2\bar{k} = 2n+2k-2\bar{p}$) is given by
\begin{eqnarray}
\Delta\Gamma^{2\bar{k}}_{\beta_{1}^{'}\gamma_{1}} = \sum_{n,k}^{2a_{j}^{max}}\sum_{\rho_{1}}\lbrace\Gamma^{2n}_{\beta_{1}^{'}\rho_{1}}G^{2\bar{p}}_{\rho_{1}\rho_{1}^{'}}\Gamma^{2k}_{\rho_{1}^{'}\gamma_{1}^{'}}\rbrace^{(j)}~.
\label{vertexflow2}
\end{eqnarray}
As observed previously, the phase $sgn(\gamma_{1},\gamma_{1}',\beta_{1}')$ in eq.\eqref{coefficient renormalization} is obtained via counting the electrons exchanged via the $2\bar{k}$  point scattering vertex. Importantly, eq.\eqref{coefficient renormalization} relates the RG flow of the many-body state space to that of the effective Hamiltonian (through the vertex flow equation). We now arrive at an important result. When the final fixed point of the vertex tensor network RG flow is reached, i.e., when $\Delta\Gamma^{2\bar{k},(j\ast)}_{\beta_{1}^{'}\gamma_{1}}=0~,~N^{(j\ast)}=1$, the RG flow of the coefficient tensor also ceases, $\Delta C^{i,(j\ast)}_{\beta_{1}}=0$. Note that the renormalization of the coefficient tensors is responsible for the renormalization of the many-particle entanglement features; this implies that the vertex tensor network RG flow guides the entanglement RG. Thus, the entanglement RG fixed points and vertex tensor RG fixed points are attained concurrently. In a recent work~\cite{mukherjee2020}, we have shown the connection between the nonlocal unitary disentangler based entanglement renormalization group and the entanglement holographic mapping (EHM) of Ref.\cite{qi2013}. An EHM is a tensor network formed via a stacking of unitary transformation layers, where each such layer disentangles a certain set of qubits. The input electronic states/nodes comprise the boundary layer describing the UV theory, and the unitary map generates the bulk of the EHM such that the IR fixed point theory is obtained deep in the bulk. In this way, eq.\eqref{coefficient renormalization} above shows that the vertex tensor network RG generates the EHM.
\pin\\
{\bf \textit{Mitigating the fermion sign problem through URG flow}}\\ 
In this section, we will show that by applying the URG to models of interacting electrons, certain classes of stable fixed points are obtained from the RG flow in the IR that are free from the signatures that arise from electronic exchanges. A system of interacting electrons with translational invariance can very generally be described by the Hamiltonian
\begin{eqnarray}
\hat{H}=\sum_{\mathbf{k}}(\epsilon_{\mathbf{k}}-\mu)\hat{n}_{\mathbf{k}\sigma}+\sum_{\mathbf{k},\mathbf{k}',\mathbf{p}}V^{\sigma\sigma'}_{\mathbf{k}-\mathbf{k}'}c^{\dagger}_{\mathbf{k},\sigma}c^{\dagger}_{\mathbf{p}-\mathbf{k},\sigma'}c_{\mathbf{p}-\mathbf{k}',\sigma'}c_{\mathbf{k}',\sigma}~.
\end{eqnarray} 
The $\mathbf{k}$ are wave-vectors belonging to the first Brillouin zone. We consider that the Fermi surface of the non-interacting part, defined as a collection of wave-vectors $\mathbf{k}_{F}$ such that $\epsilon_{\mathbf{k}_{F}}=E_{F}=\mu$, to be described as an extended object in the Brillouin zone. Further, we explore the sub-parameter space of $H$ where (i) all opposite-spin electron exchange scattering vertices are attractive: $V^{\sigma,-\sigma}_{\mathbf{k}-\mathbf{k}'\neq 0}<0$, as well as Hartree terms  $V^{\sigma,\sigma'}_{0}>0$ and the same-spin electron exchange scattering vertices $V^{\sigma,\sigma'}_{\mathbf{k}-\mathbf{k}'\neq 0}>0$ are repulsive. 
\pin
In applying the URG method to this problem, we adopt a RG scheme where the states farthest away from the Fermi surface are disentangled first. 
Following Refs.\cite{anirbanmotti,anirbanmott2}, this is carried out by defining curves \textit{parallel} to the Fermi surface. The wave-vectors $\mathbf{k}_{\Lambda\hat{s}}=\mathbf{k}_{F}(\hat{s})+\Lambda\hat{s}$ are represented in terms of the distance ($\Lambda$) normal from the Fermi surface and the unit normal vector, $\hat{s}=\nabla\epsilon_{\mathbf{k}}/|\nabla\epsilon_{\mathbf{k}}||_{\epsilon_{\mathbf{k}}=E_{F}}$. At each RG step, the entire isogeometric curve at a distance $\Lambda_{j}$ from the Fermi surface is disentangled via a product of unitary operations $U_{j}=\prod_{l}U_{(j,l)}$, such that $U_{j,l}=\sqrt{2^{-1}}[1+\eta_{j,l}-\eta^{\dagger}_{j,l}]$ disentangles the electronic state $|j,l\rangle = |\mathbf{k}_{\Lambda_{j}\hat{s}},\sigma\rangle$ from the rest. This iterative disentanglement procedure leads to the URG flow equation $H_{(j-1)}=U_{(j)}H_{(j)}U^{\dagger}_{(j)}$.
\pin
URG generates the $2n$-point vertex tensor RG equation hierarchy eq.\eqref{RG_flow_heirarchy}. We initially restrict ourselves to studying just the four-point vertex RG flow equations
\begin{eqnarray}
\Delta\Gamma^{\sigma,\sigma',(j)}_{\mathbf{q},\mathbf{p}}=\frac{\Gamma^{\sigma,\sigma',(j)}_{\mathbf{q}_{1},\mathbf{p}}\Gamma^{\sigma,\sigma',(j)}_{\mathbf{q}_{2},\mathbf{p}}}{\omega-\epsilon_{j,a}-\epsilon_{j,a'}-\frac{1}{4}\Gamma^{(j)}_{0,\mathbf{p}}}
\end{eqnarray}
where the labels $(j,a):=\mathbf{k}_{\Lambda_{j}\hat{s}},\sigma$, $(j,a'):=\mathbf{p}-\mathbf{k}_{\Lambda_{j}\hat{s}},\sigma'$. The momentum transfer wave-vectors $\mathbf{q}_{1}=\mathbf{k}-\mathbf{k}_{\Lambda_{j}\hat{s}}$ and $\mathbf{q}_{2}=\mathbf{k}_{\Lambda_{j}\hat{s}}-\mathbf{k}'$, such that $\mathbf{q}=\mathbf{q}_{1}+\mathbf{q}_{2}$. In the discussion below, we will be asking the following question: can a excited pair of electrons with momenta $\mathbf{k}$ and $\mathbf{p}-\mathbf{k}$ residing outside the Fermi surface ($\epsilon_{\mathbf{k}}$, $\epsilon_{\mathbf{p}-\mathbf{k}}>E_{F}$) have a condensation energy lower than the Fermi energy? This is the primary ingredient for bound-state condensation. If the answer is yes, can the effective theories describing the IR fixed points reached from URG analysis be free of fermion exchange signatures? 
\pin
To proceed further, we work in  the regime of quantum fluctuation energyscales: $\omega<2^{-1}(\epsilon_{\mathbf{k}}+\epsilon_{\mathbf{p}-\mathbf{k}})$. As the electronic states $\mathbf{k}$ and $\mathbf{p}-\mathbf{k}$ are both occupied and are the primary two-particle excitations with $\epsilon_{\mathbf{k}}$, $\epsilon_{\mathbf{p}-\mathbf{k}}>E_{F}$, we have 
\begin{eqnarray}
|\omega-2^{-1}(\epsilon_{\mathbf{k}}+\epsilon_{-\mathbf{k}})|<|\omega-2^{-1}(\epsilon_{\mathbf{k}}+\epsilon_{\mathbf{p}-\mathbf{k}})|<0~.\label{regime-I}
\end{eqnarray}   
Along the URG flow within the regime of eq.\eqref{regime-I}, the $\mathbf{p}=0$-momentum electron exchange scattering vertex tensors $|\Delta\Gamma^{(j)}_{\mathbf{q},0}|$ are the most dominant among all finite-momentum pairs 
\begin{eqnarray}
|\Delta\Gamma^{(j)}_{\mathbf{q},0}|>|\Delta\Gamma^{(j)}_{\mathbf{q},\mathbf{p}}|~.\label{dominant}
\end{eqnarray}
As a resulting, the RG flow of the 6-point vertices $\Gamma^{6,(j)}$ 
in eq.\eqref{RG_flow_heirarchy} are also sub-dominant, as they arise from the interplay between different pair-momentum vertices. On the other hand, the repulsive Hartree terms and the same-spin electron exchange scattering vertices are RG irrelevant:~$\Delta\Gamma^{\sigma,\sigma',(j)}_{0,\mathbf{p}}$~,~$\Delta\Gamma^{\sigma,\sigma',(j)}_{\mathbf{q},\mathbf{p}}<0$~.
\begin{figure}[h!]
\centering
\includegraphics[width=0.4\textwidth]{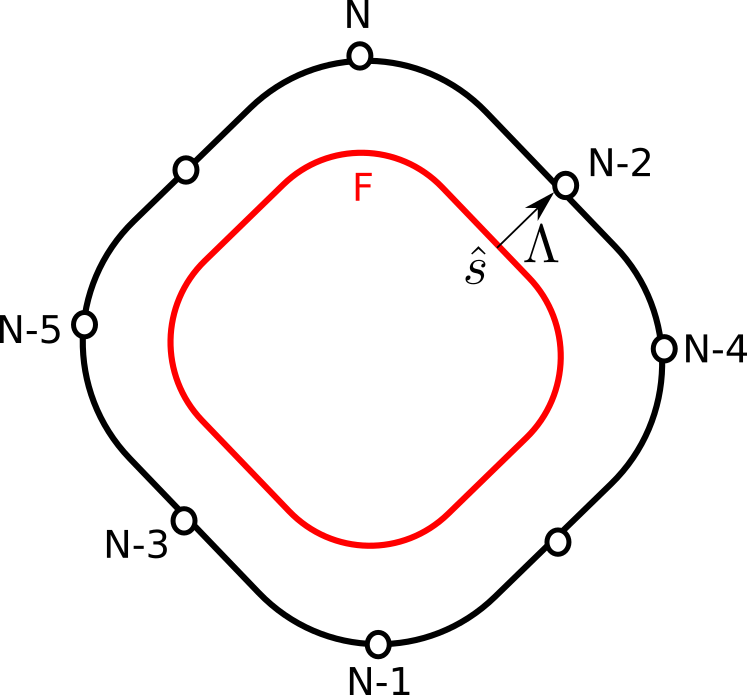}
\caption{Schematic representation of Fermi sea for representing the electron counting scheme. The labels $N$ and $N-1$ correspond to the partner electronic states $\mathbf{k}_{\Lambda\hat{s}},\uparrow$ and $-\mathbf{k}_{\Lambda\hat{s}},\downarrow$ of the opposite-spin zero momentum pair. $N-1$, $N-2$ are the next pair of such electronic states. The red curve represents the Fermi surface.}\label{countingScheme}
\end{figure}
\pin
Armed with this insight, we now explore the wavefunction coefficient tensor flow equation (eq.\eqref{coefficient renormalization}) while taking into account only the RG dominant four-point vertices. For this, we define the ordering scheme of the electronic states shown in Fig.\ref{countingScheme}. The electronic state on the isogeometric curve farthest from the Fermi surface is defined as $N:=\mathbf{k}_{\Lambda_{0},\hat{s}_{0}},\uparrow$, while the electronic state with opposite-spin residing on the diametrically opposite position (and on the same isogeometric curve) is labelled $N-1:=\mathbf{k}_{\Lambda_{0},-\hat{s}_{0}},\downarrow$. The electronic state along the next normal direction on the same isogeometric curve $\Lambda_{0}$ is labelled $N-2:=\mathbf{k}_{\Lambda_{0},\hat{s}_{1}},\uparrow$, while the diametrically opposite state is labelled $N-3:=\mathbf{k}_{\Lambda_{0},-\hat{s}_{1}},\downarrow$, and so on. In this way, all the states are labelled by progressively decreasing integers as they approach the Fermi surface. The states on the Fermi surface (red curve in Fig.\ref{countingScheme}) are marked as $2N_{F}:k_{F,\hat{s}_{0}},\uparrow$, $2N_{F}-1:k_{F,-\hat{s}_{0}},\downarrow$, leading down to the last two state $2:k_{F,\hat{s}_{N_{F}}},\uparrow$ and $1:k_{F,-\hat{s}_{N_{F}}},\downarrow$. Here, $N_{F}$ is the number of normal directions $\hat{s}$'s on the Fermi surface. The coefficient tensor flow equation for the eigenstates $|\Psi^{i}_{(j)}\rangle$ of $H_{(j)}$ is then given by 
\begin{eqnarray}
 \Delta C^{i,(j)}_{\beta_{1}}&=& (\sqrt{N^{(j)}}-1)C^{i,(j)}_{\beta_{1}}+\sqrt{N^{(j)}}\sum_{\gamma_{1},\gamma_{1}',\beta_{1}'}\frac{sgn(\beta_{1}',\gamma_{1},\gamma_{1}')\Delta\Gamma^{\sigma,\sigma',(j)}_{\mathbf{q},\mathbf{p}}}{\omega-\frac{1}{2}(\epsilon_{\mathbf{k}_{\Lambda\hat{s}}}+\epsilon_{\mathbf{p}-\mathbf{k}_{\Lambda\hat{s}}})-\frac{1}{4}\Gamma^{(j),\sigma,\sigma'}_{0,\mathbf{p}}}C^{i,(j)}_{\gamma_{1}^{'}}~,~~~~~~~
\end{eqnarray}
 where the index $\gamma_{1}'=\lbrace l\rbrace$ refers to a collection of labels where the electronic state is occupied. The indices $\gamma_{1}=\lbrace a:=(\mathbf{k}_{\Lambda\hat{s}},\sigma),b:=(\mathbf{p}-\mathbf{k}_{\Lambda\hat{s}},\sigma') \rbrace$ and $\beta_{1}'=\lbrace c:=(\mathbf{k}_{\Lambda\hat{s}}+\mathbf{q},\sigma),d:=(\mathbf{p}-\mathbf{k}_{\Lambda\hat{s}}-\mathbf{q},\sigma')\rbrace$ respectively. Keeping only the dominant RG flow contribution from $\Delta\Gamma^{\sigma,-\sigma,(j)}_{\mathbf{q},0}$ in eq.\eqref{dominant}, the coefficient RG equation simplifies to
\begin{eqnarray}
 \Delta C^{i,(j)}_{\beta_{1}}&=& (\sqrt{N^{(j)}}-1)C^{i,(j)}_{\beta_{1}}+\sqrt{N^{(j)}}\sum_{\mathbf{q},\mathbf{k}_{\Lambda\hat{s}}}\frac{sgn(\beta_{1}',\gamma_{1},\gamma_{1}')\Delta\Gamma^{\sigma,-\sigma,(j)}_{\mathbf{q},0}}{\omega-\epsilon_{\mathbf{k}_{\Lambda\hat{s}}}-\frac{1}{4}\Gamma^{(j),\sigma,-\sigma}_{0,0}}C^{i,(j)}_{\gamma_{1}^{'}}~.~
\end{eqnarray}
Here, $\gamma_{1}'$ reduces to a special class of sequences comprised of only consecutive pairs of integers $(l,l+1)$. This marks the pair of  electronic states with opposite-spins and zero net-momentum:~$\gamma_{1}'=\lbrace (l_{1},l_{1}+1),(l_{2},l_{2}+1),\ldots\rbrace$, and $\gamma_{1}=\lbrace (m,m+1)\rbrace$, $\beta_{1}'=\lbrace (n,n+1)\rbrace$.
\pin
Importantly, for this case, the fermion exchange sign function $sgn(\beta_{1}',\gamma_{1},\gamma_{1}')$ trivializes to $1$, as can be seen by recalling eq.\eqref{fermion_exchange_sign}
\begin{eqnarray}
\text{sgn}(\beta_{1}',\gamma_{1},\gamma_{1}')&=&Q_{2}Q_{1}P_{2}P_{1}\nonumber\\
&=&\exp(i\pi\sum_{i=1}^{m-1}n_{i})\exp(i\pi\sum_{i=1}^{m-1}n_{i})\exp(i\pi\sum_{i=1}^{n-1}n_{i})\exp(i\pi\sum_{i=1}^{n-1}n_{i})~=~1~.~~~~
\end{eqnarray}
In this way, we observed that in this case, the fermion string essentially counts an even number of electron exchanges for any configuration, rendering the $\mathbf{p}=0$ pair-momentum subspace free of fermion signatures. As a result, we obtain a reduced BCS effective Hamiltonian for this subspace obtained at the fixed point $\omega=\max_{\hat{s}}\epsilon_{\mathbf{k}_{\Lambda^{*}\hat{s}}}$ given by
\begin{eqnarray}
H_{\text{eff}}=\sum_{\mathbf{k}}\epsilon_{\mathbf{k}}A^{z}_{\mathbf{k}}-\sum_{\mathbf{q},\mathbf{k}}|V^{*}_{\mathbf{q},0}|A^{+}_{\mathbf{k}}A^{-}_{\mathbf{k}+\mathbf{q}}~,
\end{eqnarray} 
where $A^{+}_{\mathbf{k}}=c^{\dagger}_{\mathbf{k}\uparrow}c^{\dagger}_{-\mathbf{k}\downarrow}$ and $A^{-}_{\mathbf{k}}=c_{-\mathbf{k}\downarrow}c_{\mathbf{k}\uparrow}$, $A^{z}_{\mathbf{k}}=2^{-1}\left[A^{+}_{\mathbf{k}},A^{-}_{\mathbf{k}}\right]$ are Anderson pseudospins~\cite{anderson1958random}. Very generally, we can redefine a pair of legs $l:=(l,l+1)$ as the Anderson pseudospin $l$. 
The eigenstates of Hamiltonian $H_{\text{eff}}$ can be written as 
\begin{eqnarray}
|\Phi^{i}\rangle = \sum_{\rho}C^{i,*}_{\rho}A^{+}_{l_{1}}..A^{+}_{l_{n}}|\Downarrow...\Downarrow\rangle~,~~
\end{eqnarray}
where $\rho=\lbrace l\rbrace$ is the label for the set of Anderson pseudospins which are in the $|1_{l}1_{l+1}\rangle=|\Uparrow\rangle$ configuration. We stress that exchanging the legs of the coefficient tensor $C^{i,*}_{\rho}$ here is free of fermion exchange signatures. In this way, we have mitigated the problem of fermion exchange signatures via the URG flow to the reduced BCS theory obtained in the IR. 
\pin\\
{\bf \textit{RG flow of a geometric measure of entanglement and its relation to bound state spectral weight}}\\
The Fubini-Study distances (represented by black lines in Fig.(\ref{quantum distances and entanglement})) $d(\beta_{1},\Phi^{i}_{(j)})$ between the separable states $|\beta_{1}\rangle$ (black disk in figure) and the renormalized eigenstates $|\Phi^{i}_{(j)}\rangle$ (green disk in figure) of $H_{(j)}$ belonging to $\mathcal{A}_{(j)}$ is defined as~\cite{horodecki2009quantum,wei2003geometric}
\begin{equation}
d^{2}(\beta_{1},\Phi^{i}_{(j)}) = 1-|\langle\beta_{1}|\Phi^{i}_{(j)}\rangle|^{2} =1-|C^{i,(j)}_{\beta_{1}}|^{2}~,
%= ||~|\beta_{1}\rangle - |\Phi^{i}_{(j)}\rangle||^{2} = 1-|C^{i,(j)}_{\beta_{1}}|^{2}~.
\label{Fubini_study_quantum_dist}
\end{equation} 
where $C^{i,(j)}_{\beta_{1}}$ is the fidelity between the entangled state $|\Phi^{i}_{(j)}\rangle$ and a separable state $|\beta_{1}\rangle$~\cite{horodecki2009quantum}. If $|\Phi^{i}_{(j)}\rangle$ lies in the UV and $|\beta_{1}\rangle$ lies in the IR, the fidelity corresponds to a transition amplitude obtained from the RG flow between UV and IR~\cite{lee2016}. In general, $C^{i,(j)}_{\beta_{1}} = W_{(j)}e^{iF_{(j)}}$, i.e., $0\leq W_{(j)}\leq 1$ and $F_{(j)}$ correspond to the magnitude and phase of the fidelity respectively. Across a quantum critical point, $W_{(j)}$ is expected to display a non-monotonic behaviour~\cite{zanardi2006ground}. We will now obtain the RG evolution of the distance $d$, and observe its behaviour as the stable fixed point is obtained. 
\par\noindent
The RG equation for the Fubini-Study distances $d(\beta_{1},\Phi^{i}_{(j)})$ (eq.\eqref{Fubini_study_quantum_dist}) is obtained using the RG flow of tensor coefficient (eq.\eqref{coefficient renormalization})
\begin{eqnarray}
\Delta d^{2}(\beta_{1},\Phi^{i})^{(j)}&=&-\Delta |C^{i,(j)}|^{2} = |C^{i,(j)}_{\beta_{1}}|^{2}-|C^{i,(j-1)}_{\beta_{1}}|^{2}\nonumber\\
&=&|C^{i,(j)}_{\beta_{1}}|^{2}-|C^{i,(j)}_{\beta_{1}}+\Delta C^{i,(j)}_{\beta_{1}}|^{2}\nonumber\\
&=&-|\Delta C^{i,(j)}_{\beta_{1}}|^{2}-2Re(\bar{C}^{i,(j)}_{\beta_{1}}\Delta C^{i,(j)}_{\beta_{1}})~,
\label{Fubini-Study-distance-flows}
\end{eqnarray}
where $\bar{C}^{i,(j)}_{\beta_{1}}$ are the complex conjugate tensor coefficients.  
\begin{figure} 
\centering
\includegraphics[width=0.8\textwidth]{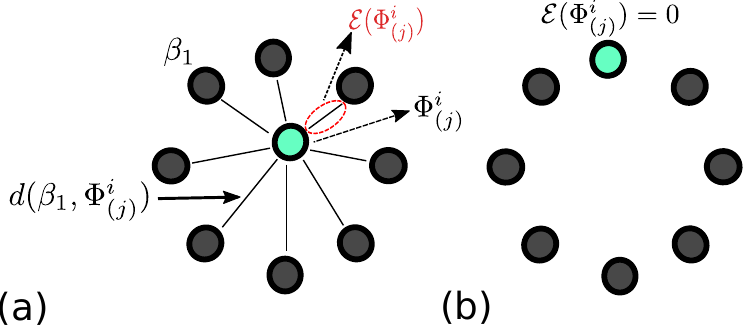}
\caption{(a) The figure represents a collection of orthogonal separable states labelled $\beta_{1}$ (black disks), and where the green disk is an entangled state. The quantum distances between the green disk and black disks $d^{2}(\beta_{1},\Phi)$ are represented by the lines connecting them. The red dashed line encircling the minimum non-zero quantum distance is a geometric measure of entanglement $\mathcal{E}(\Phi^{i}_{(j)})$. (b) The figure shows the situation where the state (green disk) is not entangled.}\label{quantum distances and entanglement}
\end{figure}
 The RG flow of the geometric measure of entanglement ($\mathcal{E}(\Phi^{i}_{(j)})$)\cite{shimony1995degree,wei2003geometric,horodecki2009quantum} for an eigenstate $|\Phi^{i}_{(j)}\rangle$ can now be obtained from the quantum distance RG equations (eq.\eqref{Fubini-Study-distance-flows}) by computing the minimum distance (red dashed circle in Fig.(\ref{quantum distances and entanglement})) of the state $|\Psi^{i}_{(j)}\rangle$ from the product states $|\phi^{(j)}_{l}\rangle$
\begin{eqnarray}
\Delta \mathcal{E}(\Phi^{i}_{(j)}) = \Delta(\text{min}_{\beta_{1}}d^{2}(\beta_{1},\Phi^{i}_{(j)}))~.
\label{entangcontentRG}
\end{eqnarray} 
The RG flow of the entanglement content $\mathcal{E}_{(j)}$ within the \textit{entire} energy eigenbasis (eq.\eqref{Eigenbasis_renormalization}) is defined as $\mathcal{E}_{(j)}=\lbrace \mathcal{E}(\Phi^{1}_{(j)}),..,\mathcal{E}(\Phi^{2^{N-j}}_{(j)})\rbrace$ is represented by the variation in the colour of the legs of tensor network from blue to red in Fig.(\ref{EHMnetwork}). 
\par\noindent
We can see from eq.\eqref{Fubini-Study-distance-flows} that the quantum distance $d^{2}(\beta_{1},\Phi^{i}_{(j^{*})})$ and the geometric measure of entanglement $\mathcal{E}(\Phi^{i}_{(j^{*})})$ will reach a fixed point $j^{*}$ under RG flow, $
\Delta d^{2}(\beta_{1},\Phi^{i}_{(j*)}) = 0 = \Delta \mathcal{E}(\Phi^{i}_{(j*)})$, simultaneously along with the tensor coefficients $\Delta C^{i,(j*)}_{\beta_{1}}=0$.  Given a fluctuation scale $\omega$, the RG relevance and irrelevance of various $n$-particle vertices are guided by the signature in the denominator of the Green's function present in the vertex flow eq.\eqref{vertexflow2}. Figure \ref{quantum distances and entanglement}(a) represents the quantum geometric distances ($d(\beta_{1},\Phi^{i}_{(j)})$) and entanglement measure ($\mathcal{E}(\Phi^{i}_{(j)})$) for RG flow that leads to an emergent subspace with finite entanglement content at the final fixed point. On the other hand, Fig.\ref{quantum distances and entanglement}(b) represents the case when the final low-energy subspace is disentangled. As discussed in an earlier section, the fixed point is determined among the various RG flow equations by considering those in which there is a signature change in the denominator coming from level crossing of fluctuation scale $\omega$ and the renormalized $n$-particle self/correlation energies. In eq.\eqref{quantum_fluc_switch_off}, we saw that the fixed point condition is accompanied by vanishing of the off-diagonal block with respect to state $j*$ (i.e., those terms that change the occupation number of state $j$). At the final fixed point $j^{*}$, there are $2^{j^{*}}$ configurations in $\mathcal{A}_{(j*)}$. A non trivial fixed point with remnant fluctuation in $\mathcal{A}_{(j^{*})}$ implies that these configurations describe a condensate of composite degrees of freedom protected by a many body gap. The Fubini-Study distances between separable states $|\phi^{(j*)}_{l}\rangle$ projected onto the subspace of coupled states $\lbrace 1,\ldots j*\rbrace$ and the entangled eigenstate configurations $|\Phi^{i}_{(j*)}\rangle\in\mathcal{A}_{(j*)}$ of $H_{(j^{*})}(\omega)$ form a \textit{squared-distance} matrix $D$, whose elements are $D_{(l)^{1}_{m},\Phi^{i}_{(j^{*})}}= d^{2}((l)^{1}_{m},\Phi^{i}_{(j*)})$. $D$ has dimensions $dim(D) = 2^{j*}\times 2^{j*}$, and accounts for the dynamical spectral weight transfer at the fixed point (i.e., the electronic spectral weight that has converted to bound states). This can be seen from the following relation connecting
the net Friedel's phase shift (or change in Luttinger volume $\Delta N$ for systems with translational invariance, as will be discussed in more detail in Sec.\ref{bound_state_form}) to $dim(D)$
\begin{eqnarray}
\Delta N = \log_{2}\sqrt{dim(D)}=j^{*}.~\label{dimension_of_distance_matrix_and_spectral_weight}
\end{eqnarray} 
\par\noindent\\
{\bf \textit{Relation between geometric measure of entanglement and composite p-h residues}}\\
The squared minimum distance between separable states $|\phi^{(j)}_{(l)^{1}_{m}}\rangle$ and eigenstate $|\Phi^{i}_{(j)}\rangle\in\mathcal{A}_{(j)}$ (in the $m$ particle-$(p-m)$ hole projected subspace) is given by
 \begin{eqnarray}
\mathcal{E}(\kappa,\Phi^{i}_{(j)})=\text{min}_{\beta}(1-|\langle \beta|\prod_{i=1}^{m}n_{l_{i}}\prod_{i=m+1}^{p}(1-n_{l_{i}})|\Phi^{i}_{(j)}\rangle|^{2})~,~~~
\end{eqnarray} 
where $\kappa=\lbrace (l_{1},1),\ldots,(l_{m},1),(l_{m+1},0),\ldots,(l_{p},0)\rbrace$ is a collection of pairwise indices defined similarly to $\alpha$.
By decomposing the projection operator as a product of composite excitation and de-excitation operators,~$\prod_{i=1}^{m}n_{l_{i}}\prod_{i=m+1}^{p}(1-n_{l_{i}}) = M_{j}^{+}(m,p)M_{j}^{-}(m,p)$, we are able to relate $\mathcal{E}$ to the spectral weight/residue ($Z_{{j}}(\kappa,i)$) of the composite $m$ electron-$(p-m)$ hole associated with the cluster excitation operator $M_{j}^{-}(m,p,\mathcal{P})$
\begin{eqnarray}
\mathcal{E}(\kappa,\Phi^{i}_{(j)}) = 1-Z_{{j}}(\kappa,i)~.\label{GEM_spectral_weight}
\end{eqnarray}
\par\noindent
This results in the quantum fluctuation scale-dependent  renormalization of $m$ particle-$(p-m)$ hole spectral weight: $\Delta Z_{{j}}(\kappa,i)=-\Delta \mathcal{E}(\kappa,\Phi^{i}_{(j)})$.
This relation clearly demonstrates the dynamical nature of the renormalization group: it shows the connection between the phenomenon of UV-IR mixing~\cite{minwalla2000noncommutative} and dynamical spectral weight transfer (as observed, for instance,  between the lower and upper Hubbard bands of a Mott insulator~\cite{phillips2010colloquium}). On the other hand, passage under RG from a fixed point at which $Z_{1}=1~,~\mathcal{E}=0$ (Fermi liquid) to one at which $Z_{1}\to 0~,~\mathcal{E}\to 1$ (non-Fermi liquid) signals a quantum phase transition in which the ground state fidelity (in terms of one-particle excitations) vanishes. This is a realisation of the Anderson orthogonality catastrophe~\cite{anderson1967infrared,yamada1979orthogonality,zanardi2006ground}.
\par\noindent
For the case of an fluctuation scale $\omega$ at which all $n$-particle off-diagonal vertices are RG irrelevant, i.e. $\Delta\Gamma^{2n,(j)}_{(l,\mu)^{1}_{2n}}<0$, the RG flow leads to a number-diagonal Hamiltonian $H^{D}_{(j)}(\omega)$. In fermionic systems with translational invariance, such RG flows approach the Fermi surface. By this, we mean that states at a distance $\Lambda_{j}$ from the Fermi surface(FS) are successively decoupled leading to a more number-diagonal Hamiltonian, i.e. with a lower magnitude of the off-diagonal coefficients. Further, among the $n$-particle self/correlation energies, if only the single-particle self-energy is relevant: $\Delta\Sigma^{2,(j)}_{l}>0$, $\Delta\Sigma^{2n,(j)}_{l}<0~ \forall ~ n>2$, the Fermi liquid fixed point is reached at the FS with a growth of the one-particle residue
\begin{eqnarray}
\lim_{\Lambda_{j}\to 0}
Z_{j}(k_{\Lambda_{j}\hat{s}},\omega)=1-\mathcal{E}(\mathbf{k}_{\Lambda_{j}\hat{s}},\omega)\to 1~,\label{quasi_particle_residue_FL_entanglement}
\end{eqnarray} 
where $\hat{s}$ denotes the directions normal to the FS.
This relations shows the decay of the entanglement measure $\mathcal{E}$ at the Fermi liquid fixed point, resulting in a separable state (green disk in Fig.(\ref{quantum distances and entanglement})(b)). Similar arguments can also be formulated for RG flows that approach a fixed point corresponding to a gapless non-Fermi liquid.
\pin
Finally, the quasiparticle residue defined precisely on the Fermi surface, $Z_{F}=(\langle\hat{n}_{\mathbf{k}_{F\hat{s}}}\rangle)^{2}=1$, allows us to recast Volovik's topological invariant $N_{1}$~\cite{volovik2003universe} along every normal direction $\hat{s}$ for every point on FS (see discussion in Sec.\ref{Fermi_surface}) in terms of the entanglement measure
\begin{eqnarray}
N_{1}= \sqrt{1-\mathcal{E}(\mathbf{k}_{F\hat{s}},\omega)}~.\label{Volovik invariant_entanglement}
\end{eqnarray} 
Remarkably, this relation links the topological stability of the FS to an entanglement property of the FS.
\pin\\
{\bf \textit{Evolution of the Fubini Study metric under RG flow}}\\
The Hilbert space geometry of many-body eigenstates can be quantified via a Fubini-Study metric defined in the space of parameters: polar and azimuthal angles-$(\theta_{l},\phi_{l})$ for electronic states labelled by $l$ ranging between $1$ and $N$. The rotation $\theta_{l},\phi_{l}$ for electronic state $l$ is a ray on a unit Bloch sphere constructed in the occupancy basis: $\lbrace 1_{l},0_{l}\rbrace$. We will now show that the unitary RG evolution of the eigenbasis eq.\eqref{Eigenbasis_renormalization} yields a RG flow of the Fubini-Study metric, thus describing the holographic renormalization of Hilbert space geometry within the bulk of the EHM (see discussion below eq.~\ref{coefficient renormalization}). To begin with, the Fubini-Study distance between a separable state $|\theta,\phi\rangle$ and the many-body eigenstate $|\Psi\rangle$ is given by
\begin{eqnarray}
d^{2}(\theta,\phi,\Psi)=1-|\langle\theta,\phi|\Psi\rangle|^{2}~.
\end{eqnarray}   
We note that $|\theta,\phi\rangle$ is a many-particle separable state whose entanglement arises purely from fermionic statistics. Here, $\theta=\lbrace\theta_{1},\ldots,\theta_{N}\rbrace$ and $\phi=\lbrace \phi_{1},\ldots,\phi_{N}\rbrace$ are a collection of polar of azimuthal angles respectively. The state $|\theta,\phi\rangle$ is constructed by applying a direct product of local unitary rotations in the space of occupied/unoccupied electron states
\begin{eqnarray}
|\theta,\phi\rangle = U(\theta_{1},\phi_{1})\otimes\ldots \otimes U(\theta_{N},\phi_{N})|0\rangle~,
\end{eqnarray}
where 
\begin{equation}
U_{l}=\exp\left(i\frac{\theta}{2}\boldsymbol{\sigma}\dot\hat{n}\right), \hat{n}=cos\phi\hat{x}+sin\phi\hat{y}~,
\end{equation}
and $\boldsymbol{\sigma}=2^{-1}(c^{\dagger}_{l}+c_{l})$,~$2^{-1}i(c^{\dagger}_{l}-c_{l})$, $n_{l}-1/2$.
Upon performing a variation of the distance by infinitesimal variations of $\theta$ and $\phi$, we obtain
\begin{eqnarray}
\delta d^{2}(\theta,\phi,\Psi^{i})&=&\sum_{i,j=1}^{N}g_{\theta_{i}\theta_{j}}\delta\theta_{i}\delta\theta_{j}+g_{\theta_{i}\phi_{j}}\sin\theta_{j}\delta\theta_{i}\delta\phi_{j}+g_{\theta_{j}\phi_{i}}\sin\theta_{j}\delta\theta_{i}\delta\phi_{j}\nonumber\\
&+&g_{\phi_{i}\phi_{j}}\sin\theta_{i}\sin\theta_{j}\delta\phi_{i}\delta\phi_{j}~,
\end{eqnarray}
where the metric 
\begin{eqnarray}
g_{ij}=\langle\partial_{ij},(\theta,\phi)|\Psi\rangle-\langle\partial_{i},(\theta,\phi)|\Psi\rangle\langle\partial_{j},(\theta,\phi)|\Psi\rangle~.
\end{eqnarray}
In the above equation for the metric $g_{ij}$, the labels $(i,j)$ belong to the four possible pairs of parameters, i.e., $(\theta_{i},\theta_{j})$, $(\theta_{i},\phi_{j})$, $(\phi_{i},\theta_{j})$ and $(\phi_{i},\phi_{j})$. The holographic renormalization of the metric is then obtained by incorporating the state space renormalization of eq.\eqref{coefficient renormalization} in the RG relation for the metric 
\begin{eqnarray}
\Delta g^{(l)}_{ij}&=&\langle\partial_{ij},(\theta,\phi)|U_{(j)}|\Psi_{(j)}\rangle-\langle\partial_{i},(\theta,\phi)|U_{(j)}|\Psi_{(j)}\rangle\langle\partial_{j},(\theta,\phi)|U_{(j)}|\Psi_{(j)}\rangle \nonumber\\
&-&\langle\partial_{ij},(\theta,\phi)|\Psi_{(j)}\rangle+\langle\partial_{i},(\theta,\phi)|\Psi_{(j)}\rangle\langle\partial_{j},(\theta,\phi)|\Psi_{(j)}\rangle ~,
\end{eqnarray}
where $l$ is the RG step number.
\pin
Note that the state $|\theta,\phi\rangle$ can be written down as a superposition of occupation number configurations
\begin{eqnarray}
|\theta,\phi\rangle = \sum_{\beta}C_{\beta}(\theta,\phi)|\beta\rangle~.
\end{eqnarray} 
The coefficients are constrained such that for any given bipartition of the state, the Schmidt rank is one~\cite{sperling2011schmidt}. This is the criterion for the separability of the state $|\theta,\phi\rangle$. With the above representation in place, we can write down the RG flow for the quantum metric in terms of the wavefunction coefficient RG flow eq.\eqref{coefficient renormalization}
\begin{eqnarray}
\Delta g^{(l)}_{ij}&=&\sum_{\beta}\partial_{ij}C_{\beta}(\theta,\phi)\Delta C^{(l)}_{\beta}\nonumber\\
&-&\sum_{\beta,\beta'}\partial_{i}C_{\beta}(\theta,\phi)\partial_{j}C_{\beta'}(\theta,\phi)(\Delta C^{(l)}_{\beta}\Delta C^{(l)}_{\beta'}+C^{(l)}_{\beta}\Delta C^{(l)}_{\beta'}+\Delta C^{(l)}_{\beta}\Delta C^{(l)}_{\beta'})~.~~~~
\end{eqnarray}
Note that since the coefficient RG flow $\Delta C_{\beta}^{(j)}$ is generated via vertex renormalization $\Delta\Gamma^{(j)}$ (as seen in eq.\eqref{coefficient renormalization}), the RG flow of the quantum metric is also governed by that of the vertices. This interplay is another important finding of our work, as it provides an explicit demonstration of the holographic principle (or holographic renormalisation) for the case of correlated electrons. Finally, we also note that, upon tracking the geodesic in this metric space, we obtain the RG flow of the geometric measure of entanglement~\cite{shimony1995degree,wei2003geometric} given earlier in eq.\eqref{entangcontentRG}. For the case of the Fermi liquid metal with a gapless Fermi surface discussed earlier in eqs.\eqref{GEM_spectral_weight}-\eqref{Volovik invariant_entanglement}, we find that the journey from UV to IR establishes adiabatic continuity with the non-interacting Fermi gas via the disentanglement of all degrees of freedom in momentum-space. This reflects the self-similarity of the state space generated by the RG flow to an integrable, quantum critical IR theory.
\par\noindent\\
{\bf \textit{Change in entanglement entropy generated in disentangling one electronic state per RG step}}\\
We end this section by accounting for the change in entanglement entropy generated by the process of disentangling a single electronic state at every step of the RG. We begin by writing the state $|\Phi^{i}_{(j)}\rangle$ as a superposition of two many-body entangled states
\begin{eqnarray}
|\Phi^{i}_{(j)}\rangle = \sqrt{a^{i}_{(j)}}|\Phi^{i,1_{j}}_{(j)}\rangle +\sqrt{b^{i}_{(j)}}|\Phi^{i,0_{j}}_{(j)}\rangle, \label{two-state-superpose}
\end{eqnarray}
where $|\Phi^{i,1_{j}}_{(j)}\rangle, |\Phi^{i,0_{j}}_{(j)}\rangle$ are orthogonal many-body states with electron occupancy and non-occupancy for the state $j$ defined following eq.\eqref{superposition_two_sets}
\begin{eqnarray}
|\Phi^{i,1_{j}}_{(j)}\rangle &=& \frac{1}{\sqrt{a^{i}_{(j)}}}\sum_{\alpha_{1}}C^{i,(j)}_{\alpha_{1}}|\alpha_{1}\rangle, a^{i}_{(j)} =\sum_{\alpha_{1}}|C_{\alpha_{1}}^{i,(j)}|^{2},\nonumber\\
|\Phi^{i,0_{j}}_{(j)}\rangle &=& \frac{1}{\sqrt{b^{i}_{(j)}}}\sum_{\alpha_{1}}C^{i,(j)}_{\alpha_{1}}|\alpha_{1}\rangle, b^{i}_{(j)} =\sum_{\beta_{1}}|C_{\beta_{1}}^{i,(j)}|^{2}~.
\end{eqnarray}
We observe that the 
criterion for the distentanglement of the state $j$ involves the vanishing of one of the coefficients of the above linear superposition, say, $a^{i}_{(j-1)}=0$. By placing eq.\eqref{two-state-superpose} into the unitary flow eq.\eqref{unitary_evolution_eqn_state} for the state $|\Phi^{i}_{(j)}\rangle$, and using the constraint equation eq.\eqref{tensor_eqn_constraint}, the vanishing of the coefficient $a^{i}_{(j-1)}$ then leads to
\begin{eqnarray}
\frac{a^{i}_{(j)}}{b^{i}_{(j)}} = \frac{\sum_{\alpha_{1}}\left|\sum^{a_{j}^{max}}_{k=1}
\sum_{\gamma_{1}}\lbrace\Gamma^{2k}_{\alpha_{1}^{'}\gamma_{1}}
G^{4k-2\bar{p}}_{\gamma_{1}\gamma_{1}^{'}}C^{i}_{\gamma_{1}^{'}}\rbrace^{(j)}\right|^{2}}{\sum_{\beta_{1}}|C^{i}_{\beta_{1}}|^{2}}.\label{ratio_of_superposition_coefficients}
\end{eqnarray}
\pin
The reduced single-electron density matrix prior to the RG step $j$ can be computed from eq.\eqref{two-state-superpose} via partial tracing over the states $\lbrace 1,j-1\rbrace = \bar{j}$
\begin{eqnarray}
\rho_{(j),j} &=& Tr_{\bar{j}} (|\Phi^{i}_{(j)}\rangle\langle \Phi^{i}_{(j)}|)\nonumber\\
 &=& |a^{i}_{(j)}|^{2}|1_{j}\rangle\langle 1_{j}|+|b^{i}_{(j)}|^{2}|0_{j}\rangle\langle 0_{j}|~,\nonumber
\end{eqnarray} 
which is clearly a mixed state. Upon disentanglement via the RG step $j$ (see also discussion below eq.\eqref{disentanglement_step}), the single-electron density matrix becomes pure $\rho_{(j-1),j} = |0_{j}\rangle\langle 0_{j}|$. The change in entanglement entropy of the state $j$ is $\Delta S_{EE,(j)}=-Tr(\rho_{(j),j}\log\rho_{(j),j})$. The difference of this entropy gain and the maximum entropy gain possible from the process of disentanglement ($\ln 2$) gives us a measure of probing the quantum entanglement from the perspective of the decoupled states
\begin{equation}
S_{1} = \Delta S_{EE,(j)} -\ln 2 ~.
\end{equation}
The $\ln 2$ is a signature of a maximally mixed single-electron density matrix, i.e., denoting states that were maximally entangled prior to the process of disentanglement. Thus, the quantity $S_{1}$ is a measure of the quantum entanglement content of the decoupled states.
\section{The gapless Fermi surface: Fermi liquid and beyond}
\label{Fermi_surface}
In a strongly correlated electronic system, the electronic spectral weight is widely distributed across various inter-electron interaction-induced scattering channels. Indeed, the phenomena of spectral weight transfer has a long history in the context of Mott Hubbard systems~\cite{harris1967single,eskes1991anomalous,hybertsen1992model, meinders1993spectral,phillips2010colloquium}. 
These studies indicate the breakdown of Landau's paradigm of adiabatic continuity (between the non-interacting electron and the electronic quasiparticle for the Fermi liquid~\cite{landau1959theory}) for the normal state of the Mott-Hubbard system, owing to the strong mixing of spectral weight between ultraviolet and infrared degrees of freedom. 
Instead, in the Mott insulating state at $T=0$, a (Luttinger) surface of zeros is observed for the single particle Green's function from both numerical and analytical techniques~\cite{georges2001quantum,stanescu2006fermi,stanescu2007theory,
sakai2009evolution}. Further, a non-Fermi liquid nature has been proposed for the normal metallic state of such Mott-Hubbard systems, and attributed to the phenomena of UV-IR mixing~\cite{mandal2015ultraviolet}. This appears to be consistent with findings from cluster variants of the dynamical mean-field theory (e.g., CDA+DMFT \cite{vidhyadhiraja2009quantum,mikelsons2009thermodynamics,khatami2010quantum}) and CDMFT~(\cite{gull2010momentum})). As mentioned earlier in Sec.\ref{RG_flow_strategy}, the RG method proposed by us can account for such UV-IR mixing. Thus, in this section, we employ our method in unveiling the physics leading to the breakdown of the Landau quasiparticle in the presence of strong correlations. 
For this, we depict the usage of the unitary decoupling operation eq.\eqref{decoupling condition} towards identifying a composite degree of freedom that can replace the quasiparticles of the Fermi liquid in the \emph{normal state} of strongly correlated systems. The propagator associated with the composite degree of freedom will be shown to preserve the Luttinger volume for the Fermi surface (FS)~\cite{dzyaloshinskii2003some,seki2017topological} (as long as there are no instabilities of the FS). Further, the geometry of the FS will also be shown to be affected by the presence of such composite objects in its immediate neighbourhood. Finally, we will demonstrate the need for a full-fledged RG treatment in deciding whether or not Landau quasiparticles populate the low-energy neighbourhood of the FS.
\newpage
\pin\\
\label{fate_single_p_exc}
{\bf \textit{Fate of single-particle excitations}}\\
Let $|\psi\rangle$ be an eigenstate of a many-particle Hamiltonian, $\hat{H}|\psi\rangle = E|\psi\rangle$, such that 
adding an electronic excitation of momentum $\mathbf{k}$ and spin $\sigma$ to it leads to the following many-body state
\begin{eqnarray}
|\psi_{1_{\mathbf{k}\sigma}}\rangle &=&Z_{1}^{-1/2}c^{\dagger}_{\mathbf{k}\sigma}|\psi\rangle~,~\label{one_electron_exc}
\end{eqnarray}
where $Z_{1}$ is the wavefunction renormalisation known as the quasiparticle residue (and identical to $Z_{F}(1,1)$ in the previous section). If the single-particle occupation number operator $\hat{n}_{\mathbf{k}\sigma}$ commutes with the Hamiltonian, $[H,\hat{n}_{\mathbf{k}\sigma}]=0$, the state $|\psi_{1_{\mathbf{k}\sigma}}\rangle$ is also an eigenstate of $H$ but with a shifted energy. On the other hand, the case of $[H,\hat{n}_{\mathbf{k}\sigma}]\neq 0$ denotes the existence of quantum fluctuations (QF) given by
\begin{eqnarray}
(H-E)|\psi_{1_{\mathbf{k}\sigma}}\rangle = [H,c^{\dagger}_{\mathbf{k}\sigma}]
 c_{\mathbf{k}\sigma}|\psi_{1_{\mathbf{k}\sigma}}\rangle~.\nonumber
\end{eqnarray}
The expression on the R.H.S can, very generally, be decomposed into number diagonal (energy shift) and off-diagonal (QF) parts with respect to the state $|\psi_{1_{\mathbf{k}\sigma}}\rangle$
\begin{eqnarray}
&& (i)~\hat{n}_{\mathbf{k}\sigma}[H,c^{\dagger}_{\mathbf{k}\sigma}]
 c_{\mathbf{k}\sigma}\hat{n}_{\mathbf{k}\sigma}|\psi_{1_{\mathbf{k}\sigma}}\rangle = \Delta  E_{\mathbf{k}\sigma} |\psi_{1_{\mathbf{k}\sigma}}\rangle~,~\\
&&(ii)~(1-\hat{n}_{\mathbf{k}\sigma})[H,c^{\dagger}_{\mathbf{k}\sigma}]
 c_{\mathbf{k}\sigma}\hat{n}_{\mathbf{k}\sigma}|\psi_{1_{\mathbf{k}\sigma}}\rangle = C |\psi_{0_{\mathbf{k}\sigma}}\rangle~,~\label{definition_QF}
 \end{eqnarray}
where  $\langle\psi_{1_{\mathbf{k}\sigma}}|\psi_{0_{\mathbf{k}\sigma}}\rangle =0$. We see that the number diagonal term shifts the energy of the many-body configuration, but preserves the spectral weight of the single particle excitation. The off-diagonal term encodes QF in the  
occupation number space of the state 
$|\psi_{1_{\mathbf{k}\sigma}}\rangle$, preventing the single- particle excitation from being infinitely long-lived. Given the presence of such QF terms, we present below 
the qualitative outcome of the creation of a single particle excitation on a many-body eigenstate in a very general setting of a system of interacting fermions with lattice translational symmetry.
\par
We begin by defining a single particle excitation Hamiltonian (SEH) using the e-h scattering terms,
\begin{eqnarray}
H_{[\mathbf{k}\sigma]} &=& \frac{1}{2}\left([H,c^{\dagger}_{\mathbf{k}\sigma}]c_{\mathbf{k}\sigma}+h.c.\right)~.~~\nonumber
\label{SEH}
\end{eqnarray}
It is important to note that the Hamiltonian $H$ of a system of electrons with four-fermi interactions and lattice translation symmetry can be built 
from the Hamiltonian $H_{[\mathbf{k}\sigma]}$ 
\begin{eqnarray}
H&=&\sum_{\mathbf{k}\sigma}\bigg[Tr_{\mathbf{k}\sigma}(H_{[\mathbf{k}\sigma]}\hat{n}_{\mathbf{k}\sigma})\hat{n}_{\mathbf{k}\sigma}+ \frac{1}{2}\left(c^{\dagger}_{\mathbf{k}\sigma}Tr_{\mathbf{k}\sigma}(H_{[\mathbf{k}\sigma]}c_{\mathbf{k}\sigma})+h.c.\right)\bigg]~,~\hspace*{0.6cm}
\end{eqnarray}
where the first and second terms denote the \textit{energy shift} and \textit{QF} terms associated with the state $|\psi_{1_{\mathbf{k}\sigma}}\rangle$
sum up to give the various scattering terms of the entire Hamiltonian. Thus, for the generic case of a single band of strongly correlated electrons with four-fermionic interactions 
\begin{equation}
H_{SFIM}=\sum_{\mathbf{k}}\epsilon_{\mathbf{k}}\hat{n}_{\mathbf{k}\sigma}+\sum_{\mathbf{k}\mathbf{k}'\mathbf{q}}V^{\sigma\sigma'}_{\mathbf{k}\mathbf{k'}\mathbf{q}}c^{\dagger}_{\mathbf{k}+q\sigma}c^{\dagger}_{\mathbf{k}'-q\sigma'}c_{\mathbf{k}'\sigma'}c_{\mathbf{k}\sigma}~,\label{single_band_general_four_fermion_interacting_model}
\end{equation}
the single-particle excitation Hamiltonian $H_{[\mathbf{k}\sigma]}$ has the form
\begin{eqnarray}
H_{[\mathbf{k}\sigma]}&=&
(\epsilon_{\mathbf{k}}+\sum_{\mathbf{k}'\sigma'}V^{\sigma\sigma'}_{\mathbf{k}\mathbf{k'} 0}\hat{n}_{\mathbf{k}'\sigma'})\hat{n}_{\mathbf{k}\sigma}+\sum_{\mathbf{k}'\sigma',\mathbf{q}}V^{\sigma\sigma'}_{\mathbf{k}\mathbf{k'}\hat{q}}c^{\dagger}_{\mathbf{k}\sigma}c^{\dagger}_{\mathbf{k}'\sigma'}c_{\mathbf{k}'-\mathbf{q}\sigma'}c_{\mathbf{k}+\mathbf{q}\sigma}~.~
\label{SEH_general_four_fermion_interacting_model}
\end{eqnarray}
\pin
The number diagonal and off-diagonal contributions of the SEH, $H_{[\mathbf{k}\sigma]}$, can be written 
in the occupation number representation of the state $|\psi_{1_{\mathbf{k}\sigma}}\rangle$ as
\begin{eqnarray}
H_{[\mathbf{k}\sigma]} &=&\begin{pmatrix}
H_{[\mathbf{k}\sigma],e} & c^{\dagger}_{\mathbf{k}\sigma}T_{[\mathbf{k}\sigma],e-h}\\
T^{\dagger}_{[\mathbf{k}\sigma],e-h}c_{\mathbf{k}\sigma} & H_{[\mathbf{k}\sigma],h}
\end{pmatrix}~,\label{OP_ham}
\end{eqnarray} 
where the energy shifts are obtained from 
\begin{equation}
H_{[\mathbf{k}\sigma],e} = Tr_{\mathbf{k}\sigma}(H_{[\mathbf{k}\sigma]}\hat{n}_{\mathbf{k}\sigma}), H_{[\mathbf{k}\sigma],h} = Tr_{\mathbf{k}\sigma}(H_{[\mathbf{k}\sigma]}(1-\hat{n}_{\mathbf{k}\sigma}))~,
\end{equation}
and the QF are indicated by the term
\begin{equation}
T_{[\mathbf{k}\sigma],e-h} = Tr_{\mathbf{k}\sigma}(H_{[\mathbf{k}\sigma]}c_{\mathbf{k}\sigma})~.
\end{equation}\label{TTrace}
The off-diagonal (QF) elements in eq.\eqref{OP_ham} are associated with the 
mixing between UV and IR degrees of freedom via the occupation number fluctuations of the state $|\psi_{1_{\mathbf{k}\sigma}}\rangle$ (eq.\eqref{definition_QF}).
By solving the decoupling equation eq.\eqref{decoupling condition}, the Hamiltonian eq.\eqref{OP_ham} can be brought into a block diagonal form 
\begin{eqnarray}
H_{[\mathbf{k}\sigma]} = U^{\dagger}_{[\mathbf{k}\sigma]}\begin{pmatrix}
\tilde{H}_{[\mathbf{k}\sigma]}^{e} & 0\\
0 & \tilde{H}_{[\mathbf{k}\sigma]}^{h}
\end{pmatrix}U_{[\mathbf{k}\sigma]}~,~\label{block_diagonal_FS}
\end{eqnarray} 
and where the form of decoupling unitary operator is given by eq.\eqref{Unitary operator}. Following the decomposition shown in eq.\eqref{D-X decomposition}, the block diagonal forms $\tilde{H}_{[\mathbf{k}\sigma]}^{e/h}$ can be resolved into number diagonal piece ($D$) and off-diagonal ($X$) pieces
\begin{eqnarray}
&&\tilde{H}_{[\mathbf{k}\sigma]}^{e/h} - H_{[\mathbf{k}\sigma]}^{e/h} = \Delta H^{e/h,D}_{[\mathbf{k}\sigma]}+\Delta H^{e/h,X}_{[\mathbf{k}\sigma]}~,\label{changed_Hamiltonian}
\end{eqnarray}
where the number diagonal term $H^{e/h,D}_{[\mathbf{k}\sigma]}$ contains the renormalised energy shift terms (i.e. shifts in both self and correlation energies). On the other hand, the off-diagonal term $H^{e/h,X}_{[\mathbf{k}\sigma]}$ contain the renormalised scattering vertices with respect to the remaining coupled single particle states. By putting 
eq.\eqref{block_diagonal_FS} into the Hamiltonian $H$ eq.\eqref{single_band_general_four_fermion_interacting_model},  
and using the following unitary transformation relations for the creation ($c^{\dagger}_{\mathbf{k}\sigma}$) and number ($\hat{n}_{\mathbf{k}\sigma}$) operators 
\begin{eqnarray}
U_{[\mathbf{k}\sigma]}\hat{n}_{\mathbf{k}\sigma}U^{\dagger}_{[\mathbf{k}\sigma]} &=& \frac{1}{2}\left[1+\eta_{[\mathbf{k}\sigma]}+\eta^{\dagger}_{[\mathbf{k}\sigma]}\right] \label{unitary_trans_fermion_number_op}~,~~~\\
U_{[\mathbf{k}\sigma]}c^{\dagger}_{\mathbf{k}\sigma}U^{\dagger}_{[\mathbf{k}\sigma]} &=& \frac{1}{2}c^{\dagger}_{\mathbf{k}\sigma}-\frac{1}{2}[\eta_{[\mathbf{k}\sigma]},c^{\dagger}_{\mathbf{k}\sigma}]-\frac{1}{2}\eta_{[\mathbf{k}\sigma]}c^{\dagger}_{\mathbf{k}\sigma}\eta_{[\mathbf{k}\sigma]}~,~~~~~\label{unitary_trans_fermion_annhilation_op}
\end{eqnarray} 
we obtain the 
renormalized Hamiltonian for the occupied/unoccupied (e/h) block as
\begin{eqnarray}
\tilde H^{e/h} = H + \sum_{\mathbf{k}\sigma}(\Delta H^{e/h,D}_{[\mathbf{k}\sigma]}+\Delta H^{e/h,X}_{[\mathbf{k}\sigma]})~.\label{changed_block}
\end{eqnarray}
\par
The detailed derivation of this renormalisation procedure is presented in the Appendix \ref{Appendix-rearrangement}. Here, the e-h transition operator $\eta_{[\mathbf{k}\sigma]}$  is defined by putting the block-matrix representation of the excitation Hamiltonian (eq.\eqref{OP_ham}) $H_{[k\sigma]}$ in eq.\eqref{e-h transition operator1} and has the form
\begin{eqnarray}
\eta_{[\mathbf{k}\sigma]} = G_{[\mathbf{k}\sigma],h}\Gamma^{4,(0)}_{\mathbf{k}\sigma,\alpha}\tilde{c}^{\dagger}_{\alpha}c_{\mathbf{k}\sigma}~,
\end{eqnarray}
where the Green's function  $G^{4}_{[\mathbf{k}\sigma],h}$ is associated with the intermediate many-body configurations and is given by
\begin{eqnarray}
G_{[\mathbf{k}\sigma],h} = \frac{1}{\hat{\omega} +\epsilon_{\mathbf{k}}\tau_{\mathbf{k}\sigma}+V^{\sigma\sigma_{1}}_{\mathbf{k}\mathbf{k}_{1}}\tau_{\mathbf{k}\sigma}\tau_{\mathbf{k}_{1}\sigma_{1}}}~.\label{green_function}
\end{eqnarray} 
Note that we have used the Einstein summation convention on the indexes $(\mathbf{k}_{1},\sigma_{1})$. In the above Green's function operator eq.\eqref{green_function}, $\tau_{\mathbf{k}\sigma}=\hat{n}_{\mathbf{k}\sigma}-\frac{1}{2}$ is the occupation number operator defined about the electron/hole symmetric point. The operator $\hat{\omega}$ is the \textit{quantum fluctuation operator} defined in eq.\eqref{quantum fluctuation scale}, whose spectral decomposition (given by eq.\eqref{quantum fluctuation eigenvalue}) corresponds to quantum fluctuation energy eigenvalues. These quantum fluctuation energy scales are the correlation/self energies of the number-diagonal configurations of the coupled states as seen from the cluster expansion of the Green's function (eq.\eqref{n-particle_green_fn}). The action of the unitary operator on the single-particle creation/annihilation operator leads to the expansion given in eq.\eqref{unitary_trans_fermion_annhilation_op}, where the commutator between the e-h transition operator and the creation operator appears as the first higher order term
\begin{equation}
U_{[\mathbf{k}\sigma]}c^{\dagger}_{\mathbf{k}\sigma}U^{\dagger}_{[\mathbf{k}\sigma]}\to \frac{1}{2}[\eta_{[\mathbf{k}\sigma]},c^{\dagger}_{\mathbf{k}\sigma}]
 =\frac{1}{2}\tau_{\mathbf{k}\sigma}G_{[\mathbf{k}\sigma],h}\Gamma^{4,(0)}_{\mathbf{k}\sigma,\alpha}\tilde{c}^{\dagger}_{\alpha}~,
\end{equation}
where $\alpha = \lbrace (\mathbf{k}'\sigma',0),((\mathbf{k}'-\mathbf{q})\sigma',1),((\mathbf{k}+\mathbf{q})\sigma,1)\rbrace$. Here $\tilde{c}^{\dagger}_{\alpha}$ represents a 2-electron 1-hole correlated excitation.
\par
\begin{figure}
\centering
\includegraphics[width=0.6\textwidth]{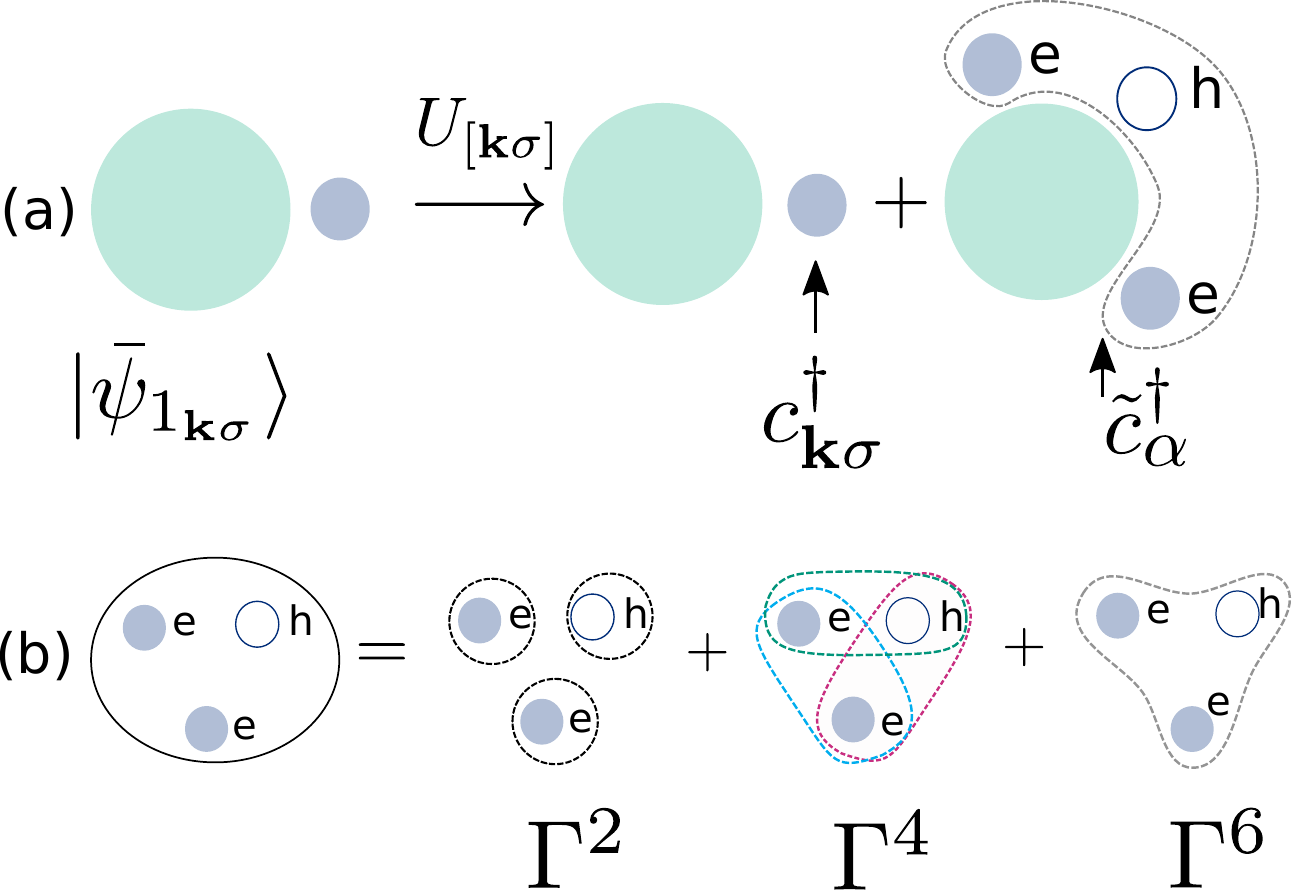} 
\caption{(a) Schematic diagram representing decay of the single-electron spectral weight into the one-electron channel and 2 electron-1 hole channel. (b) A pictorial depiction of the composition of a three- particle dispersion, i.e., one particle self-energy ($\Gamma^{2}$), two-particle correlation energy ($\Gamma^{4}$) and three-particle correlation energy ($\Gamma^{6}$).}\label{fate_spectral_weight}
\end{figure}

We note that a similar expansion was obtained in Ref.\cite{eskes1994spectral} in the context of decoupling total doublon-number subspaces in the 2D Hubbard model. Note that we have used the Einstein summation convention on the indexes $(l,\mu)^{1}_{3}$. In the expression above, $\Gamma^{4,(0)}_{\mathbf{k}\sigma,\alpha}$ is the bare 2-particle (or 4-point) vertex $V_{\mathbf{k}\mathbf{k}'\mathbf{q}}$ in Hamiltonian eq.\eqref{single_band_general_four_fermion_interacting_model}. The two-particle vertex is connected to the leading correlated excitation  $\tilde{c}^{\dagger}_{\alpha}$ (defined in Sec.\ref{RG}), which corresponds here to a two-electron and one-hole (2e-1h) creation operator with indices $(l,\mu)^{1}_{3}$ given by
\begin{equation}
 \alpha = \lbrace (\mathbf{k}'\sigma',0),((\mathbf{k}'-\mathbf{q})\sigma',1),((\mathbf{k}+\mathbf{q})\sigma,1)\rbrace~.
\end{equation}
The members of the set $\alpha$ are constrained indexes that manifest the translation symmetry of the Hamiltonian $H_{1}$, and are responsible for pairwise momentum conservation. This 2e-1h excitation 
configuration is therefore the primary decay channel for the  single-electron excitation (Fig.\ref{fate_spectral_weight}a), as can be seen by the unitary map of the many body state $|\psi_{1_{\mathbf{k}\sigma}}\rangle$ (eq.\eqref{one_electron_exc}) 
\begin{eqnarray}
U_{[\mathbf{k}\sigma]}|\psi_{1_{\mathbf{k}\sigma}}\rangle &=&Z_{1}^{-1/2}\left(U_{[\mathbf{k}\sigma]}c^{\dagger}_{\mathbf{k}\sigma}U^{\dagger}_{[\mathbf{k}\sigma]}\right)U_{[\mathbf{k}\sigma]}|\psi\rangle\nonumber\\
				&=&-\frac{1}{2}Z_{1}^{-1/2}\tau_{\mathbf{k}\sigma}G_{[\mathbf{k}\sigma],h}\Gamma^{4,(0)}_{\mathbf{k}\sigma,\alpha}\tilde{c}^{\dagger}_{\alpha}U_{[\mathbf{k}\sigma]}|\psi\rangle+\frac{1}{2}c^{\dagger}_{\mathbf{k}\sigma}U_{[\mathbf{k}\sigma]}|\psi\rangle~.\label{expansion_excitation_wvfn}
\end{eqnarray}
The rotated state $U_{[\mathbf{k}\sigma]}|\psi\rangle$ is in the hole-occupation subspace corresponding to the label $\mathbf{k}\sigma$, and annihilated by the third term in eq.\eqref{unitary_trans_fermion_annhilation_op}. We can, therefore drop the third term. From this demonstration, we conclude that the spectral weight transfer naturally happens from the single particle excitation to the next term in the expansion of the unitary transformed electron creation eq.\eqref{unitary_trans_fermion_annhilation_op}: the 2-electron 1-hole composite. As shown in eq.\eqref{changed_block}, these changes are also brought about concomitantly in the effective Hamiltonian blocks at a given \textit{QF} scale.  
Thus, the dispersion of these composite objects is given by the change in the number-diagonal part of the Hamiltonian $\Delta H_{e}^{D}(\omega)$
\begin{eqnarray}
\Delta H_{e}^{D}(\omega) = G_{[\mathbf{k}\sigma],h}(\omega)(V^{\sigma\sigma'}_{\mathbf{k}\mathbf{k}'\mathbf{q}})^{2}(1-\hat{n}_{\mathbf{k}'\sigma'})\hat{n}_{\mathbf{k}+\mathbf{q}\sigma}\hat{n}_{\mathbf{k}'-\mathbf{q}\sigma'}~,\label{three particle dispersion}\nonumber\\
\end{eqnarray}
where the 2e-1h projector is equal to the product of the composite e-h excitation/de-excitation operators 
\begin{equation}
(1-\hat{n}_{\mathbf{k}'\sigma'})\hat{n}_{\mathbf{k}+\mathbf{q}\sigma}\hat{n}_{\mathbf{k}'-\mathbf{q}\sigma'} = \tilde{c}^{\dagger}_{\alpha}\tilde{c}_{\alpha}~.
\end{equation}
We recall that such three-particle terms were studied on  phenomenological grounds in Refs.\cite{ruckenstein1991theory} and \cite{ho1998marginal} towards explaining the the linear resistivity of the marginal Fermi liquid. The associated new three-particle number off- diagonal scattering terms that are generated in $\Delta H^{X}_{[\mathbf{k}\sigma],e}$ provide the source of three-particle bound-state formation~\cite{ho1998marginal}.
\pin  
From the cluster decomposition of $\Delta\tilde{H}^{D}$ (see Sec.(\ref{cluster_decomposition})), we obtain the self/correlation energies as
\begin{eqnarray}
\Delta H^{D}(\omega)&=& \sum_{n=1}^{3}\Delta\Gamma^{2n}_{(\mathbf{k}\sigma)^{1}_{n}}\prod_{l=1}^{n}\tau_{\mathbf{k}_{l}\sigma_{l}}~.\label{contributions_cluster}
\end{eqnarray}
This cluster decomposition 
reveals the one-, two- and three-particle contributions (Fig. \ref{fate_spectral_weight}b) to the 2e-1h composite. 
\par\noindent\\
{\bf \textit{One-particle self-energy}}\\
From eq.\eqref{contributions_cluster}, the one-particle components in the 2e-1h composite leads to 
the leading order one-particle self-energy ($\hat{\Sigma}^{I}_{\mathbf{k}'\sigma'}=\Gamma^{2}_{\mathbf{k}'\sigma'}$) in the form of an energy shift of the kinetic energy 
\begin{eqnarray}
\hat{\Sigma}^{I}_{\mathbf{k}'\sigma'} &=& \left[\sum_{\mathbf{k}\sigma}Tr\left(\Delta H^{D}_{[\mathbf{k}\sigma],e}\tau_{\mathbf{k}'\sigma'}\right)\right]\tau_{\mathbf{k}'\sigma'}~.~\hspace*{1cm}\label{self_energy}
\end{eqnarray}                                                                                                                                                                                                                                                                                                                                                                          
For the Hamiltonian eq.\eqref{single_band_general_four_fermion_interacting_model}, the self energy $\Sigma_{\mathbf{k}'\sigma'}$ for the state $\mathbf{k}'\sigma'$ is computed by taking the bare Fermi distribution $\theta(E_{F}-\epsilon_{k})$ at $T=0$ and $E_{F}$ being the Fermi energy, and by considering the contribution from all $\mathbf{k}\sigma$ states withing the energy range $\epsilon_{\mathbf{k}'}>\epsilon_{\mathbf{k}}\geq E_{F}$
\begin{eqnarray}
\hat{\Sigma}^{I}_{\mathbf{k}'\sigma'}(\omega) &=&  \sum_{\mathbf{k}\sigma,\mathbf{q}}(V^{\sigma\sigma'}_{\mathbf{k}\mathbf{k}'\mathbf{q}})^{2}f_{\mathbf{k},\mathbf{k}',q}G_{[\mathbf{k}\sigma],h}(\omega)\tau_{\mathbf{k}'\sigma'} ~.~
\label{Fermi_surface self energy}
\end{eqnarray}
In the above equation, $f_{\mathbf{k},\mathbf{k}',\mathbf{q}}$ is a function that sets the allowed energy ranges of the 2e-1h composite
\begin{eqnarray}f_{\mathbf{k},\mathbf{k}',\mathbf{q}}&=&\theta(\epsilon_{\mathbf{k}+\mathbf{q}}-E_{F})\theta(\epsilon_{\mathbf{k}'-\mathbf{q}}-E_{F})\nonumber\\&&\theta(\epsilon_{\mathbf{k'}}-\epsilon_{\mathbf{k}+\mathbf{q}})\theta(\epsilon_{\mathbf{k'}}-\epsilon_{\mathbf{k}'-\mathbf{q}})~.\nonumber
\end{eqnarray}
The \textit{QF} term asssociated with $\mathbf{q}\neq 0$ scattering terms leads to a self-energy term $\hat{\Sigma}_{\mathbf{k}'\sigma'}(\omega)$ that can be decomposed into a zeroth piece of the self-energy shift and another contribution associated with changes in the shape of the Fermi surface
\begin{eqnarray}
\hat{\Sigma}^{I}_{\mathbf{k}'\sigma'}(\omega) = \hat{\Sigma}^{I,(0)}_{\mathbf{k}'\sigma'}(\omega)+(\hat{\Sigma}^{I}_{\mathbf{k}'\sigma'}(\omega)-\hat{\Sigma}^{I,(0)}_{\mathbf{k}'\sigma'}(\omega))~,~
\end{eqnarray}
where $\hat{\Sigma}^{I,(0)}_{\mathbf{k}'\sigma'}(\omega)$ defined as (here $\Delta \epsilon_{\mathbf{k}} = \epsilon_{\mathbf{k}}-E_{F}$),
\begin{eqnarray}
\hspace*{0cm}\hat{\Sigma}^{I,(0)}_{\mathbf{k}'\sigma'}(\omega) &=& \sum_{\mathbf{k}\sigma}\frac{C^{(0)}}{\omega +\frac{1}{2}\Delta\epsilon_{\mathbf{k}} + V^{(0)}}\theta(\epsilon_{\mathbf{k}'}-\epsilon_{\mathbf{k}})\theta(\epsilon_{\mathbf{k}}-E_{F})~,~\nonumber\\
\hspace*{-2cm}C^{(0)} &=& \frac{1}{\text{Vol}^{2}}\sum_{\mathbf{k}'\mathbf{k}\mathbf{q}\sigma\sigma'} (V^{\sigma\sigma'}_{\mathbf{k}\mathbf{k}'\mathbf{q}})^{2}f_{\mathbf{k},\mathbf{k}',\mathbf{q}}~,~\nonumber\\
V^{(0)} &=& \frac{1}{\text{Vol}^{2}}\sum_{\mathbf{k}'\mathbf{k}\sigma\sigma'}V_{\mathbf{k}\mathbf{k}'}\theta(E_{F}-\epsilon_{\mathbf{k}'})~.~
\end{eqnarray}
\pin
The Fermi surface (FS) geometry is identified by the family of unit vectors $\hat{\mathbf{s}}=\mathbf{v}_{F}/|\mathbf{v}_{F}|$, where $\mathbf{v}_{F}$ are the Fermi surface velocities $\mathbf{v}_{F}=\nabla\epsilon_{\mathbf{k}}|_{\mathbf{k}=\mathbf{k}_{F}}$ at every point on the FS. The self-energy component $\Sigma^{(0)}_{\mathbf{k}'\sigma'}(\omega)$ then leads to 
\begin{eqnarray}
\tilde{\epsilon}_{\mathbf{k}'} &=& \epsilon_{\mathbf{k}'} + \Sigma^{I,(0)}_{\mathbf{k}'\sigma'}(\omega)~,~ \nonumber\\
\nabla_{\mathbf{k}'}\tilde{\epsilon}_{\mathbf{k}'} &=& \left[|\nabla_{\mathbf{k}'}\epsilon_{\mathbf{k}'}| + \frac{C^{(0)}}{\omega +\Delta\epsilon_{\mathbf{k}'}-\frac{V^{(0)}}{4}}|\nabla_{\mathbf{k}'}\epsilon_{\mathbf{k}'}|\right]\hat{s}~,~\nonumber\\
\hat{\tilde{s}} &=& \frac{\nabla_{\mathbf{k}'}\tilde{\epsilon}_{\mathbf{k}'}}{|\nabla_{\mathbf{k}'}\tilde{\epsilon}_{\mathbf{k}'}|}\vert_{\mathbf{k}' =\mathbf{k}_{F}} =\hat{s}~.
\end{eqnarray}
Thus, we find that the zeroth self-energy piece $\Sigma_{\mathbf{k}'\sigma'}^{(0)}(\omega)$ leaves the Fermi surface normal vectors $\hat{s}'s$ invariant preserving the Fermi surface geometry.
\pin\\
{\bf \textit{Universal logarithmic contribution to self-energy from the Fermi surface}}\\
Within the zeroth piece of the self-energy, $\Sigma^{I,(0)}_{\mathbf{k}'\sigma'}(\omega)$, there exists a logarithmic contribution to the \textit{energy shift} arising from the density of states $D(E) = \sum_{\epsilon_{\mathbf{k}}}\delta(E-\epsilon_{\mathbf{k}})$ at the Fermi surface  
\begin{eqnarray}
\Sigma^{I,(0)}_{\mathbf{k}'\sigma'}(\omega)  &=&  \sum_{E=E_{F}}^{E'} \frac{C^{(0)}D(E)}{\omega + (E-E_{F})  + V^{(0)}}\nonumber\\
 &=&\frac{C^{(0)}}{(\epsilon_{\mathbf{k'}}-E_{F})}D(E_{F})\log\left(1+\frac{\epsilon_{\mathbf{k'}}-E_{F}}{\omega - \frac{V^{(0)}}{4}}\right)\nonumber\\
 &+&\sum_{E=E_{F}}^{E'} \frac{C^{(0)}(D(E)-D(E_{F}))}{\omega + (E-E_{F})  - \frac{V^{(0)}}{4}}~.\hspace*{0.5cm}\label{Fermi surface log}
\end{eqnarray}
\par
For instance, for the Hubbard model with $V_{\mathbf{k}\mathbf{k}'\mathbf{q}}^{\sigma\sigma'} = U$, the self-energy term $\Sigma^{(I)}_{\mathbf{k}'\sigma'}(\omega)$ is equal to its zeroth piece
\begin{eqnarray}
\Sigma^{I}_{\mathbf{k}'\sigma'}(\omega) = \Sigma^{I,(0)}_{\mathbf{k}'\sigma'}(\omega)~,~
\end{eqnarray}
leading to the conclusion that, while there is no shape deformation of the Fermi surface caused by the Hubbard repulsion,  there is nevertheless a logarithmic contribution to the self-energy coming from the density of states at the Fermi surface eq.\eqref{Fermi surface log}. In this case, 
the logarithmic singularity of the self-energy shows that the FS is shifted from the non-interacting FS at $\omega = 0$ and $\epsilon_{\mathbf{k}} = E_{F}$ to $\omega\to V^{(0)}/4$ and $\epsilon_{\mathbf{k}}\to E_{F}$, due to the zero momentum transfer $\mathbf{q}=0$ mode forward-scattering amplitudes arising from the 2e-1h composites (eq.(\ref{contributions_cluster})).  
Such log-divergences provide a reason to turn to a renormalization group procedure that takes account of the \textit{QF} term (eq.\eqref{definition_QF}).
The \textit{QF} term can lead to two important possibilities:~(a) the destabilization of the Fermi surface through bound-state formation (as seen via generalized Luttinger surfaces of zeros of the one-particle Green's function~\cite{seki2017topological,stanescu2007theory}) and,~(b) renormalization of the 2e-1h dispersion~. We will provide results obtained for these possibilities from the RG formulation in a later section.
\par
\begin{figure}
\centering
\includegraphics[scale=1]{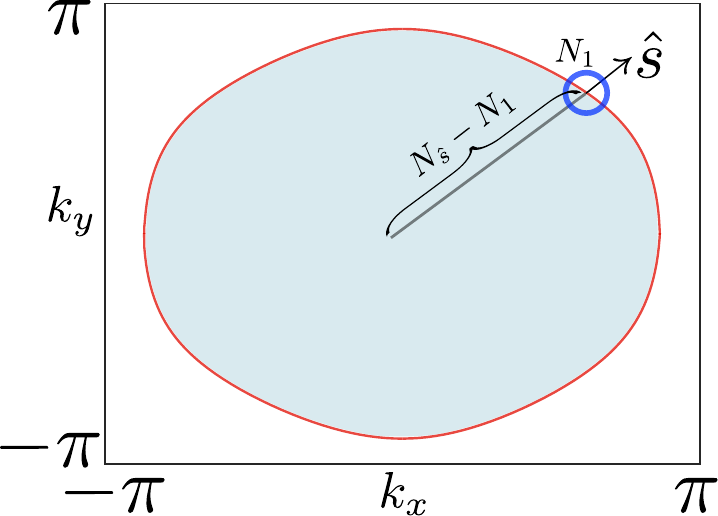}
\caption{Graphical representation of the Fermi volume (light blue region) and Fermi surface (FS, red boundary) for the triangular lattice. The small dark blue circle represents a given point on the FS, and $N_{1}$ is the Volovik invariant associated with this Fermi point (see discussion in text). The line extending from the origin of the Brillouin zone to the FS point represents the partial Luttinger volume $\hat{N}_{\hat{s}}-N_{1}$ for that particular FS point.} 
\label{Fermi_vol}
\end{figure}

Below, we assume that the self-energy contribution $\Sigma^{I}_{\mathbf{k}\sigma}$ of the non-interacting single-particle Green's function respects separate conservation laws for every direction $\hat{s}$ normal to the Fermi surface in the form of Luttinger-Ward identities. We then demonstrate the topological features of the count of occupied states along the orientation $\hat{s}$ associated with a given point on the Fermi surface. This enables the definition of a \textit{Luttinger point}, together with the notion of a 
partial Luttinger sum associated with every Luttinger point.
\par\noindent\\
{\bf \textit{Partial Luttinger sum, Luttinger points}}\\
The single-particle Green's function is given by
\begin{eqnarray}
G_{\mathbf{k}\sigma}(\omega) = \frac{1}{\omega - Tr(H_{[\mathbf{k}\sigma]}\left(\hat{n}_{\mathbf{k}\sigma}-\frac{1}{2}\right))-\Sigma_{\mathbf{k}\sigma}(\omega)}~.
\end{eqnarray}
With the set of reference normal vectors of the Fermi surface $\{\hat{s}\}$, we recast the momentum-space wave-vectors as follows
\begin{eqnarray}
\mathbf{k} &=& \mathbf{k}_{\perp\hat{s}}+\mathbf{k}_{||\hat{s}}~,~ \mathbf{k}_{\perp\hat{s}}\cdot\mathbf{k}_{||\hat{s}} = 0~,\nonumber\\
\mathbf{k}_{||\hat{s}}&=&(\mathbf{k}\cdot\hat{s})\hat{s}~,~ \mathbf{k}_{\perp\hat{s}} = \hat{s}\times(\mathbf{k}\times\hat{s})~.\label{vecs wrt FS}
\end{eqnarray}
With respect to the Fermi surface curvilinear frame of reference, we write the single-particle Green's function in the coordinates of $\hat{s}$ and the distance from FS along $\hat{s}$, $\Lambda = (\mathbf{k}_{||\hat{s}}-\mathbf{k}_{F\hat{s}})\cdot\hat{s}$, as
\begin{eqnarray}
G_{\Lambda\hat{s},\sigma}(\omega) = \frac{1}{\omega - Tr(H_{[\Lambda\hat{s}\sigma]}\left(\hat{n}_{\Lambda\hat{s}\sigma}-\frac{1}{2}\right))-\Sigma_{\Lambda\hat{s}\sigma}(\omega)}.~\hspace*{1cm}
\end{eqnarray}
\begin{theorem} 
If the Luttinger-Ward identity 
\[\partial_{\omega}\Sigma_{\Lambda\hat{s}\sigma}(\omega)+\partial_{\omega}G^{-1}_{\Lambda\hat{s}\sigma}(\omega) =1\] holds for every $\hat{s}$ normal to FS, and if 
\[I_{2,\Lambda\hat{s}} = \sum_{\omega =-\infty}^{\infty}G_{\Lambda\hat{s},\sigma}(\omega)\frac{\partial\Sigma_{\Lambda\hat{s}\sigma}(\omega)}{\partial\omega}=0~,\] then the partial Luttinger sum defined as
\[N_{\hat{s}} = \sum_{\omega,\Lambda,\sigma}G_{\Lambda\hat{s},\sigma}(\omega)\] is an integer, and corresponds to a topological winding in the energy-momentum space along $\hat{s}$.\label{theorem_Luttinger_count}\\
\textbf{Proof:} The single-particle Green's function $G$ is defined as $G^{-1}=G_{0}^{-1}-\Sigma$, where $G_{0}^{-1}= \omega - \epsilon_{\mathbf{k}} $ is the Green's function of the non- interacting problem and $\Sigma$ is the self-energy. The Luttinger Ward identity~\cite{luttinger1960ground} satisfied by the Green's function is:~$\partial_{\omega}G^{-1} + \partial_{\omega}\Sigma=1$~. Following Dzyaloshinskii~\cite{dzyaloshinskii2003some}, the summation of the Green's function over an energy-momentum space contour
\begin{eqnarray}
N &=& \sum_{\Lambda,\hat{s},\sigma}\bigg[\oint dz\frac{\partial}{\partial z}\ln G_{\Lambda,\hat{s},\sigma}(z)^{-1}+ \int dz  G_{\Lambda,\hat{s},\sigma}(z)\frac{\partial}{\partial z} \Sigma_{\Lambda\hat{s},\sigma}(z)\bigg] ~.\label{luttinger_vol}
\end{eqnarray}
equals the number of electrons $N$. Further, if the relation 
\begin{equation}
I_{2\hat{s}} = \sum_{\Lambda}\int  G_{\Lambda\hat{s},\sigma}(z)\frac{\partial}{\partial z} \Sigma_{\Lambda\hat{s},\sigma}(z) = 0 
\end{equation}
holds, $N$ corresponds to the total number of occupied states such that the count stops due to a change of sign of Green's function upon reaching the unoccupied states in energy-momentum space. $N$ can then be written as a sum of integers $\sum_{\hat{s}}N_{\hat{s}}$, where the number $N_{\hat{s}}$ is defined as follows
\begin{eqnarray}
N_{\hat{s}} = \sum_{\Lambda\sigma}\oint dz\frac{\partial}{\partial z}\ln \text{det} G_{\Lambda,\hat{s},\sigma}(z)^{-1}~.\nonumber
\end{eqnarray}
This proves that $N_{\hat{s}}$ is the count of occupied states, the \textit{partial Luttinger sum}, along the direction $\hat{s}$ normal to the Fermi surface, and corresponds to a topological winding number in energy-momentum space centered around a \textit{point on the Fermi surface point} $\mathbf{k} = \mathbf{k}_{F}(\hat{s})$, $E_{F}=\epsilon_{\mathbf{k}_{F}(\hat{s})}$. For every such point on the Fermi surface, there exists an associated Volovik invariant~\cite{volovik2003universe}
\begin{equation}
N_{1} = -\frac{i}{2\pi}\sum_{\sigma}\oint dz G^{-1}_{\Lambda,\hat{s},\sigma}(z)\partial_{z}G_{\Lambda,\hat{s},\sigma}(z)|_{\Lambda =0}~,~\label{volovik}
\end{equation}
such that the partial Luttinger count can be written as a sum of the Volovik invariant at points on the Fermi surface and the partial Luttinger volume leading upto it
\begin{equation}
N_{\hat{s}} = \sum_{\Lambda\neq 0,\sigma}\oint dz\frac{\partial}{\partial z}\ln  G_{\Lambda,\hat{s},\sigma}(z)^{-1} + N_{1}~,
\end{equation} 
where $N_{1}$ is represented by the Fermi surface point (the center of the blue circle) in Fig.\ref{Fermi_vol} and the rest of the partial volume leading upto it is represented by the black line. This completes the proof that a partial Luttinger sum is a topological winding number associated with the existence of a pole at the Fermi surface point. 
\end{theorem}
Further, following Ref.\cite{swingle2012conformal}, the existence of separate Luttinger-Ward identities~\cite{luttinger1960ground} along every direction normal direction to the FS  
allows us to visualize the FS as a collection of 1+1D chiral conformal field theories (CFTs).  
\par\noindent\\
{\bf \textit{Oshikawa's counting argument}}\\
Following Oshikawa~\cite{oshikawa2000topological}, we can now connect the topological invariant $N_{\hat{s}}$ with the change in the center of mass momentum arising from changing boundary conditions along the direction normal to the Fermi surface given by $\hat{s}$. For this, we need a twist operator that changes precisely the momentum of electronic states along $\hat{s}$ 
\begin{equation}
O_{\hat{s}} = \exp\left[2\pi i\sum_{x_{||\hat{s}}}\frac{x_{||\hat{s}}\hat{n}_{\mathbf{x}}}{L}\right]~,~\label{twist_normal_direction}
\end{equation}  
and the center of mass momentum vector along $\hat{s}$ is defined as $P_{cm,\hat{s}} = \sum_{\Lambda}\mathbf{k}_{||\hat{s}}\hat{n}_{\mathbf{k}\sigma}$. Applying the twist operator on the Hamiltonian changes $H\to O_{\hat{s}}HO^{\dagger}_{\hat{s}}$, and the state space $|\psi\rangle \to O_{\hat{s}}|\psi\rangle$. Defining $\hat{T}_{\hat{s}}=\exp[iP_{cm,\hat{s}}]$ as the translation operator, we use the identity 
\begin{eqnarray}
O_{\hat{s}} \hat{T}_{\hat{s}}O^{-1}_{\hat{s}} = \exp\left[i2\pi\sum_{x_{||\hat{s}}}\frac{\hat{n}_{\mathbf{x}}}{L}\right]\hat{T}_{\hat{s}}~,~\nonumber
\end{eqnarray}
to we see that the center of mass momentum along $\hat{s}$ changes as
\begin{equation}
P'_{cm,\hat{s}} =P_{cm,\hat{s}} +  \frac{2\pi}{L}N_{\hat{s}}~.
\end{equation}
This change arises from the fact that, for $I_{2\hat{s}}=0$, the quantity $\sum_{x_{||\hat{s}}}\hat{n}_{\mathbf{x}}$ is preserved in the presence of interactions.  
\par\noindent\\
{\bf \textit{Preservation of partial Luttinger's count}}\\ 
In Theorem (\ref{theorem_Luttinger_count}), we have shown that when $I_{2\hat{s}}=0$, the total particle number is conserved in the presence of interactions
\begin{eqnarray}
N = \sum_{\omega,\Lambda,\hat{s},\sigma}G^{0}_{\Lambda,\hat{s}}(\omega) 
= \sum_{\omega,\Lambda,\hat{s},\sigma}G^{I}_{\Lambda\hat{s}}(\omega)~,\label{Luttinger_VOL}
\end{eqnarray}
where $G^{I}_{\Lambda\hat{s}}(\omega)$ and $G^{0}_{\Lambda\hat{s}}(\omega)$ are the interacting and non-interacting single-particle Green's functions respectively.
The second of these relations is non-trivial, as 
a state count over the entire energy-momentum space for the interacting Green's function $G^{I}_{\Lambda\hat{s}}(\omega)$ involves keeping track of both its poles as well as its zeros. The unchanged Luttinger count in the presence of interactions leads to a relation for $I_{2\hat{s}}$ involving the ratio of $G_{0}$ and $G$~\cite{seki2017topological}
\begin{eqnarray}
I_{2\hat{s}} &=& \sum_{\Lambda}\int_{-\infty}^{0} dz \frac{\partial}{\partial z}\ln \left(\frac{1-\Sigma^{I}_{\Lambda\hat{s}}(\omega)G^{0}_{\Lambda\hat{s}}(\omega)}{1-\Sigma^{I*}_{\Lambda\hat{s}}(\omega)G^{0*}_{\Lambda\hat{s}}(\omega)}\right)~.\nonumber
\end{eqnarray}
We can see that the integral $I_{2\hat{s}}$ becomes equal to the difference of phase of $G_{0}/G$
\begin{eqnarray}
I_{2\hat{s}} &=& \sum_{\Lambda}[\phi_{\Lambda,\hat{s}}(-\infty)-\phi_{\Lambda,\hat{s}}(0)]~, \nonumber 
\end{eqnarray} 
where $\phi_{\Lambda,\hat{s}}(\omega) = \ln \left(\frac{1-\Sigma^{I}_{\Lambda\hat{s}}(\omega)G^{0}_{\Lambda\hat{s}}(\omega)}{1-\Sigma^{I*}_{\Lambda\hat{s}}(\omega)G^{0*}_{\Lambda\hat{s}}(\omega)}\right)$. 
\par
We can now reach some conclusions for the single-particle self-energy $\Sigma^{I}_{\mathbf{k}\sigma}(\omega)$ (eq.\eqref{Fermi_surface self energy}) computed earlier for  the generic interacting Hamiltonian (eq.\eqref{single_band_general_four_fermion_interacting_model}). As $\Sigma^{I}_{\mathbf{k}\sigma}(\omega)$ is analytic at $\omega =0$, the phase difference $[\phi_{\Lambda,\hat{s}}(-\infty)-\phi_{\Lambda,\hat{s}}(0)]=0$ and leads to $I_{2\hat{s}} =0$. The partial Luttinger sum $N_{\hat{s}}$ is then preserved for every $\hat{s}$, and we can use the individual Luttinger-Ward identity for every Fermi point in reconstructing the Luttinger sum for the entire connected Fermi surface. This is despite the fact that, upon the inclusion of two-particle interactions, the resulting three-particle effective Hamiltonian $\Delta H^{D}_{[\mathbf{k}\sigma],e}(\omega)$ (eq.\eqref{three particle dispersion}) leads to a damping of the quasiparticle peak in the single-particle Green's function. However, the concomitant appearance of logarithmic non-analyticities at finite frequencies signals the need for a renormalization group treatment in reaching a firmer conclusion. We will turn to this in a later section.
\section{RG for bound state condensation: gapping the Fermi surface}\label{bound_state_form}
In a strongly coupled electronic system, the destabilization of the gapless Fermi surface is signaled by the appearance of surfaces of zeros of the single-electron Green's function in the complex frequency vs. momentum plane~\cite{dzyaloshinskii2003some,stanescu2007theory}. This surface of zeros brings about a change in Luttinger's sum~\cite{dzyaloshinskii2003some,seki2017topological}, and is accounted for by the Friedel-Levinson phase shift~\cite{friedel1952xiv,levinson1949uniqueness} indicating the number of bound charge composites formed out of a collection of single electronic states. The change in the Luttinger sum has, for instance, been investigated in the context of Anderson impurity models~\cite{martin1982fermi}, and Kondo lattice systems~\cite{coleman2005sum} where a larger Fermi surface replaces the non-interacting Fermi surface in the heavy-electron phase. We recall that, in the context of electronic pairing via a attractive interaction potential, Cooper~\cite{cooper1956bound} had demonstrated bound-state formation out of degenerate electron pairs with zero \textit{pair-momentum} placed outside the Fermi surface. A condensation of such bound Cooper pair states 
leads to the BCS instability~\cite{bardeen1957theory} for the Fermi surface.
\par\noindent
The associated loss of electronic spectral weight in the condensation process is accounted for by the Ferrel-Glover-Tinkham sum rule~\cite{ferrell1958conductivity,tinkham2004introduction} via an addition of the zero-frequency superfluid spectral weight along with the normal state quasiparticle spectral weight at finite frequency. This addition of partial spectral weights, i.e., the spectral weight from the Fermi surface to a cut-off scale together with that beyond the cut-off scale, is required for the conservation of the f-sum rule and denotes the process of \textit{dynamical spectral weight transfer} between high and low energies~\cite{phillips2012advanced}. The cut-off scale itself emerges from some underlying microscopic mechanism, e.g., the Debye cutoff scale for phonon-driven BCS superconductivity. In this way, both Luttinger's sum rule 
and the f-sum rule 
carry signatures of bound state formation. Other examples of the gapping of the Fermi surface, and a subsequent breakdown of the Luttinger sum rule, include the high-T$_{\textrm{C}}$ superconductors~\cite{gros2006determining,sensarma2007can} and doped Mott insulators~\cite{stanescu2007theory,rosch2007breakdown}. Both examples again imply the formation of bound states in these states of matter. 
\par\noindent
Indeed, the pairing of electronic states (e.g., Cooper pairing of $\mathbf{k}\uparrow$ with $-\mathbf{k}\downarrow$) happens together with a projection of the microscopic Hamiltonian and its associated eigenbasis onto a sub-configuration space (e.g., the Anderson-pseudospin subspace with the constraint $\hat{n}_{\mathbf{k}\uparrow}=\hat{n}_{-\mathbf{k}\downarrow}$~\cite{anderson1958random}), enabling an effective description in terms of bound objects. As we shall see below, by starting from a microscopic theory, a renormalization group treatment is best suited towards generating such an effective description in a controlled manner. By starting from a microscopic Hamiltonians for electrons like eq.\eqref{single_band_general_four_fermion_interacting_model}, the renormalization group treatment we have outlined in Sec.(\ref{RG_flow_strategy}) can be used to reach effective Hamiltonians at stable fixed points in terms of paired electronic states or Anderson-like pseudospins~\cite{anderson1958random}. We note that the problem of BCS superconductivity, as well as in theories of nuclear pairing, models belonging to the Richardson class of Hamiltonians~(see Ref.\cite{dukelsky2004colloquium} and references therein) are written in terms of paired electronic state operators or generalized \textit{Anderson pseudospins}. This includes the BCS reduced Hamiltonian~\cite{bardeen1957theory} and the nuclear pairing force models~\cite{bohr1958possible,belyaev1959st}. 
\par\noindent 
In order to the set the stage for a renormalization group analysis, we will demonstrate the generalized Cooper-pairing problem for strongly correlated systems. By creating a two-electron or electron-hole excitation on the eigenstates of the Hamiltonian $H$, and then applying the unitary decoupling operation eq.\eqref{decoupling condition} on it, we will observe the phenomenon of dynamical spectral weight distribution across multiple two-electron or electron-hole pair-momenta channels. From the most singular spectral weight transfer process 
(and its associated correlation energy),  
we will find signatures of bound-state formation in the form of log-divergent T-matrix elements and an associated Friedel's scattering phase shift. This approach is similar to the calculation presented for the Kondo problem in, e.g., Ref.\cite{phillips2012advanced}). These signatures will, very generally, help in identifying the appropriate pairing-force Hamiltonian for strongly correlated electronic systems. We will then verify the connection between the total Friedel's phase shift of the electronic pairs and the change in Luttinger's volume~\cite{martin1982fermi,seki2017topological} of strongly correlated electrons, thereby revealing the \textit{Luttinger surface of zeros}~\cite{georges2001quantum,stanescu2006fermi,stanescu2007theory,
sakai2009evolution} in the reduced Hilbert space of the associated pairing-force Hamiltonians.
\par\noindent\\
{\bf \textit{Outcome of two-particle excitations}}\\
We begin by considering a two-electron (ee) or electron-hole (eh) excitation on an eigenstate $|\psi\rangle$ of a Hamiltonian $H$ with $E$ as its eigenvalue 
\begin{eqnarray}
|\bar{\psi}_{1_{\mathbf{k}\sigma}1_{\mathbf{k}'\sigma'}}\rangle &=& Z_{pp,2}^{-1/2}c^{\dagger}_{\mathbf{k}\sigma}c^{\dagger}_{\mathbf{k}'\sigma'}|\psi\rangle~,~\label{2-electron excitation} \\
|\bar{\psi}_{1_{\mathbf{k}\sigma}0_{\mathbf{k}'\sigma'}}\rangle &=& Z_{ph,2}^{-1/2}c^{\dagger}_{\mathbf{k}\sigma}c_{\mathbf{k}'\sigma'}|\psi\rangle~.\label{electron hole excitation}
\end{eqnarray} 
Our considerations are in the same spirit as Cooper's problem~\cite{cooper1956bound} of placing two electrons in proximity to the effectively noninteracting Fermi sea, but with one major difference: here, $|\Psi\rangle$ is the eigenstate of the complete Hamiltonian $H$. This being the case, the final outcome of such excitations will have contributions from strong electronic correlations present in the Hamiltonian. The action of $H$ on the state $|\bar{\psi}_{1_{\mathbf{k}\sigma}1_{\mathbf{k}'\sigma'}}\rangle$  
leads to a number diagonal two-particle energy shift of the bare energy $E$ 
\begin{eqnarray}
&&\hat{n}_{\mathbf{k}\sigma}\hat{n}_{\mathbf{k}'\sigma'}\left[H,c^{\dagger}_{\mathbf{k}\sigma}c^{\dagger}_{\mathbf{k}'\sigma'}\right]c_{\mathbf{k}'\sigma'}c_{\mathbf{k}\sigma}\hat{n}_{\mathbf{k}\sigma}\hat{n}_{\mathbf{k}'\sigma'}|\bar{\psi}_{1_{\mathbf{k}\sigma}1_{\mathbf{k}'\sigma'}}\rangle\nonumber\\
& =&\Delta E |\bar{\psi}_{1_{\mathbf{k}\sigma}1_{\mathbf{k}'\sigma'}}\rangle~,
\end{eqnarray}
as well as an off-diagonal quantum fluctuation(QF) term (similar to that present in eq.\eqref{definition_QF} for a single electron excitation) induced by two particle scattering
\begin{eqnarray}
&&(1-\hat{n}_{\mathbf{k}\sigma}\hat{n}_{\mathbf{k}'\sigma'})\left[H,c^{\dagger}_{\mathbf{k}\sigma}c^{\dagger}_{\mathbf{k}'\sigma'}\right]c_{\mathbf{k}'\sigma'}c_{\mathbf{k}\sigma}\hat{n}_{\mathbf{k}\sigma}\hat{n}_{\mathbf{k}'\sigma'}|\bar{\psi}_{1_{\mathbf{k}\sigma}1_{\mathbf{k}'\sigma'}}\rangle\nonumber\\
& =& C|\bar{\psi}^{\perp}_{\mathbf{k}\sigma,\mathbf{k}'\sigma'}\rangle~.~\label{2-state_QF}\hspace*{-2cm}
\end{eqnarray}
\begin{figure}
\centering
\includegraphics[scale=0.45]{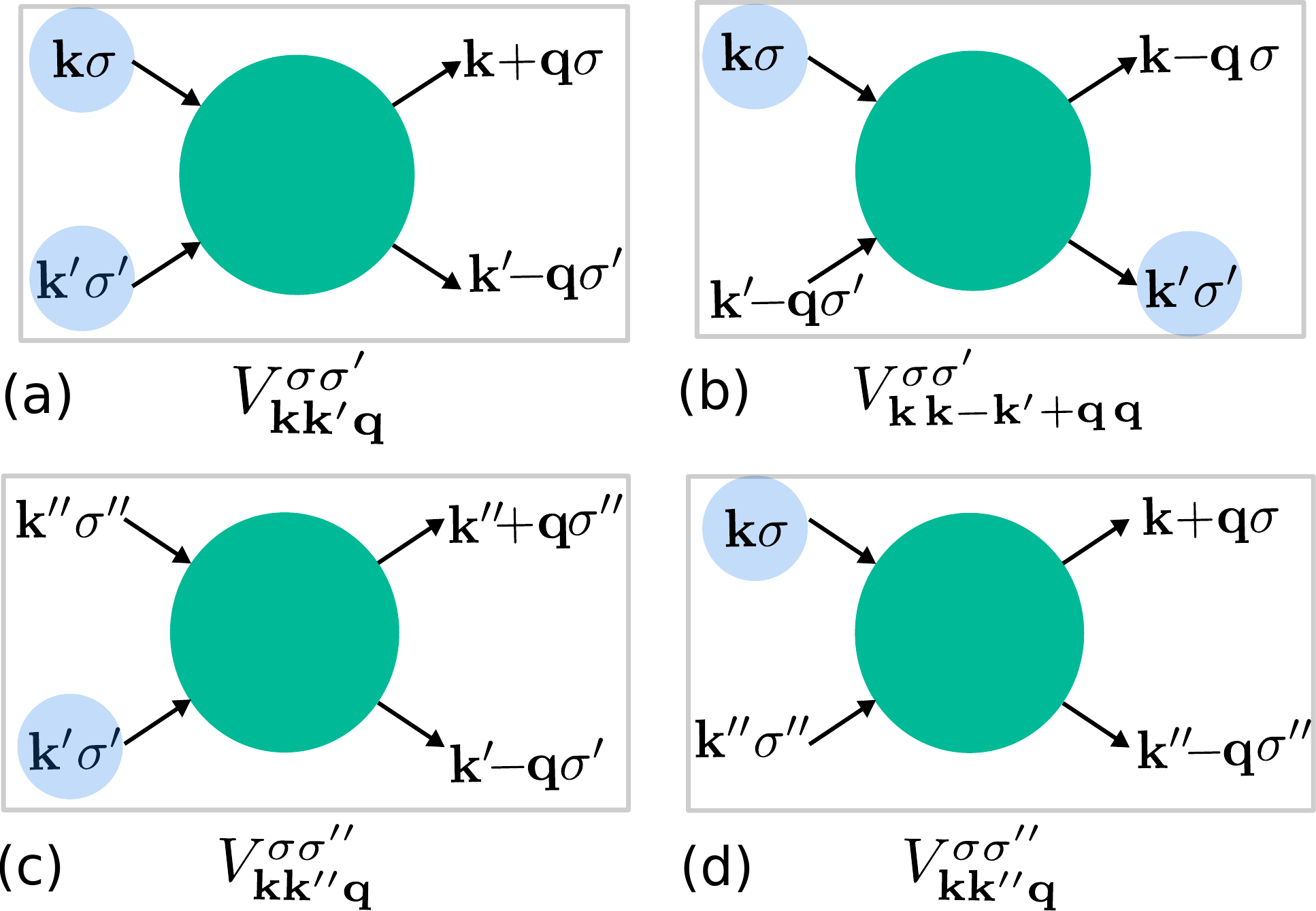}
\caption{Schematic diagram of various 2-particle (i.e., $4$-point) vertices representing the quantum fluctuation terms in the two- particle excitation Hamiltonian. (a) represents the ee/hh 2-particle scattering vertex, (b) represents the eh/he 2-particle scattering vertex and (c,d) represent 2-particle scattering vertices involving another state $\mathbf{k}''\sigma''$.}\label{2-particle_vertices}
\end{figure}
\pin 
A similar set of matrix elements exist for an eh excitation. Using the ee and eh scattering terms and their conjugate processes, the two-particle excitation Hamiltonian (TEH) $H_{[\mathbf{k}\sigma,\mathbf{k}'\sigma']}$ can be written as
\begin{eqnarray}
H_{[\mathbf{k}\sigma,\mathbf{k}'\sigma']} &=& \frac{1}{2}\bigg([H,c^{\dagger}_{\mathbf{k}\sigma}c^{\dagger}_{\mathbf{k}'\sigma'}]c_{\mathbf{k}'\sigma'}c_{\mathbf{k}\sigma}\nonumber\\
&+&[H,c^{\dagger}_{\mathbf{k}\sigma}c_{\mathbf{k}'\sigma'}]c^{\dagger}_{\mathbf{k}'\sigma'}c_{\mathbf{k}\sigma}\bigg)+h.c.~.~\hspace*{0.5cm}\label{TEH}
\end{eqnarray} 
The sub-figures in Fig.(\ref{2-particle_vertices}) represent the QF terms in TEH for the the single-band four-fermion interacting model $H_{1}$ (eq.\eqref{single_band_general_four_fermion_interacting_model}) as follows: (a) represents the ee/hh scattering vertices containing the pair of two-electron excitations (eq.\eqref{2-electron excitation}) while (b) represents the eh/he scattering vertices for electron-hole excitations (eq.\eqref{electron hole excitation}). Sub-figure (c) represents the correlated scattering of state $\mathbf{k}\sigma$ with other electronic states not including $\mathbf{k}'\sigma'$, and (d) represents the same for the state $\mathbf{k}'\sigma'$.
\par\noindent
In order to observe the effect of QF terms (e.g., eq.\eqref{2-state_QF}) on the self/correlation energies and correlated scattering terms, we proceed as in Sec.(\ref{Fermi_surface}) for the case of single-particle excitations. We begin by bringing the Hamiltonian $H_{[\mathbf{k}\sigma,\mathbf{k}'\sigma']}$ into block-diagonal form. This is accomplished by first
by writing the Hamiltonian $H_{[\mathbf{k}\sigma,\mathbf{k}'\sigma']}$ in the form of a block matrix
\begin{eqnarray}
H_{[\mathbf{k}\sigma,\mathbf{k}'\sigma']}=\begin{pmatrix}
H^{1_{\mathbf{k}\sigma}}_{[\mathbf{k}\sigma,\mathbf{k}'\sigma']} & c^{\dagger}_{\mathbf{k}\sigma}T_{[\mathbf{k}\sigma,\mathbf{k}'\sigma'],e-h}\\
T^{\dagger}_{[\mathbf{k}\sigma,\mathbf{k}'\sigma'],e-h}c_{\mathbf{k}\sigma}&H^{0_{\mathbf{k}\sigma}}_{[\mathbf{k}\sigma,\mathbf{k}'\sigma']}\end{pmatrix}~,\hspace*{0.6cm}\nonumber
\end{eqnarray}
and then by decoupling the state $\mathbf{k}\sigma$ in the TEH
\begin{eqnarray}
H_{[\mathbf{k}\sigma,\mathbf{k}'\sigma']}= U^{\dagger}_{[\mathbf{k}\sigma,\mathbf{k}'\sigma']}\begin{pmatrix}
\tilde{H}^{1_{\mathbf{k}\sigma}}_{[\mathbf{k}\sigma,\mathbf{k}'\sigma']} & 0\\
0 & \tilde{H}^{0_{\mathbf{k}\sigma}}_{[\mathbf{k}\sigma,\mathbf{k}'\sigma']} 
\end{pmatrix}U_{[\mathbf{k}\sigma,\mathbf{k}'\sigma']}~.\nonumber
\end{eqnarray}
Note that in the block matrix form of the Hamiltonian $H_{[\mathbf{k}\sigma,\mathbf{k}'\sigma']}$, the off-diagonal blocks contain the electron creation/annihilation operator in product with $T_{[\mathbf{k}\sigma,\mathbf{k}'\sigma'],e-h}$. The defintion of $T_{[\mathbf{k}\sigma,\mathbf{k}'\sigma'],e-h}$ is that given in eq.\eqref{TTrace}, and represents the associated electronic states that comprise the various $n$-particle vertices of the cluster expansion. 
As before, the unitary decoupling operator $U_{[\mathbf{k}\sigma,\mathbf{k}'\sigma']}$ is determined by solving eq.\eqref{decoupling condition}. The e-h transition operator $\eta_{\mathbf{k}\sigma}$ constituting the unitary operator, $U_{[\mathbf{k}\sigma,\mathbf{k}'\sigma']}=\sqrt{2^{-1}}[1+\eta_{\mathbf{k}\sigma}-\eta^{\dagger}_{\mathbf{k}\sigma}]$, is written down in terms of the off-diagonal occupation number fluctuation terms ($c^{\dagger}_{\mathbf{k}\sigma}T_{[\mathbf{k}\sigma,\mathbf{k}'\sigma'],e-h}$) and the number-diagonal many-body Green's function ($G^{e}_{[\mathbf{k}\sigma,\mathbf{k}'\sigma']}=(\hat{\omega} - H^{D}_{[\mathbf{k}\sigma,\mathbf{k}'\sigma']})^{-1}$, using eq.(\ref{e-h transition operator1})):
\begin{equation}
\eta_{\mathbf{k}\sigma} = \frac{1}{\hat{\omega} - H^{D}_{[\mathbf{k}\sigma,\mathbf{k}'\sigma']}}c^{\dagger}_{\mathbf{k}\sigma}T_{[\mathbf{k}\sigma,\mathbf{k}'\sigma'],e-h}~,
\end{equation}
where $\hat{\omega}$ represents the QF operator (eq.\eqref{quantum fluctuation scale}) taking account of the differences between exact energies of the TEH and its diagonal part $H^{D}_{[\mathbf{k}\sigma,\mathbf{k}'\sigma']}$.
\par
\begin{figure}
\centering
\includegraphics[scale=0.4]{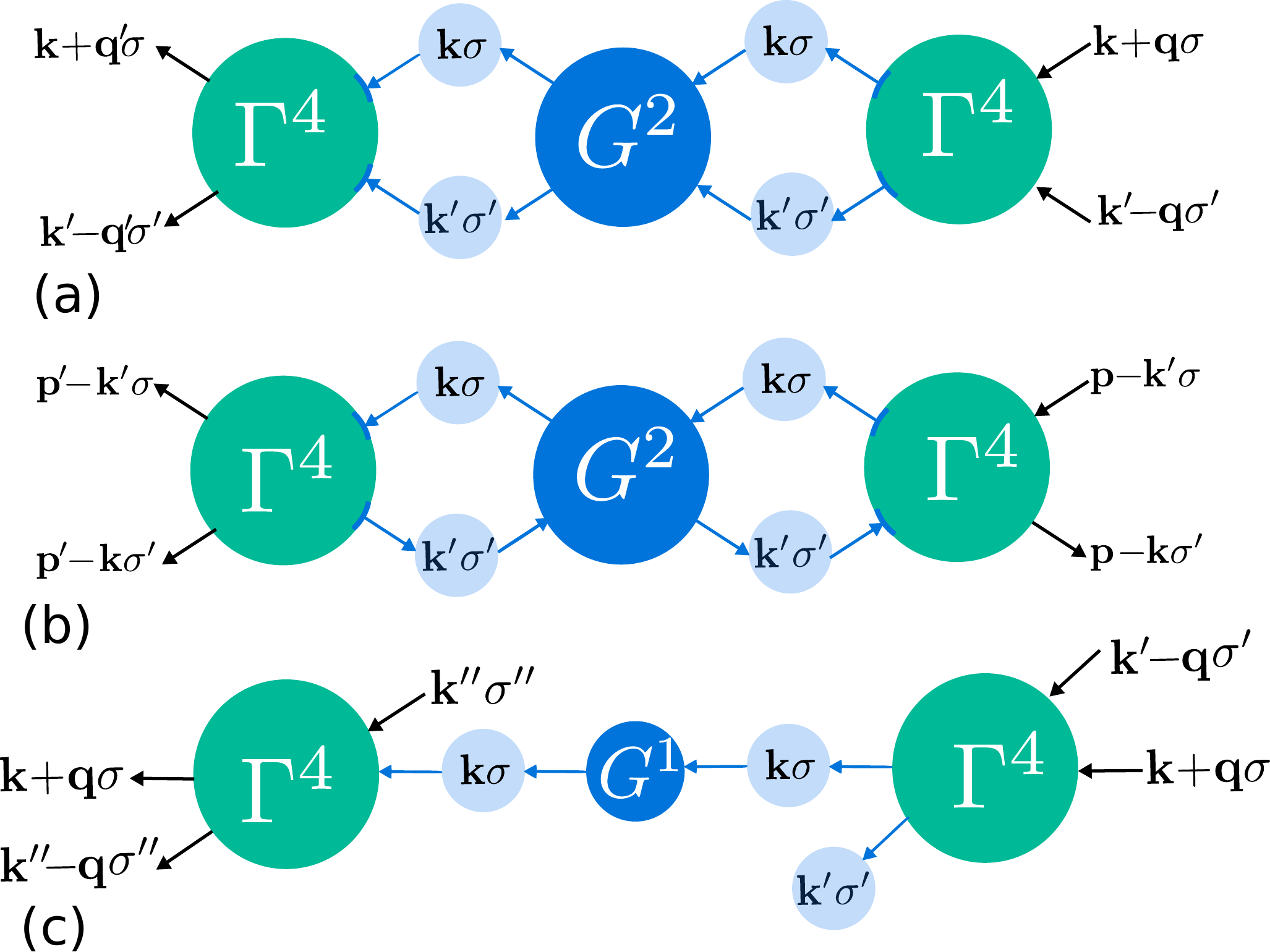}
\caption{The figures represent the 2- and 3-particle vertices that are generated via the unitary decoupling of the state $\mathbf{k}\sigma$ that block diagonalizes the TEH. (a,b) represent the 2-particle scattering vertices (green circles) generated via the ee/hh and eh/he intermediate configurations respectively of the states $\mathbf{k}\sigma$ and $\mathbf{k}'\sigma'$ (light blue circles). These processes involve the intermediate two particle propagator $G^{2}$ (dark blue circles) for the $\mathbf{k}\sigma$ and $\mathbf{k}'\sigma'$ states. (c) represents the three-particle vertex generated via the sandwiching of the single-particle propagator $G^{1}$ in the e/h-configuration of the state $\mathbf{k}\sigma$ by the ee/hh scattering vertex (for states $\mathbf{k}\sigma$, $\mathbf{k}'\sigma'$) on one side, and the scattering vertex of $\mathbf{k}\sigma$ with the state $\mathbf{k}''\sigma''$ on the other side.} \label{three_scattering_pathways}
\end{figure}

\noindent
For the four-fermion interacting model ($H_{1}$), the diagonal piece of TEH is given by 
\begin{equation}
H^{D}_{[\mathbf{k}\sigma,\mathbf{k}'\sigma']} = \epsilon_{\mathbf{k}}\tau_{\mathbf{k}\sigma}+ \epsilon_{\mathbf{k}'}\tau_{\mathbf{k}'\sigma'}+V_{\mathbf{k}\mathbf{k}'}^{\sigma\sigma'}\tau_{\mathbf{k}\sigma}\tau_{\mathbf{k}'\sigma'}~,
\end{equation}
containing both the individual kinetic energy and correlation energy terms. The operator $\tau_{\mathbf{k}\sigma}$ is the occupation number operator $\hat{n}_{\mathbf{k}\sigma}$ defined in a manifestly particle-hole symmetric manner. The one-step renormalization of the two decoupled blocks in the block-diagonal Hamiltonian can be decomposed generically into number-diagonal and number off-diagonal parts
 \begin{eqnarray}
 \tilde{H}^{1_{\mathbf{k}\sigma}}_{[\mathbf{k}\sigma,\mathbf{k}'\sigma']} - H^{1_{\mathbf{k}\sigma}}_{[\mathbf{k}\sigma,\mathbf{k}'\sigma']}&=&\Delta H^{D,1_{\mathbf{k}\sigma}}_{[\mathbf{k}\sigma,\mathbf{k}'\sigma']}+\Delta H^{X,1_{\mathbf{k}\sigma}}_{[\mathbf{k}\sigma,\mathbf{k}'\sigma']} \label{D and X}~,~~~~
 \end{eqnarray}
containing contributions due to QFs in occupation number of state $\mathbf{k}\sigma$ that are generated via 2-particle scattering processes given by Fig.(\ref{2-particle_vertices}(a,b,d)). The two-particle scattering (off-diagonal) and energy shift terms (diagonal) terms present in $\Delta H$ (the RHS of eq.\eqref{D and X}) possess contributions from three classes of processes. The first two of these are: (i) the $ee$ or $hh$ mediated scattering (Fig.\ref{three_scattering_pathways}(a)) with occupied/unoccupied configurations of the states $\mathbf{k}\sigma$ and $\mathbf{k}'\sigma'$ and involving diagram Fig.\ref{2-particle_vertices}(a) and, (ii) the $eh$ or $he$ mediated scattering (Fig.\ref{three_scattering_pathways}(b)) with only one among the states $\mathbf{k}\sigma$ and $\mathbf{k}'\sigma'$ being occupied and involving diagram Fig.\ref{2-particle_vertices}(b). This two processes generate one step renormalization of the two particle vertices. The third process mixes ee/hh and eh/he configurations. This process proceeds as follows: first, an ee/hh (or eh/he) pair of $\mathbf{k}\sigma$ and $\mathbf{k}'\sigma'$ states are created by the 2-particle vertices Fig.\ref{2-particle_vertices}(a) (or (b)) in the intermediate occupation number configuration. Then, that pair is broken due to $\mathbf{k}\sigma$ scattering with other electronic states like $\mathbf{k}''\sigma''$, as shown in Fig.\ref{2-particle_vertices}(d). 
This involves an intermediate single-electron Green's function $G^{1}$ (for the state $\mathbf{k}\sigma$), and the result three particle scattering process is shown in Fig.\ref{three_scattering_pathways}(c). The detailed formulation of these scattering processes is presented in Appendix \ref{Appendix-two_particle_eff}.
\par
\begin{figure}
\centering
\includegraphics[width=0.7\textwidth]{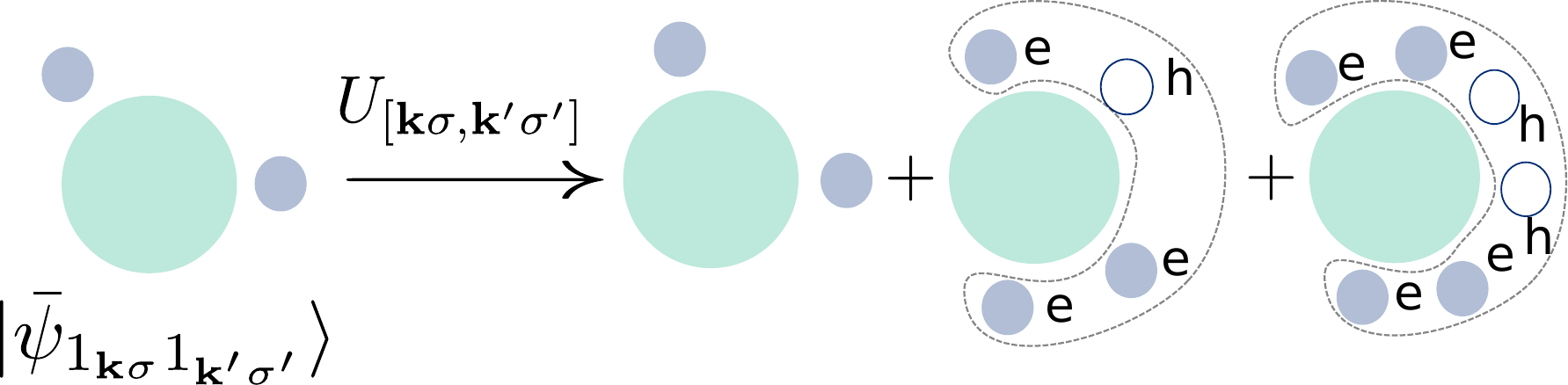}
\caption{The decay channels of the many-body wave function $|\bar{\psi}_{1_{\mathbf{k}\sigma}1_{\mathbf{k}'\sigma'}}\rangle$ containing the two-electron excitation configuration.}\label{2-electron_exc_decay}
\end{figure}

\noindent
The processes (i), (ii) and (iii) described above lead to new energy costs and quantum dynamics of various (ee/hh), (eh/he) and higher ($n$-particle, $m$-hole) composite objects into which the two-electron or electron-hole configurations decay. This can, for instance, be seen from the application of the unitary operator on the state space of the ee/hh excitations (eq.\eqref{2-electron excitation})
\begin{eqnarray}
U_{[\mathbf{k}\sigma,\mathbf{k}'\sigma']}|\bar{\psi}_{1_{\mathbf{k}\sigma}1_{\mathbf{k}'\sigma'}}\rangle &=& Z^{-1}_{pp,2}\tilde{c}^{\dagger}_{\mathbf{k}\sigma}\tilde{c}^{\dagger}_{\mathbf{k}'\sigma'}U_{[\mathbf{k}\sigma,\mathbf{k}'\sigma']}|\psi\rangle~, \label{dyn_spec_weight}
\end{eqnarray}
where $\tilde{c}^{\dagger}_{\mathbf{k}\sigma}=U_{[\mathbf{k}\sigma,\mathbf{k}'\sigma']}c^{\dagger}_{\mathbf{k}\sigma}U^{\dagger}_{[\mathbf{k}\sigma,\mathbf{k}'\sigma']}$ represents the rotated electron creation operators. This rotated operator can be recast in the a cluster expansion as in eq.\eqref{expansion_excitation_wvfn} (see Sec.\ref{cluster_decomposition}). Then, the cluster expansion of the single creation operator will contain (as before) a $1$-electron creation operator and a $2$-electron$+$ $1$-hole (a three-fermion) creation operator. Therefore, the cluster expansion of the product of the rotated two e-creation operator will contain a $2$-e configuration, a $3$-e+$1$-h configuration and a $4$-e+$2$-h configuration. The $4$-e+$2$-h configuration appears at a next-to-leading order in the bare interaction vertex $(\Gamma^{4,(N)})^{2}$, making the $3$-e+$1$-h configuration the leading decay channel for the two-particle excitation (Fig.\ref{2-electron_exc_decay}). This $3$-e+$1$-h excitation is composed of an ee excitation together with an eh excitation, and manifests in the dynamical mixing of the pairs with different net electronic charge. We will now detail this process.
\newpage
\par\noindent\\
{\bf \textit{Dynamical mixing between ee/hh, eh/he pairs}}\\ 
The feedback of the three-particle scattering process (Fig. \ref{three_scattering_pathways}(c)) on ee/hh and eh/he scattering processes (Fig.\ref{three_scattering_pathways}(a,b)) is an outcome of the non-commutativity between two operators. The first of these is the composite-electron creation operator $(1-\hat{n}_{\mathbf{k}\sigma})c^{\dagger}_{\mathbf{k}'\sigma'}$,which is dependent on the occupation of the state $\mathbf{k}'\sigma'$ (as can be seen from the blue $\mathbf{k}'\sigma'$ circle adjacent to the green circle in Fig.  \ref{three_scattering_pathways}(c)), and the second are the ee-hh/eh-he pseudospin operators~\cite{anderson1958random}, $c^{\dagger}_{\mathbf{k}\sigma}c^{\dagger}_{\mathbf{k}'\sigma'}$ and $c^{\dagger}_{\mathbf{k}\sigma}c_{\mathbf{k}'\sigma'}$. The leading contributions of the three-particle vertices resulting from Fig. \ref{three_scattering_pathways}(c) can then be included into the two- particle vertices (Fig.\ref{three_scattering_pathways}(a,b)) by performing a rotation in the space of operators described above. This rotation induces a probabilistic superposition between these two kinds of pairs, with $p$ being the probability coefficient. 
\par\noindent
The composite-electron operator carries 1 unit each of electronic charge and spin, the ee/hh pair operator has a 2 units of charge and the eh/he pair operator has a 0 unit charge. The \textit{spin-charge hybridized} pseudospin excitations resulting out  of the rotation are then given by
\begin{eqnarray}
c^{\dagger}_{\mathbf{k}\sigma}\gamma^{p\dagger}_{\mathbf{k}'\sigma'} =\sqrt{p}c^{\dagger}_{\mathbf{k}\sigma}c^{\dagger}_{\mathbf{k}'\sigma'}+\sqrt{1-p}(c^{\dagger}_{\mathbf{k}\sigma}c_{\mathbf{k}'\sigma'})c^{\dagger}_{\mathbf{k}'\sigma'}~,\nonumber\\
c^{\dagger}_{\mathbf{k}\sigma}\nu^{p\dagger}_{\mathbf{k}'\sigma'} = -\sqrt{1-p}c^{\dagger}_{\mathbf{k}\sigma}c^{\dagger}_{\mathbf{k}'\sigma'}+\sqrt{p}(c^{\dagger}_{\mathbf{k}\sigma}c_{\mathbf{k}'\sigma'})c^{\dagger}_{\mathbf{k}'\sigma'}~,
\end{eqnarray}
and describe \textit{mixed valence configurations} arising out of electronic correlations. Such mixed valence regimes are known to exist in the heavy fermion systems, where they arise from \textit{quantum fluctuations} between different electron occupation number configurations mixing spin and charge degrees of freedom~\cite{Coleman2015}. The basis states that are obtained via rotations of the empty configurations of $\mathbf{k}\sigma$ and $\mathbf{k}'\sigma'$ are (using Appendix \ref{Appendix-two_particle_eff})
\begin{eqnarray}
|1_{\mathbf{k}\sigma}\psi_{\mathbf{k}'\sigma', p}\rangle &=& \sqrt{p}|1_{\mathbf{k}\sigma}1_{\mathbf{k}'\sigma'}\rangle +\sqrt{1-p}|1_{\mathbf{k}\sigma}0_{\mathbf{k}'\sigma'}\rangle \nonumber\\
|1_{\mathbf{k}\sigma}\psi^{\perp}_{\mathbf{k}'\sigma', p}\rangle &=&  \sqrt{1-p}|1_{\mathbf{k}\sigma}1_{\mathbf{k}'\sigma'}\rangle -\sqrt{p}|1_{\mathbf{k}\sigma}0_{\mathbf{k}'\sigma'}\rangle ~.~\label{mixed_valence_configuration}
\end{eqnarray}
\par\noindent 
The spectral decomposition (eq.\eqref{quantum fluctuation eigenvalue}) of the spin-charge hybridized pseudospin Green's functions can now be written down in this basis. Here, we present the Green's function corresponding to the configuration $|1_{\mathbf{k}\sigma}\psi_{\mathbf{k}'\sigma', p}\rangle$
\begin{eqnarray}
G^{e}_{[\mathbf{k}\sigma,\mathbf{k}'\sigma'],p}(\omega) = \frac{1}{\omega -p\epsilon^{ee}_{\mathbf{k}\sigma,\mathbf{k}'\sigma'}-p'\epsilon^{eh}_{\mathbf{k}\sigma,\mathbf{k}'\sigma'}-\frac{1}{4}V^{\sigma\sigma'}_{\mathbf{k},\mathbf{k}'}}~,\label{mixed_valence_green_function}~~~
\end{eqnarray}
where $\epsilon^{ee/eh}_{\mathbf{k}\sigma,\mathbf{k}'\sigma'}=2^{-1}(\tilde{\epsilon}_{\mathbf{k}\sigma}\pm\tilde{\epsilon}_{\mathbf{k}'\sigma'})$ and $V^{\sigma\sigma'}_{\mathbf{k},\mathbf{k}'}$ represent the ee/hh and eh/he pairwise kinetic energies and pair correlation energy respectively, and the probability $p'=1-p$. The energy $\tilde{\epsilon}_{\mathbf{k}\sigma}$ is the electronic dispersion  measured from Fermi energy($E_{F}$), i.e., $\tilde{\epsilon}_{\mathbf{k}\sigma}= \epsilon_{\mathbf{k}\sigma}-E_{F}$, such that states lying outside/inside Fermi sea has positive/negative energy. The magnitude of the spin-charge hybridization term $p\equiv p(\omega)$ is determined by maximizing the two-electron Green's function contribution at a given quantum fluctuation scale $\omega$(i.e., the eigenvalue of the $\hat{\omega}$ operator) in the spin-charge hybridized second-quantized basis of the operators $c^{\dagger}_{\mathbf{k}\sigma}\gamma^{p\dagger}_{\mathbf{k}'\sigma'}$ and $c^{\dagger}_{\mathbf{k}\sigma}\nu^{p\dagger}_{\mathbf{k}'\sigma'}$. With this set up in place, we will determine the two particle self-energies.
\par\noindent\\
{\bf \textit{Self-energy hybridized by ee-eh pair mixing}}\\
The two-electron spin-charge hybridized Green's function $G^{e}_{[\mathbf{k}\sigma,\mathbf{k}'\sigma'],p}$ sandwiched between off-diagonal two-particle scattering vertices (Appendix \ref{Appendix-two_particle_eff}) results in the (ee/hh)/(eh/he) hybridized self- energies. Taking account of the hybridized pseudospin correlation terms present in the renormalized TEH, $\Delta H^{D,1_{\mathbf{k}\sigma}}_{[\mathbf{k}\sigma,\mathbf{k}'\sigma']}$, we obtain the two-particle self-energy
\begin{eqnarray}
\hat{\Sigma}^{2}_{\mathbf{k}\sigma,\mathbf{k}'\sigma',p}(\omega) &=& \sum_{\mathbf{q}\neq 0}G^{e}_{[\mathbf{k}+\mathbf{q}\sigma,\mathbf{k}'-\mathbf{q}\sigma'],p}(V^{\sigma\sigma'}_{\mathbf{k}\mathbf{k}'\mathbf{q}})^{2}f_{\mathbf{k}\mathbf{k}'\mathbf{q}}\tau_{\mathbf{k}\sigma}\tau_{\mathbf{k}'\sigma'}~.\nonumber
\end{eqnarray}
Here, $f_{\mathbf{k}\mathbf{k}'\mathbf{q}}$ (eq.\eqref{Fermi_surface self energy}) represents the restriction of the scattered states energies as follows: $\epsilon_{\mathbf{k}'},\epsilon_{\mathbf{k}}>\epsilon_{\mathbf{k}+\mathbf{q}},\epsilon_{\mathbf{k}'-\mathbf{q}}\geq E_{F}$. The self-energy $\hat{\Sigma}_{\mathbf{k}\sigma , \mathbf{k}'\sigma'}(\omega)$ can be decomposed into a two-particle correlation energy shift ($\hat{\Sigma}^{2,(0)}_{\mathbf{k}\sigma,\mathbf{k}'\sigma',p}$) and terms that are dependent on the lattice geometry
\begin{eqnarray}
\hat{\Sigma}^{2}_{\mathbf{k}\sigma,\mathbf{k}'\sigma',p} = \hat{\Sigma}^{2,(0)}_{\mathbf{k}\sigma,\mathbf{k}'\sigma',p}+(\hat{\Sigma}^{2}_{\mathbf{k}\sigma,\mathbf{k}'\sigma',p}-\hat{\Sigma}^{2,(0)}_{\mathbf{k}\sigma,\mathbf{k}'\sigma',p})~,
\end{eqnarray}
with $\hat{\Sigma}^{2,(0)}_{\mathbf{k}\sigma,\mathbf{k}'\sigma',p}(\omega)$ given by
\begin{eqnarray}
\Sigma^{2,(0)}_{\mathbf{k}\sigma,\mathbf{k}'\sigma',p}(\omega) &=& \sum_{\mathbf{q}\neq 0}\frac{C^{(0)}_{\mathbf{k}\sigma,\mathbf{k}'\sigma'}}{\omega - E^{p}_{\mathbf{k}\mathbf{k}'\mathbf{q}}-\frac{1}{4}V^{\sigma\sigma'(0)}_{\mathbf{k}\mathbf{k}'}}~.\label{2-particle_zeroth_self_energy}
\end{eqnarray}
In the above, $C^{(0)}_{\mathbf{k}\sigma,\mathbf{k}'\sigma'}= N^{-1}\sum_{\mathbf{q}}(V^{\sigma\sigma'}_{\mathbf{k}\mathbf{k}'\mathbf{q}})^{2}f_{\mathbf{k}\mathbf{k}'\mathbf{q}}$ and $V^{\sigma\sigma'(0)}_{\mathbf{k}\mathbf{k}'} = N^{-1}\sum_{\mathbf{q}} V^{\sigma\sigma'}_{\mathbf{k}+\mathbf{q}\mathbf{k}'-\mathbf{q}}$. The hybridized pairwise-energy is given by  $E^{p}_{\mathbf{k}\mathbf{k}'\mathbf{q}} = p\tilde{\epsilon}^{pp}_{\mathbf{k}+\mathbf{q},\mathbf{k}'-\mathbf{q}}+p'\tilde{\epsilon}^{ph}_{\mathbf{k}+\mathbf{q},\mathbf{k}'-\mathbf{q}}$.
\par\noindent
We will show below that one part of the zeroth piece of the hybridized pairwise correlation energy $\Sigma^{2,(0)}_{\mathbf{k}\sigma,\mathbf{k}'\sigma',p}(\omega)$ has a generic logarithmic form in the vicinity of the erstwhile Fermi surface, 
enabling the observation of a pairing instability of the Fermi surface associated with the formation of two-particle bound condensates.  
\par\noindent\\
{\bf \textit{Bound state formation near the Fermi surface}}\\
The $T=0$ Fermi distribution functions in $f_{\mathbf{k}\mathbf{k'}\mathbf{q}}$ (eq.\eqref{Fermi_surface self energy}) cuts off the momentum-space states through a lower cutoff $\mathbf{q}_{min} = 0$, and an upper $\mathbf{q}_{max}$ cutoff given by
\begin{eqnarray}
\text{if }~\epsilon_{\mathbf{k}}<\epsilon_{\mathbf{k}'}\to \mathbf{q}_{max} = \mathbf{k}_{F}-\mathbf{k}~\text{ else }~\mathbf{q}_{max} = \mathbf{k}'-\mathbf{k}_{F}~.\hspace*{1cm}
\end{eqnarray} 
The kinetic energy associated with the wavevectors $\mathbf{q}_{max},\mathbf{q}_{min}$, $\mathbf{k}$ and $\mathbf{k}'$ measures how close excitations can approach Fermi energy, 
as seen from the the hybridized kinetic energy $E_{\mathbf{k}\mathbf{k}'\mathbf{q}_{max}}^{p}$ given by
\begin{eqnarray}
E_{\mathbf{k}\mathbf{k}'\mathbf{q}_{max}}^{p} &=& p(\epsilon_{\mathbf{k}+\mathbf{k}'-\mathbf{k}_{F}}-E_{F})+(1-p)(E_{F}-\epsilon_{\mathbf{k}'+\mathbf{k}-\mathbf{k}_{F}})~,\nonumber\\
E_{\mathbf{k}\mathbf{k}'\mathbf{q}_{min}}^{p} &=& p(\epsilon_{\mathbf{k}}+\epsilon_{\mathbf{k}'}-2E_{F})+(1-p)(\epsilon_{\mathbf{k}}-\epsilon_{\mathbf{k}'})~. 
\end{eqnarray}
Thus, one finds that the constraints on the summation over $\mathbf{q}'s$ in eq.\eqref{Fermi_surface self energy} 
are given by
\begin{eqnarray}
E_{F}<\epsilon_{\mathbf{k}+\mathbf{q}}<\epsilon_{\mathbf{k}}~,~
E_{F}<\epsilon_{\mathbf{k}'-\mathbf{q}}<\epsilon_{\mathbf{k}'}~.
\end{eqnarray}
Using the definition of $\mathbf{q}_{max}$, we then write the summation eq.\eqref{2-particle_zeroth_self_energy} as
\begin{eqnarray}
\hspace*{-1cm}
\Sigma^{2,(0)}_{\mathbf{k}\sigma,\mathbf{k}'\sigma',p}(\omega) &=& \sum_{E^{p}_{\mathbf{k}\mathbf{k}'\mathbf{q}_{min}=0}}^{E^{p}_{\mathbf{k}\mathbf{k}'\mathbf{q}_{max}}} \frac{C^{(0)}_{\mathbf{k}\sigma,\mathbf{k}'\sigma'}g_{\mathbf{k}\mathbf{k}'\mathbf{q}}}{\omega -E^{p}_{\mathbf{k}\mathbf{k}'\mathbf{q}}-\frac{1}{4}V^{\sigma\sigma'(0)}_{\mathbf{k}\mathbf{k}'}}~,
\end{eqnarray}
where $g_{\mathbf{k}\mathbf{k}'\mathbf{q}} = D(\epsilon_{\mathbf{k}+\mathbf{q}})\theta(\epsilon_{\mathbf{k}'}-\epsilon_{\mathbf{k}}) +D(\epsilon_{\mathbf{k}'-\mathbf{q}})\theta(\epsilon_{\mathbf{k}}-\epsilon_{\mathbf{k}'})$. The density of states (DOS) $D(E)$ is defined as usual: $D(E) = \sum_{\mathbf{k}}\delta(E-\epsilon_{\mathbf{k}})$. Writing the DOS about the Fermi surface as $D(E) = D(E_{F}) +D(E)-D(E_{F})$, we have
\begin{eqnarray}
\Sigma^{2,(0)}_{\mathbf{k}\sigma,\mathbf{k}'\sigma',p}(\omega) &=& \sum_{E^{p}_{\mathbf{k}\mathbf{k}'\mathbf{q}_{min}=0}}^{E^{p}_{\mathbf{k}\mathbf{k}'\mathbf{q}_{max}}} \frac{C^{(0)}_{\mathbf{k}\sigma,\mathbf{k}'\sigma'}D(E_{F})}{\omega -E^{p}_{\mathbf{k}\mathbf{k}'\mathbf{q}}-\frac{1}{4}V^{\sigma\sigma'(0)}_{\mathbf{k}\mathbf{k}'} }+\sum_{E^{p}_{\mathbf{k}\mathbf{k}'\mathbf{q}_{min}=0}}^{E^{p}_{\mathbf{k}\mathbf{k}'\mathbf{q}_{max}}} \frac{C^{(0)}_{\mathbf{k}\sigma,\mathbf{k}'\sigma'}(g_{\mathbf{k}\mathbf{k}'\mathbf{q}}-D(E_{F}))}{\omega -E^{p}_{\mathbf{k}\mathbf{k}'\mathbf{q}}-\frac{1}{4}V^{\sigma\sigma'(0)}_{\mathbf{k}\mathbf{k}'} }.\hspace*{0.7cm}~~~~\label{2 particle self energy}
\end{eqnarray} 
The  first summation in eq.\eqref{2 particle self energy} gives a logarithm contribution to the 2-particle self-energy
\begin{eqnarray}
\Sigma^{2,(0)}_{\mathbf{k}\sigma,\mathbf{k}'\sigma',p}(\omega) &\approx &\frac{C^{(0)}_{\mathbf{k}\sigma,\mathbf{k}'\sigma'}}{E^{p}_{\mathbf{k}\mathbf{k}'\mathbf{q}_{min}}-E^{p}_{\mathbf{k}\mathbf{k}'\mathbf{q}_{max}}}D(E_{F})\nonumber\\
&&\times\log\left(1+\frac{E^{p}_{\mathbf{k}\mathbf{k}'\mathbf{q}_{min}}-E^{p}_{\mathbf{k}\mathbf{k}'\mathbf{q}_{max}}}{\omega -\frac{1}{4}V^{\sigma\sigma'}_{\mathbf{k}\mathbf{k}'}}\right)~.~~\label{dressed_self_energy_2}
\end{eqnarray}
This calculation shows that $\Sigma^{2,(0)}_{\mathbf{k}\sigma,\mathbf{k}'\sigma',p}(\omega)$ has a logarithmic non-analyticity at $\omega \to V^{\sigma\sigma'}_{\mathbf{k}\mathbf{k}'}$ and $E_{\mathbf{k}\mathbf{k}'\mathbf{q}_{min}}^{p} = E_{\mathbf{k}\mathbf{k}'\mathbf{q}_{max}}^{p}$~. The leading contribution to this non analyticity exists for total momenta $\mathbf{k}+\mathbf{k'}$ pairs whose energy is \textit{resonant} with the Fermi energy $E_{F}$, satisfying the condition
\begin{eqnarray}
E_{\mathbf{k}\mathbf{k}'\mathbf{q}_{max}}^{p} = 0 \equiv \mathbf{k}+\mathbf{k}'= \mathbf{k}_{F}+\mathbf{k}'_{F}~, \label{mom_constraint}
\end{eqnarray} 
where $\mathbf{k}'_{F}$ is a general Fermi wave vector not necessarily the same as $\mathbf{k}_{F}$. Such a logarithmic term signals an instability of the Fermi surface via a four-fermion interaction with the above pair-momentum constraint (eq.\eqref{mom_constraint}). As mentioned earlier, this is a generalized version of Cooper's pairing instability~\cite{cooper1956bound} for attractive interactions on a circular Fermi surface. We remind the reader that the \textit{dynamical spectral weight transfer} along channels such a logarithmic instability  is observed in the decay of the two-particle excitation (Fig.\ref{2-electron_exc_decay}) from the action of the unitary operator on the two-particle excitation subspace (eq.\eqref{dyn_spec_weight}). 
\par\noindent\\
{\bf \textit{Bound states, Friedel's phase shift and RG flows}}\\
We will now show that the formation of pairwise spin-charge hybridized composites is accompanied by a change in Luttinger's volume~\cite{dzyaloshinskii2003some,stanescu2007theory} via the appearance of surfaces of Luttinger zeros. This change in Luttinger volume is quantified by the Freidel-Levinson phase shift~\cite{langer1961friedel,martin1982fermi}, and can be seen naturally through a scattering-matrix formulation of the above problem. The emergence of pseudospin pairing will, in general, be restricted to a energy-momentum shell $\Lambda^{*}$ around the erstwhile Fermi surface of the non-interacting problem, where $\Lambda^{*}$ is the normal displacement from the Fermi sea (described in text below eq.\eqref{vecs wrt FS}). The $\Lambda^{*}$ momentum-space scale ought to arise from a stable fixed point theory attained via renormalization group procedure implemented on the microscopic model. The RG procedure we have detailed in an earlier section can be used to reach a final stable fixed point theory owing to a frequency dependent self energy feedback in the RG flow equations (eq.\eqref{n-particle_green_fn}), leading to the emergence of the momentum scale $\Lambda^{*}$. 
\par\noindent
From the cluster- and spectral- decompositions of the Hamiltonian RG relation $H_{(j-1)}=U_{(j)}H_{(j)}U^{\dagger}_{(j)}$ and state space renormalization $|\Psi_{(j-1)}\rangle =U_{(j)}|\Psi_{(j)}\rangle$, 
the 4-point vertex flow equation (using eq.\eqref{RG_flow_heirarchy}) and the 2-particle excitation flow equations (using eq.\eqref{2-electron excitation}) can be obtained. 
The cluster expansion of the excitations about the momentum-space number-diagonal configurations is given by
 \begin{eqnarray}
 |\Psi^{i}_{(j)}\rangle = \sum_{n=1}^{a^{j}_{max}}c^{n,(j)}_{\alpha}\tilde{c}^{\dagger}_{\alpha}|\Psi_{D,(j)}^{i}\rangle~,\label{cluster_expansion_wvfn}
 \end{eqnarray}
here $\alpha$ is a set of electronic state labels which are in occupied configuration, and $c^{n,(j)}_{\alpha}$ is the coefficient of the $n$-body cluster. Using eq.\eqref{cluster_expansion_wvfn} with the cluster expansion of the Hamiltonian eq.\eqref{cluster decomposition} at every step of the RG, we find the RG flow equations for the 4-point vertex ($\Gamma^{4}_{\alpha\beta}$) and the coefficient of the 2-body cluster ($c^{2}_{\alpha}$) as
\begin{eqnarray}
\Delta \Gamma^{4,(j)}_{\alpha\beta}(\omega^{i})&=&\sum_{p_{1},p_{3}}^{2a_{j}^{max}}\sum_{\gamma,\gamma'}\{\Gamma^{p_{1}}_{\alpha\gamma}G^{2p_{2}}_{\gamma\gamma'}\Gamma^{p_{3}}_{\gamma'\beta}\}^{(j)}(\omega^{i}),\nonumber\\
\Delta c^{2,(j)}_{\alpha} &=& \frac{3}{4}c^{2,(j)}_{\alpha}+\frac{1}{2}\bigg[G^{4,(j)}_{\beta\beta'}\Gamma^{4,(j)}_{\beta\alpha}c^{2,(j)}_{\beta}+\frac{1}{2}\Delta\Gamma^{4,(j)}_{\alpha\beta}c^{2,(j)}_{\beta}\bigg]~.~~~~~\label{2-electron_flows}
\end{eqnarray}
\par\noindent
At quantum fluctuation energy scales ($\omega$) in the regime 
\begin{equation}
\sum_{i=1}^{n}\epsilon_{l_{i}} >\epsilon^{ee/eh}_{\mathbf{k}\sigma,\mathbf{p}'\sigma'}>\omega > \epsilon^{ee/eh}_{\mathbf{k}\sigma,\mathbf{p}\sigma'}, (\mathbf{k}\sigma,\mathbf{p}\sigma')=(\mathbf{k}\sigma,\mathbf{p}-\mathbf{k}\sigma')~,
\end{equation}
the signature of the Green's function $G^{4}_{\beta\beta'}$ (eq.\eqref{n-particle_green_fn}) is negative, leading to the RG irrelevance 
of all vertices greater than the 4-point vertex: $\Gamma^{2n}$, $n>2$. 
The RG flow equations can, therefore, be simplified to contain only the 4-point vertices with pairing-momentum $\mathbf{p}$. Subsequently, the two-particle Green's function can be resolved in the spin-charge hybridized mixed valence basis (eq.\eqref{mixed_valence_green_function}). Concomitantly, the leading contributer to the state-space renormalization (eq.\eqref{2-electron_flows}) are the two-electron/electron-hole \textit{pseudospins} for $\mathbf{p}$ net momentum, as $\Delta\Gamma^{4,(j)}_{\mathbf{p}}$ has relevant contributions only from 4-point vertices. As the denominator in the hybridized ee/eh Green's function (eq.\eqref{mixed_valence_green_function}) within the flow equations eq.\eqref{2-electron_flows} vanishes, the quantum fluctuation energy scale $\omega$ obtains the exact eigenvalue of the paired electronic states, and we attain a stable fixed point pairing force pseudospin Hamiltonian\cite{dukelsky2004colloquium} along with its renormalized Hilbert space. This will be seen in more detail in a accompanying work for the effective Hamiltonians reached from the four-fermion interacting model eq.\eqref{single_band_general_four_fermion_interacting_model}.
\par\noindent
In the vicinity of the Fermi energy, these pseudospin pairs condense independently along every pairwise normal directions $(\hat{s},\hat{s}')$, as seen from the constraint in eq.(\ref{mom_constraint}). Following this process for every normal direction $\hat{s}$, and at the quantum fluctuation scale $\omega$, a momentum scale $\Lambda^{*}(\omega,\hat{s})$ is generated at the stable fixed point. This corresponds to the low-energy window formed around the Fermi surface (FS) associated with the condensation phenomenon. 
Using the unitary decoupling operator's connection to the scattering matrix (Appendix \ref{unitary_scattering}), we can define the T-matrix at the final fixed point theory. This T-matrix satisfies the generalized optical theorem~\cite{schwartz2014quantum}, as shown in the appendix. We resolve this T-matrix within the low-energy window in the ee/eh mixed-valence configuration of pairwise states $(\mathbf{k},\mathbf{k}')$ (eq.\eqref{mixed_valence_configuration}) along pairwise normal directions $(\hat{s},\hat{s}')$ at a distance $\Lambda$ from the FS fulfilling constraint eq.\eqref{mom_constraint}. The backscattering T-matrix thus obtained has the form 
\begin{eqnarray}
T^{\Lambda}_{\hat{s},\hat{s}'\to -(\hat{s},\hat{s}')} (\omega)
&=& i\frac{V^{\sigma\sigma',\text{eff}}_{\hat{s},\hat{s}'\to-(\hat{s},\hat{s}')}(\omega)}{\omega -E^{p}_{\mathbf{k}\mathbf{k}'}-\Sigma^{2,(0)}_{\mathbf{s}\mathbf{s}'}(\omega)}~,\nonumber\\
T^{\Lambda}_{\hat{s},\hat{s}'\to -(\hat{s},\hat{s}')} (\omega) &=& -(T^{\Lambda}_{\hat{s},\hat{s}'\to -(\hat{s},\hat{s}')} (\omega))^{*}~,~\Lambda < \Lambda^{*}(\omega,\hat{s})~,\nonumber
\end{eqnarray}
where we have carried out a spectral-decomposition of the unitary operator using eq.\eqref{e-h transition operator1}. The backscattering diagrams are present in $H$ if $\mathbf{k}_{F,\hat{s}}+\mathbf{k}_{F,\hat{s}'}=\mathbf{k}_{F,-\hat{s}}+\mathbf{k}_{F,-\hat{s}'}$, or there is an offset in the pair-momentum equal to a reciprocal-lattice vector.  Within a $2\times 2$ subspace of four fermionic states (but with two states occupied), i.e., $|1_{\hat{s}\sigma}1_{\hat{s}'\sigma'}0_{-\hat{s}\sigma}0_{-\hat{s}'\sigma'}\rangle$, $|0_{\hat{s}\sigma}0_{\hat{s}'\sigma'}1_{-\hat{s}\sigma}1_{-\hat{s}'\sigma'}\rangle$, the T-matrix can be written as
\begin{eqnarray}
\hat{T}^{\Lambda}_{\hat{s},\hat{s}'\to -(\hat{s} ,\hat{s}')} = \begin{pmatrix}
 0 & T_{\hat{s}\hat{s}'\to -\hat{s} -\hat{s}'}\\
 T^{*}_{\hat{s}\hat{s}'\to -\hat{s} -\hat{s}'} & 0
\end{pmatrix}~.\nonumber
\end{eqnarray}
In the eigenbasis of bonding (+) and antibonding(-) states, the T matrix elements are given by
\begin{eqnarray}
\hat{T}^{\pm}_{\hat{s}\hat{s}',-\hat{s} -\hat{s}'}(\omega) &=& \pm\frac{1}{2}\bigg\vert\frac{V^{\sigma\sigma',\text{eff}}_{\mathbf{s},\mathbf{s}'\to-(\hat{s},\hat{s}')}(\omega)}{\omega -E^{p}_{\mathbf{k}\mathbf{k}'}-\Sigma^{2,(0)}_{\mathbf{s}\mathbf{s}'}(\omega)}\bigg\vert~.\label{T matrix diags}
\end{eqnarray}
A similar T-matrix calculation for the Kondo problem 
is presented in Ref.\cite{phillips2012advanced}. The change in the Luttinger volume $\Delta N$~\cite{martin1982fermi,dzyaloshinskii2003some} is known to be connected to the Friedel's phase shift. In the same way, the change in the partial Luttinger volume for every normal $\hat{s}$ defined in Theorem \ref{theorem_Luttinger_count} can be connected to the net Friedel's phase shift for states within the low energy window along a normal $\hat{s}$\cite{seki2017topological}
\begin{eqnarray}
N_{\hat{s}}-\bar{N}_{\hat{s}} = -\frac{i}{\pi}\sum_{\Lambda<\Lambda^{*}(\omega,\hat{s})}Tr \ln S^{\Lambda}_{\hat{s},\hat{s}'\to -(\hat{s} ,\hat{s}')}(\omega)~,\label{Friedel's phase shift-formula}
\end{eqnarray} 
where the scattering matrix for the paired states 
\begin{equation}
S^{\Lambda}_{\hat{s},\hat{s}'\to -(\hat{s} ,\hat{s}')}=1+iT^{\Lambda}_{\hat{s},\hat{s}'\to -(\hat{s} ,\hat{s}')}=e^{i\hat{\delta}^{\Lambda}_{\hat{s}\hat{s}'}}
\end{equation}
is written in terms of the T-matrix. The opposite signatures and equal magnitudes of the T matrix elements in eq.\eqref{T matrix diags} leads to a net phase shift $\delta_{\Lambda,\hat{s},\hat{s}',+}+\delta_{\Lambda,\hat{s},\hat{s}',-}=0\text{ or }2\pi$, where $2\pi$ originates from summing the phases $\delta_{\Lambda,\hat{s},\hat{s}'+}=\theta$ from a given Riemann sheet and $2\pi-\theta$ from the next Riemann sheet. These phase shifts are the eigenvalues of the phase operator $\hat{\delta}^{\Lambda}_{\hat{s}\hat{s}'}$. The net phase shift leads to a integer change of $2$ in the Luttinger volume for every pair of electrons at a given distance $\Lambda$ (and involving the pair of normal directions ($\hat{s}, \hat{s}'$)). The change in partial Luttinger sum is given by
\begin{equation}
\Delta N_{\hat{s}}=N_{\hat{s}}-\bar{N}_{\hat{s}} = \sum_{\Lambda\leq\Lambda^{*}(\omega,\hat{s})}2~,\label{Friedel's phase shift-Int}
\end{equation}
such that half of $\Delta N_{\hat{s}}$ counts the number of bound states formed along the normal direction $\hat{s}$. 
\par\noindent
This patch of Luttinger zeros along the normal direction $\hat{s}$ can also be seen through the sensitivity towards boundary conditions by the adiabatic application of a twist operator that affects electronic states along a normal direction (eq.\eqref{twist_normal_direction}) within the fixed point low-energy window. The change in partial Luttinger volume is given by the non-commutativity between twist and translation operator ($T$)  \begin{equation}
\Delta N_{\hat{s}} = \frac{i}{\pi}Tr\ln(T\hat{O}^{L}_{\hat{s},\Lambda^{*}}T^{\dagger}\hat{O}^{\dagger L}_{\hat{s},\Lambda^{*}})~,\label{Friedel's phase shift}
\end{equation}
offering an equivalent topological characteristic observed through an argument involving invariance under a large gauge transformation. $L$ is the total number of states along $\hat{s}$. Summing up this partial Friedel's phase shift for all pair of normal directions, we get the change in Luttinger volume~\cite{seki2017topological}, $\Delta N = \sum_{\hat{s}}\Delta N_{\hat{s}}$. In this way, we find a non-perturbative signature of a connected Luttinger surface of zeros~\cite{georges2001quantum} describing a gapped phase in a strongly correlated system of electrons. 
Further note that in an earlier section Sec.~\ref{eig_base_renorm} we had shown that the pairing of electrons into bound states mitigates the Fermion sign present in the electronic model. Via URG The Hilbert space morphs from an fermionic Hilbert space to a SU(2) spin 1/2 Hilbert space.
The mechanism outlined here displays how a collection of 1+1D chiral conformal field theories (CFTs) composing a Fermi surface~\cite{swingle2012conformal} 
breaks down due to the emergent momentum scale generated via the RG. In a companion work~\cite{anirbanurg2}, we have performed the RG treatment on various microscopic strongly correlated electronic models, with a view towards obtaining therefrom simpler effective models from the stable fixed points of the RG flow. In some of these effective models, we will demonstrate the existence of 
(i) bound state formation with Luttinger zero surfaces, and (ii) two electron 1 hole composite degrees of freedom about a gapless Fermi surface which preserve the Luttinger volume. 

\section{Conclusions and discussions}\label{conclusions}
\pin
The present work formalises as well as extends substantially the unitary renormalisation group (URG) procedure introduced in Refs.\cite{anirbanmotti,anirbanmott2,pal2019} for a finite system of interacting electrons on a lattice, and described by the Hamiltonian framework. In doing so, we obtain a hierarchy of $2n$-point vertex RG flow equations, where all loop-contributions are resummed. By relating the $2n$-point vertices to vertex tensors, we interpret the Hamiltonian renormalization as a vertex tensor network RG scheme. The RG flow for the many-particle eigenspace of the Hamiltonian is generated via the action of the same unitary operations on the eigenstates. This is seen via the RG flow for the coefficient tensors comprising the many body eigenstates, 
generating the entanglement renormalization in the form of an entanglement holographic mapping (EHM). We have recently demonstrated in Ref.\cite{mukherjee2020} an entanglement renormalization group scheme/EHM constructed for the Mott liquid ground state of Ref.\cite{anirbanmotti}. Further, we showed in Ref.\cite{mukherjee2020} the validity of the Ryu-Takayanagi relation for the EHM constructed for the Mott liquid: the entanglement entropy of a subsystem is bounded from above by the area of the minimal surface isolating it from the rest of the system.
\pin 
This brings us to the main result of the present work. The unified RG formalism presented here for the Hamiltonian and its eigenspace, is a mathematical realisation of the holographic principle, i.e., a demonstration of how the entanglement renormalization seen via the EHM~ \cite{mukherjee2020,lee2016,qi2013} is generated from the scattering vertex tensor network RG for the Hamiltonian. In order to understand the EHM better, we define a metric space associated with the Fubini-Study distances~\cite{provost1980riemannian} between a given many-body eigenstate and all possible separable states. The vertex tensor RG is observed to generate holographically the renormalization of the Fubini-Study metric in the bulk of the EHM. In Ref.\cite{chapman2018toward}, the geodesic on the Fubini-Study metric space is shown to be related to circuit complexity. A future direction would be check the ``circuit complexity=volume" conjecture of holographic complexity~\cite{jefferson2017circuit,krishnan2018,hackl2018circuit} for the URG formalism.
\pin 
Importantly, the renormalization of the geodesic on the Fubini-Study metric space is also related to the RG flow for the geometric measure of entanglment~\cite{shimony1995degree,wei2003geometric}. We argue that for gapped phases associated with bound state formation, the renormalization group flow for the geometric measure of entanglement attains a fixed point at a finite value. This describes the remnant entanglement content within the low-energy eigenstates of the IR stable fixed point. On the other hand, the geometric measure vanishes for gapless phases, as momentum-space coordinates remain good quantum numbers under the RG flow. In turn, this implies that the Hilbert space geometry generated along the holographic direction in the IR is very different for gapless as against gapped phases. We have also demonstrated this distinction of entanglement space-time using entanglement based measures (e.g., entanglement entropy, mutual information etc.) in Ref.\cite{mukherjee2020} in a specific case of the Mott liquid phase of the 2D Hubbard model. It may be possible to further extend our study of the quantum geometry of the many-particle Hilbert space by following the methods developed in Refs.\cite{hassan2018,hassan2019}. In a companion work~\cite{anirbanurg2}, we show that the vertex tensor network RG generates a Hamiltonian gauge theory described in terms of nonlocal Wilson loop operators. In this way, we obtain an ab-initio perspective of the gauge/gravity~(holographic spacetime) duality~\cite{maldacena1999large,witten1998anti} from the URG framework. 
\pin 
Another important outcome is that the URG framework offers a renormalisation group perspective of the fermion sign problem, i.e., the appearance of sign factors in the wavefunction coefficient tensor network from the exchange of electrons in the vertex renormalisation functions. First, the fermion exchange sign factors  signifies the complex evolution of multipartite entanglement within the many-particle wavefunction. We find that the stable fixed point theories obtained in the IR are generically free of all fermion exchange sign factors. This mitigation of the fermion sign factors appears to indicate a novel topological mechanism guiding the URG flows from UV to IR~\cite{iazzi2016topological}. Thus, the URG provides a pathway for the discovery of effective Hamiltonians and eigenbases that are fermion-sign free even in problems (i.e., bare Hamiltonians) that possess them~\cite{troyer2005}. Following the strategy adopted in Ref.\cite{mukherjee2020} likely also paves the way for learning the many-particle content of theories that possess fermion signs.
\pin
We have also provided a preview  of the usage of URG towards detecting composite degrees of freedom in problems of correlated electrons. These excitations either replace the Fermi liquid phase by another gapless phase, or generate a many-body gap via the destabilization of the Fermi surface. Both possibilities are explicitly demonstrated as obeying important spectral sum-rules. Specifically, we show that due to strong forward scattering processes, a 2-electron 1-hole degree of freedom can replace the Landau quasiparticle as the excitation proximate to the Fermi surface, and leads to a non-Fermi liquid metal. The nature of such excitations in the Marginal Fermi liquid phase of the 2D Hubbard model has been studied by us in Ref.\cite{anirbanmotti,anirbanmott2}. We have also demonstrated the destabilisation of the Fermi surface towards the formation of bound states. The condensation of such bound states has, for instance, been shown to lead to the Mott liquid phase of the 2D Hubbard model in Ref.\cite{anirbanmotti}. In a companion work~\cite{anirbanurg2}, we perform the URG for two generic models of strong correlated electronic models, one with translational invariance and the other without, in order to demonstrate the emergence of composite degrees of freedom and the effective theories that describe their dynamics. The results obtained from those studies help concretise the URG framework presented in this work. They also offer fresh insight into the criticality of correlated fermions, and the novel states of quantum matter that are emergent therefrom.
\pin\\
\textbf{Acknowledgments}\\
The authors thank R. K. Singh, A. Dasgupta, S. Patra, A. Taraphder, N. S. Vidhyadhiraja, S. Pal and P. Majumdar for several discussions and feedback. A. M. thanks the CSIR, Govt. of India for funding through a junior and senior research fellowship. S. L. thanks the DST, Govt. of India for funding through a Ramanujan Fellowship during which a part of this work was carried out.
\appendix
\renewcommand{\thesection}{\Alph{section}}
\section{Block matrix representation of fermionic operators in single fermion number occupancy basis}\label{block matrix}
\pin
The block matrix representation of fermionic operators using partial trace operations will be demonstrated here. Partial trace operations are prone to fermion sign ambiguities, as shown by Montero and Martinez\cite{montero2011fermionic}, as well as Friis et al~\cite{friis2013fermionic}. We will show how we take care of fermion sign issues, and obtain a block matrix form for fermionic operators in the occupation number basis. 
A general number ordered (N.O.) operator in a $2^{N}$ dimensional fermionic Fock space created out of $N$ single-particle number occupancy spaces labeled by $l\in [1,N]$ is represented as
\begin{eqnarray}
\hat{B}&=&\sum_{i}\hat{B}_{i}~,~ \hat{B}_{i} = \prod_{j=1}^{p_{i}}c^{\dagger}_{l^{i}_{e,j}} \prod_{j=1}^{q_{i}}c_{l^{i}_{h,j}}~,~\nonumber\\
\prod_{j=1}^{p_{i}}c^{\dagger}_{l^{i}_{e,j}} &:=& c^{\dagger}_{l^{i}_{e,1}}c^{\dagger}_{l^{i}_{e,2}}\ldots c^{\dagger}_{l^{i}_{e,p_{i}}}~,~
\end{eqnarray} 
where the indices $l^{i}_{e,j}$ and $l^{i}_{h,j}$ are the state labels acted upon by the electron creation and annihilation operators contained within the $i$th operator $\hat{B}_{i}$.
\par\noindent
\begin{theorem}-
With respect to the single particle number occupancy space labelled by $l$, the operator $\hat{B}$  can be resolved into the following block form d
\begin{eqnarray}
\hat{B} &=& \hat{n}_{l}\otimes U_{l} + (I_{2}\otimes V_{l})(c_{l}\otimes I_{2^{N-1}})\nonumber\\
&+& (c^{\dagger}_{l}\otimes I_{2^{N-1}})(I_{2}\otimes W_{l})+(I_{2}\otimes X_{l})((1-\hat{n}_{l})\otimes I_{2^{N-1}})\nonumber\\
&=&\begin{pmatrix}
U_{l} & W_{l}\\
V_{l} & X_{l}
\end{pmatrix}~.~~~~~
\end{eqnarray}
Above $I_{2}=\hat{n}_{l}+1-\hat{n}_{l}$ represent the $2\times 2$ identity matrix and fermion operators $\hat{n}_{l}$, $c^{\dagger}_{l}$ and $c_{l}$ has the following matrix representation ,
\begin{eqnarray}
\hat{n}_{l}:=\begin{pmatrix}
1 & 0\\
0 & 0
\end{pmatrix}, c^{\dagger}_{l}:=\begin{pmatrix}
0 & 1\\
0 & 0
\end{pmatrix}, c_{l}:=\begin{pmatrix}
0 & 0\\
1 & 0
\end{pmatrix} 
\end{eqnarray}
and $I_{2^{N-1}}$ is the $2^{N-1}\times 2^{N-1}$ identity matrix.\\
\textbf{Definition}:~The partial trace of $\hat{O}$ with respect to state $l$ is defined as ,
\begin{eqnarray}
&&\hspace*{3cm}Tr_{l}(\hat{B})=\sum_{i}Tr_{l}(\hat{B}_{i})\nonumber\\
&&\text{where} ~,~\nonumber\\
&&Tr_{l}(\hat{B}_{i}) = 2\bigg(1-\sum_{j=1}^{p_{i}}\delta_{l^{i}_{e,j},l}\bigg)\bigg(1-\sum_{k=1}^{q_{i}}\delta_{l^{i}_{e,k},l}\bigg)\hat{B}_{i}\nonumber\\
&&\hspace*{-0.5cm}+\sum_{\substack{j'=1,\\k'=1}}^{p_{i},q_{i}}\delta_{l^{i}_{e,j'},l}\delta_{l^{i}_{h,k'},l}\times e^{i\pi[(j'-1)+(q_{i}-k')]}\times\prod_{\substack{j=1,\\j\neq j'}}^{p_{i}}c^{\dagger}_{l^{i}_{e,j}}\prod_{\substack{k=1,\\k\neq k'}}^{q_{i}}c_{l^{i}_{h,j}}.\hspace*{0.5cm}\label{PT_def}
\end{eqnarray}
For the rest of this appendix, we will represent the fermionic operators in the shorthand notation as follows $c^{\dagger}_{l}:=c^{\dagger}_{l}\otimes I_{2^{N-1}}$,
 $c_{l}:=c_{l}\otimes I_{2^{N-1}}$, $\hat{n}_{l}:=\hat{n}_{l}\otimes I_{2^{N-1}}$. Also, the operators $U, V, W$ and $X$ have the following definitions $U_{l}:=I_{2}\otimes U_{l}$, $W_{l}:=I_{2}\otimes W_{l}$, $V_{l}:=I_{2}\otimes V_{l}$, $X_{l}:=I_{2}\otimes X_{l}$. Using the above definition eq.\eqref{PT_def}, the following three identities can be derived
\begin{eqnarray}
\hspace*{-0.8cm}&&
\hat{n}_{l}Tr_{l}(\hat{B}_{i}\hat{n}_{l})=e^{i\pi(p_{i}+q_{i})}\bigg[\bigg(1-\sum_{j=1}^{p_{i}}\delta_{l^{i}_{e,j},l}\bigg)\bigg(1-\sum_{k=1}^{q_{i}}\delta_{l^{i}_{e,k},l}\bigg)\hat{n}_{l}\nonumber\\
\hspace*{-0.8cm}&+&\sum_{\substack{j'=1,\\k'=1}}^{p_{i},q_{i}}\delta_{l^{i}_{e,j'},l}\delta_{l^{i}_{h,k'},l}\bigg]\hat{B}_{i}~,~\label{first_identity}\\
\hspace*{-0.8cm}&&Tr_{l}(c^{\dagger}_{l}\hat{B}_{i})c_{l}= \bigg(1-\sum_{j'=1}^{p_{i}}\delta_{l^{i}_{e,j'},l}\bigg)\sum_{k'=1}^{q_{i}}\delta_{l^{i}_{h,k'},l}\hat{B}_{i}~,~\label{second_identity}\\
\hspace*{-0.8cm}&&c^{\dagger}_{l}Tr_{l}(\hat{B}_{i}c_{l})=\bigg(1-\sum_{k'=1}^{q_{i}}\delta_{l^{i}_{h,k'},l}\bigg)\sum_{j'=1}^{p_{i}}\delta_{l^{i}_{e,j'},l}\hat{B}_{i}~.~\label{three_identities}
\end{eqnarray}
The above three identities lead to the following fourth relation as a corollary
\begin{eqnarray}
Tr_{l}(\hat{B}_{i}(1-\hat{n}_{l}))(1-\hat{n}_{l}) &=& \bigg(2-e^{i\pi(p_{i}+q_{i})}\bigg)\bigg(1-\sum_{j=1}^{p_{i}}\delta_{l^{i}_{e,j},l}\bigg)\nonumber\\
&&\times\bigg(1-\sum_{k=1}^{q_{i}}\delta_{l^{i}_{e,k},l}\bigg)\hat{B}_{i}(1-\hat{n}_{l})~.~~~~\label{fourth_identity}
\end{eqnarray}
The operator $\hat{B}_{i}$ can now be reconstructed by using the partial traced operators (with respect to the state $l$) and multiplied by the triad of operators $\hat{n}_{l}-\frac{1}{2},c^{\dagger}_{l},c_{l}$ using eq.\eqref{first_identity} - eq.\eqref{fourth_identity}\vspace*{-0.2cm}
\begin{eqnarray}
\hat{B}_{i} &=& e^{i\pi(p_{i}+q_{i})}Tr_{l}(\hat{B}_{i}\hat{n}_{l})\hat{n}_{l}+Tr_{l}(c^{\dagger}_{l}\hat{B}_{i})c_{l}+c^{\dagger}_{l}Tr_{l}(\hat{B}_{i}c_{l})\nonumber\\
&+&\bigg(2-e^{i\pi(p_{i}+q_{i})}\bigg)^{-1}Tr_{l}(\hat{B}_{i}(1-\hat{n}_{l}))(1-\hat{n}_{l})~.\nonumber
\end{eqnarray}
Hence, any arbitrary N.O. fermionic operator can be reconstructed in terms of partial traced operators and the triad $\hat{n}_{l}-\frac{1}{2},c^{\dagger}_{l},c_{l}$ as follows
\begin{eqnarray}
\hat{B} &=& \sum_{i}\bigg[e^{i\pi(p_{i}+q_{i})}Tr_{l}(\hat{B}_{i}\hat{n}_{l})\hat{n}_{l}+Tr_{l}(c^{\dagger}_{l}\hat{B}_{i})c_{l}+c^{\dagger}_{l}Tr_{l}(\hat{B}_{i}c_{l})\nonumber\\
&+& \bigg(2-e^{i\pi(p_{i}+q_{i})}\bigg)^{-1}Tr_{l}(\hat{B}_{i}(1-\hat{n}_{l}))(1-\hat{n}_{l})\bigg]~.\label{operator_decomposition}
\end{eqnarray}
The operator decomposition proved above allows for a block matrix representation of the operator $\hat{B}$
\begin{eqnarray}
\hat{B} = \begin{pmatrix}
\sum_{i}e^{i\pi(p_{i}+q_{i})}Tr_{l}(\hat{B}_{i}\hat{n}_{l}) & Tr_{l}(\hat{B}_{i}c_{l}) \\
& \\
Tr_{l}(c^{\dagger}_{l}\hat{B}_{i}) & \sum_{i}\frac{Tr_{l}(\hat{B}_{i}(1-\hat{n}_{l}))}{2-e^{i\pi(p_{i}+q_{i})}}
\end{pmatrix}~.~~~~
\end{eqnarray}
For $\hat{B}$ containing only even number of fermion operators, it has a block matrix form
\begin{eqnarray}
\hat{B} = \begin{pmatrix}
Tr_{l}(\hat{B}_{i}\hat{n}_{l}) & Tr_{l}(\hat{B}_{i}c_{l}) \\
& \\
Tr_{l}(c^{\dagger}_{l}\hat{B}_{i}) & Tr_{l}(\hat{B}_{i}(1-\hat{n}_{l}))
\end{pmatrix}~.\label{block_form_even}
\end{eqnarray}
\end{theorem}
\section{Connection to the continuous unitary transformation (CUT) RG}\label{URGconnectionCUT}
The complete number diagonal Hamiltonian is attained in $n$-steps given by
\begin{eqnarray}
H_{(0)} &=&[U_{(1)}\ldots U_{(N)}]H_{(N)}[U_{(1)}\ldots U_{(N)}]^{\dagger}~.\hspace*{1cm}\label{Unitary_rep_diag}
\end{eqnarray}
This number diagonal Hamiltonian commutes with $N$ local Hermitian operators \[[H^{(0)},\hat{n}_{j}]=0 ,\forall j\in [1,N]\]~, leading to a complete set of local integrals of motion~\cite{you2016entanglement}.
The logarithm of the total unitary operation can be taken to obtain the generator of the complete rotation
\begin{eqnarray}
\hat{G} = -i\log\prod_{j=1}^{N}U_{(j)}~.\label{net_unitary_operation}
\end{eqnarray}
Now, the total unitary transformation can also be carried out as a product of infinitesimal rotations $\delta\theta$ on the configuration space as follows
\begin{eqnarray}
\prod_{j=1}^{N}U_{(j)} &=& \lim_{L\to \infty }\left[\hat{U}(\delta\theta)\right]^{L} =\lim_{L\to \infty }\left[1+\delta\theta \hat{G}\right]^{L},~L\delta\theta =1,~\hat{U}(\delta\theta) =\exp\left[i\delta\theta\hat{G}\right]~.~~~~~~~
\end{eqnarray} 
The generator of the infinitesimal unitary operation $\hat{G}$ can now be related to the canonical generator of continuous unitary transformations based RG~\cite{glazekWilson1993,glazekWilson1994,wegner1994}
\begin{eqnarray}
H(\delta\theta) &=& \hat{U}(\delta\theta)\hat{H} \hat{U}^{\dagger}(\delta\theta)= \hat{H} +i\delta\theta\left[\hat{G},\hat{H}\right]~\Rightarrow \frac{dH(\theta)}{d\theta} = i\left[\hat{G},\hat{H}(\theta)\right]~.
\end{eqnarray} 
In this implementation, all the single electron states become partially disentangled at every RG step via an infinitesimal amount of rotation in the associated Hilbert space.
\section{Highest n-particle vertex at the RG step $j$}
\label{n-particle vertex highest}
Let $a_{N}$ be the order of the highest $n$-particle (i.e., $2n$-point) vertex $\Gamma^{a_{N},(N)}$ in the bare Hamiltonian. In the Hamiltonian RG iteration procedure, the next highest off-diagonal scattering element is generated by the sandwiching of two $\Gamma^{a_{N},(N)}$ with the smallest possible overlap (i.e., one single-electron state). Therefore, its magnitude is determined as
\begin{eqnarray}
a_{N-1}= 2(a_{N}-1)~.
\end{eqnarray}
This suggests that, at the RG step $j$, the highest $n$-particle scattering vertex (appearing in eq.\eqref{RG_flow_heirarchy}) is given by
\begin{eqnarray}
a_{j-1}=2(a_{j}-1)~.\label{highest_vertex}
\end{eqnarray} 
However, as the RG proceeds, integrals of motion are generated and lesser single-particle states participate in the off-diagonal scattering processes. 
At the RG step $j$, there are $j$ single-particle states that are coupled with each other, such that the highest $n$-particle vertex is correctly determined by the relation
\begin{eqnarray}
a_{j} = \text{min}\lbrace 2^{j}a_{0}-2^{j+1}+2,j\rbrace~.\label{highest_n-particle_vertex}
\end{eqnarray}
For an arbitrary Hamiltonian with highest bare-level vertex $a_{0}$, the total number of $1$, $2$, $3$-particle ... off-diagonal terms at the RG step $j$ is given by
\begin{eqnarray}
K_{j} = \sum_{l=1}^{a_{j}/2}\binom{j}{2l}< 2^{j-1}~.\label{tot. no. off-diag}
\end{eqnarray}
\section{Rearrangement Scheme for generating the effective Hamiltonian}\label{Appendix-rearrangement}
Using the unitary operators $U_{\mathbf{k}\sigma}$ together with the block diagonalized SEH $\tilde{H}_{[\mathbf{k}\sigma]}$ (eq.\eqref{block_diagonal_FS}) for every $\mathbf{k}\sigma$, we write down the Hamiltonian $H_{SFIM}$ in eq.\eqref{single_band_general_four_fermion_interacting_model} as
\begin{eqnarray}
H &=& \sum_{\mathbf{k}\sigma}Tr_{\mathbf{k}\sigma}(U^{\dagger}_{[\mathbf{k}\sigma]}\tilde{H}_{[\mathbf{k}\sigma]}U_{[\mathbf{k}\sigma]}\hat{n}_{\mathbf{k}\sigma})\hat{n}_{\mathbf{k}\sigma} + \frac{1}{2}\left(c^{\dagger}_{\mathbf{k}\sigma}Tr_{\mathbf{k}\sigma}(U^{\dagger}_{[\mathbf{k}\sigma]}\tilde{H}_{[\mathbf{k}\sigma]}U_{[\mathbf{k}\sigma]}c_{\mathbf{k}\sigma})+h.c.\right)~.~~~~~\label{complete Hamiltonian}
\end{eqnarray}
Using the form of the unitary operator, we obtain the following relations
\begin{eqnarray}
U_{[\mathbf{k}\sigma]}\hat{n}_{\mathbf{k}\sigma}U^{\dagger}_{[\mathbf{k}\sigma]} &=& \frac{1}{2}\left[1+\eta_{\mathbf{k}\sigma}+\eta^{\dagger}_{\mathbf{k}\sigma}\right] ~,~\nonumber\\
U_{[\mathbf{k}\sigma]}c_{\mathbf{k}\sigma}U^{\dagger}_{[\mathbf{k}\sigma]} &=& \frac{1}{2}c_{\mathbf{k}\sigma}-\frac{1}{2}[\eta^{\dagger}_{\mathbf{k}\sigma},c_{\mathbf{k}\sigma}]-\frac{1}{2}\eta^{\dagger}_{\mathbf{k}\sigma}c_{\mathbf{k}\sigma}\eta^{\dagger}_{\mathbf{k}\sigma}~.~\hspace*{1cm}
\end{eqnarray}
Now, by putting these relations for $U$ back in eq.\eqref{complete Hamiltonian}, we obtain a rearrangement of the terms in Hamiltonian $H_{SFIM}$ as follows
\begin{eqnarray}
H &=& \sum_{\mathbf{k}\sigma}\bigg[\frac{1}{2}Tr_{\mathbf{k}\sigma}(\tilde{H}_{[\mathbf{k}\sigma]})\hat{n}_{\mathbf{k}\sigma}-\frac{1}{4}\left(c^{\dagger}_{\mathbf{k}\sigma}Tr_{\mathbf{k}\sigma}(\tilde{H}_{[\mathbf{k}\sigma]}[\eta^{\dagger}_{\mathbf{k}\sigma},c_{\mathbf{k}\sigma}])+h.c.\right)\bigg]~.
\label{rearranged Hamiltonian}
\end{eqnarray}
The block diagonalized SEH ($Tr_{\mathbf{k}\sigma}(\tilde{H}_{[\mathbf{k}\sigma]})$, eq.\eqref{block_diagonal_FS}) in the rearranged Hamiltonian eq.\eqref{rearranged Hamiltonian} can be written as the sum of the blocks projected onto the electron-occupied ($\tilde{H}^{e}_{[\mathbf{k}\sigma]}$) and hole-occupied subspaces ($\tilde{H}^{h}_{[\mathbf{k}\sigma]}$)
\begin{eqnarray}
Tr_{\mathbf{k}\sigma}(\tilde{H}_{[\mathbf{k}\sigma]}) &=& Tr_{\mathbf{k}\sigma}(\tilde{H}_{[\mathbf{k}\sigma]}\hat{n}_{\mathbf{k}\sigma})+Tr_{\mathbf{k}\sigma}(\tilde{H}_{[\mathbf{k}\sigma]}(1-\hat{n}_{\mathbf{k}\sigma}))~.\hspace*{0.5cm}
\end{eqnarray} 
Under block diagonalization, the partial trace operation $Tr_{\mathbf{k}\sigma}()$ remains preserved, implying that the changes induced by the quantum fluctuation terms in the block Hamiltonians for the electron occupied and hole occupied blocks are constrained as follows
\begin{eqnarray}
\Delta\tilde{H}_{[\mathbf{k}\sigma],e} &=& -\Delta\tilde{H}_{[\mathbf{k}\sigma],h}~,~\Delta\tilde{H}_{[\mathbf{k}\sigma]e/h}= \tilde{H}_{[\mathbf{k}\sigma]e/h}-H_{[\mathbf{k}\sigma]e/h}~,~\hspace*{1cm}\label{energy_shift_adjustment}
\end{eqnarray} 
where the changes in the e and h block Hamiltonians are defined as 
\begin{equation}
\Delta\tilde{H}_{[\mathbf{k}\sigma],e}=Tr_{\mathbf{k}\sigma}(\Delta\tilde{H}_{[\mathbf{k}\sigma]}\hat{n}_{\mathbf{k}\sigma})~,~\Delta\tilde{H}_{[\mathbf{k}\sigma],h}=Tr_{\mathbf{k}\sigma}(\Delta\tilde{H}_{[\mathbf{k}\sigma]}(1-\hat{n}_{\mathbf{k}\sigma}))~.
\end{equation}
The block diagonal Hamiltonian, $\tilde{H}_{[\mathbf{k}\sigma]}$, upon being projected onto the electron occupation subspace, i.e, $Tr_{\mathbf{k}\sigma}(\tilde{H}_{[\mathbf{k}\sigma]}
\hat{n}_{\mathbf{k}\sigma})$, contains new energy shift and quantum fluctuations terms with respect to the rest of coupled states. Then, summing over all such $\mathbf{k}\sigma$ states leads to an effective Hamiltonian
\begin{eqnarray}
\tilde{H}_{e} &=& H + \sum_{\mathbf{k}\sigma}(\Delta H^{D}_{[\mathbf{k}\sigma],e}+\Delta H^{X}_{[\mathbf{k}\sigma],e})~.
\end{eqnarray}
Here, $\Delta H^{D}_{[\mathbf{k}\sigma],e}$ accounts for the self energy/ correlation energy terms that are number diagonal, and $\Delta H^{X}_{[\mathbf{k}\sigma],e}$ contains the renormalization of the scattering terms.
\section{Constituents of the effective two-particle excitation Hamiltonian}\label{Appendix-two_particle_eff}
The changes in the block Hamiltonian  due to inter-particle scattering mediated via intermediate electron-electron or electron-hole configurations of Fig.\ref{2-particle_vertices}(a,b) are described by
\begin{eqnarray}
\Delta H^{1}_{[\mathbf{k}\sigma,\mathbf{k}'\sigma']}(\omega) &=& \sum_{\mathbf{q}\mathbf{q}'}V^{\sigma\sigma'}_{\mathbf{k}\mathbf{k'}\mathbf{q}}G_{\mathbf{k}\sigma,\mathbf{k}'\sigma'}V^{\sigma,\sigma''}_{\mathbf{k}\mathbf{k'}\mathbf{q}'}\tau_{\mathbf{k}\sigma}\tau_{\mathbf{k}'\sigma'}
\times c^{\dagger}_{\mathbf{k}+\mathbf{q}'\sigma}c^{\dagger}_{\mathbf{k}'-\mathbf{q}'\sigma'}c_{\mathbf{k}'-\mathbf{q}\sigma'}c^{\dagger}_{\mathbf{k}+\mathbf{q}\sigma}~,~~~~~
\label{ee-hh/eh-he scatt}
\end{eqnarray}
where the Green's function operator $G_{\mathbf{k}\sigma,\mathbf{k}'\sigma'}$ is given by
\begin{eqnarray}
G_{\mathbf{k}\sigma,\mathbf{k}'\sigma'} = (\omega -\epsilon_{\mathbf{k}}\tau_{\mathbf{k}\sigma}-\epsilon_{\mathbf{k}'}\tau_{\mathbf{k}'\sigma'}-V^{\sigma\sigma'}_{\mathbf{k}\mathbf{k'}0}\tau_{\mathbf{k}\sigma}\tau_{\mathbf{k}'\sigma'})^{-1}~.~~~~~
\end{eqnarray}
The product of the $\tau_{\mathbf{k}\sigma}=\tau_{\mathbf{k}'\sigma'}=\pm\frac{1}{2}$ operators corresponds to the intermediate electron-electron/hole-hole configuration entering the scattering process of Fig.\ref{2-particle_vertices}(a). Similarly, the product of the $\tau_{\mathbf{k}\sigma}=\pm\tau_{\mathbf{k}'\sigma'}=\pm\frac{1}{2}$ operators corresponds to an intermediate electron-hole configuration entering the scattering process of Fig.\ref{2-particle_vertices}(b). The third scattering process, Fig.(\ref{2-particle_vertices}(c)), arises out of the mixing between \textbf{ee} ($1_{\mathbf{k}\sigma}1_{\mathbf{k}'\sigma'}$)  and \textbf{eh} ($1_{\mathbf{k}\sigma}0_{\mathbf{k}'\sigma'}$) pairs, leading to effective three-particle scattering terms given by 
\begin{eqnarray}
\Delta H^{2}_{[\mathbf{k}\sigma,\mathbf{k}'\sigma']} &=& \sum_{\mathbf{q}\mathbf{q}'\mathbf{k}''}V^{\sigma\sigma'}_{\mathbf{k}\mathbf{k}'\mathbf{q}}G_{\mathbf{k}\sigma}V^{\sigma\sigma''}_{\mathbf{k}\mathbf{k}''\mathbf{q}'}\tau_{\mathbf{k}\sigma}c^{\dagger}_{\mathbf{k}'\sigma'}+c_{\mathbf{k}'+\mathbf{q}'\sigma'}c_{\mathbf{k}-\mathbf{q}'\sigma}c^{\dagger}_{\mathbf{k}+\mathbf{q}\sigma}c^{\dagger}_{\mathbf{k}''-\mathbf{q}\sigma''}c_{\mathbf{k}''\sigma''}~,~~~~~
\label{mixing}
\end{eqnarray}
where the effective 1-particle Green's function is given by
\begin{equation}G_{\mathbf{k}\sigma}=(\omega -\epsilon_{\mathbf{k}}\tau_{\mathbf{k}\sigma}-\sum_{\mathbf{k}''}V^{\sigma\sigma''}_{\mathbf{k}\mathbf{k}''}\tau_{\mathbf{k}\sigma}\tau_{\mathbf{k}''\sigma''})^{-1}.
\end{equation}  
The three-fermionic operators $\tau_{\mathbf{k}\sigma}c^{\dagger}_{\mathbf{k}'\sigma'}$ in eq.\eqref{mixing} lead to dynamical hybridization of the ee/hh creation operators ($c^{\dagger}_{\mathbf{k}\sigma}c^{\dagger}_{\mathbf{k}'\sigma'}$, $c_{\mathbf{k}\sigma}c_{\mathbf{k}'\sigma'}$) and eh/he  creation operators ($c^{\dagger}_{\mathbf{k}\sigma}c_{\mathbf{k}'\sigma'}$, $c^{\dagger}_{\mathbf{k}'\sigma'}c_{\mathbf{k}'\sigma'}$). In order to represent this mixing, let us first define the three-fermionic and two-fermionic creation operators as follows
\begin{eqnarray}
\mu^{\dagger}_{\mathbf{k}\sigma,\mathbf{k}'\sigma'} &=& \hat{n}_{\mathbf{k}\sigma}c^{\dagger}_{\mathbf{k}'\sigma'}~,~\rho^{\dagger}_{\mathbf{k}\sigma,\mathbf{k}'\sigma'} = (1-\hat{n}_{\mathbf{k}\sigma})c^{\dagger}_{\mathbf{k}'\sigma'}~,\nonumber\\
C^{+}_{\mathbf{k}\sigma,\mathbf{k}'\sigma'} &=& c^{\dagger}_{\mathbf{k}\sigma}c^{\dagger}_{\mathbf{k}'\sigma'}~,~ S^{+}_{\mathbf{k}\sigma,\mathbf{k}'\sigma'} = c^{\dagger}_{\mathbf{k}\sigma}c_{\mathbf{k}'\sigma'}~,
\end{eqnarray}
where 
$\mu^{\dagger}_{\mathbf{k}\sigma,\mathbf{k}'\sigma'}-\rho^{\dagger}_{\mathbf{k}\sigma,\mathbf{k}'\sigma'} = 2\tau_{\mathbf{k}\sigma}c^{\dagger}_{\mathbf{k}'\sigma'}$. Similarly, we can define $\mu^{\dagger}_{\mathbf{k}'\sigma',\mathbf{k}\sigma} = \hat{n}_{\mathbf{k}'\sigma'}c^{\dagger}_{\mathbf{k}\sigma}$ and $\rho^{\dagger}_{\mathbf{k}'\sigma',\mathbf{k}\sigma} = (1-\hat{n}_{\mathbf{k}'\sigma'})c^{\dagger}_{\mathbf{k}\sigma}$. 
The source of the dynamical hybridization resides in the non-commutativity of the three-fermionic operator $\hat{n}_{\mathbf{k}\sigma}c^{\dagger}_{\mathbf{k}'\sigma'}$ with the $c^{\dagger}_{\mathbf{k}\sigma}c^{\dagger}_{\mathbf{k}'\sigma'}$,$c_{\mathbf{k}'\sigma'}c_{\mathbf{k}\sigma}$, $c^{\dagger}_{\mathbf{k}\sigma}c^{\dagger}_{\mathbf{k}'\sigma'}$ and $c^{\dagger}_{\mathbf{k}'\sigma'}c_{\mathbf{k}\sigma}$ type operators
\begin{eqnarray}
[C^{+}_{\mathbf{k}\sigma,\mathbf{k}'\sigma'},\mu_{\mathbf{k}\sigma,\mathbf{k}'\sigma'}] = \rho^{\dagger}_{\mathbf{k}'\sigma',\mathbf{k}\sigma}~,~\left[\rho^{\dagger}_{\mathbf{k}'\sigma' ,\mathbf{k}\sigma},\mu^{\dagger}_{\mathbf{k}\sigma ,\mathbf{k}'\sigma'}\right]= C^{+}_{\mathbf{k}\sigma,\mathbf{k}'\sigma'}~,~\left[\rho_{\mathbf{k}'\sigma',\mathbf{k}\sigma},C^{+}_{\mathbf{k}\sigma,\mathbf{k}'\sigma'}\right]&=& \mu^{\dagger}_{\mathbf{k}\sigma,\mathbf{k}'\sigma'}~.~~~~~~~
\label{dynamical mixing} 
\end{eqnarray}
Similar commutation relations are found for the $S^{+}_{\mathbf{k}\sigma,\mathbf{k}'\sigma'} (\equiv c^{\dagger}_{\mathbf{k}\sigma}c_{\mathbf{k}'\sigma'})$ operators as well. 
\pin
The effective three-particle term $\Delta H^{2}_{[\mathbf{k}\sigma,\mathbf{k}'\sigma']}$ does not, therefore, commute with the effective two particle term $\Delta H^{1}_{[\mathbf{k}\sigma,\mathbf{k}'\sigma']}$. In order to take account of this non-commutativity, we perform a unitary rotation in the space of the operators defined above 
 \begin{eqnarray}
 c^{\dagger}_{\mathbf{k}\sigma}\begin{pmatrix}
 \gamma^{p\dagger}_{\mathbf{k}'\sigma'} \\ \\  \nu^{p\dagger}_{\mathbf{k}'\sigma'}
 \end{pmatrix} &=& \begin{pmatrix}
 \sqrt{p} & \sqrt{1-p} \\
 & \\
 -\sqrt{1-p} & \sqrt{p}
 \end{pmatrix}\begin{pmatrix}
C^{+}_{\mathbf{k}\sigma,\mathbf{k}'\sigma'} \\ \\ \rho^{\dagger}_{\mathbf{k}'\sigma',\mathbf{k}\sigma}
 \end{pmatrix}~,~~~~\label{ph superposition}
 \end{eqnarray}
where $0\leq p\leq 1$ represents the probability for the electronic state $\mathbf{k}'\sigma'$ to be occupied. Our goal is to rotate into a particular particle-hole superposition channel (say, $p^{*}$), such that the scattering amplitude of the particle-particle or particle-hole pairs in $\Delta H^{1}_{[\mathbf{k}\sigma,\mathbf{k}'\sigma']}$ are \emph{bigger} compared to the three-particle scattering amplitudes in $\Delta H^{2}_{[\mathbf{k}\sigma,\mathbf{k}'\sigma']}$. In this way, we will incorporate the effects of three-particle scattering physics and the spin-charge interplay phenomenon. To fulfil this, we perform the state-space rotation using the rotated operators of eq.\eqref{ph superposition}, and compute the contribution of the $\Delta H^{2}_{[\mathbf{k}\sigma,\mathbf{k}'\sigma']}$ and $\Delta H^{1}_{[\mathbf{k}\sigma,\mathbf{k}'\sigma']}$ matrix elements in the rotated basis. First, we note that the rotated particle-hole superposition operators of eq.\eqref{ph superposition}, $\gamma^{p\dagger}_{\mathbf{k}'\sigma'}~,~\nu^{p\dagger}_{\mathbf{k}'\sigma'}$, satisfy the following completeness relation
\begin{eqnarray}
&&\gamma^{p\dagger}_{\mathbf{k}'\sigma'}\gamma^{p}_{\mathbf{k}'\sigma'}+\nu^{p\dagger}_{\mathbf{k}'\sigma'}\nu^{p}_{\mathbf{k}'\sigma'} = 1~.
\label{completeness_rel}
\end{eqnarray}
The action of the operators $\gamma^{p\dagger}_{\mathbf{k}'\sigma'},\bar{\gamma}^{p\dagger}_{\mathbf{k}'\sigma'}$ on the basis states $|1_{\mathbf{k}'\sigma'}\rangle$ and $|0_{\mathbf{k}'\sigma'}\rangle$ are given by
\begin{eqnarray}
&&\gamma^{p\dagger}_{\mathbf{k}'\sigma'}|1_{\mathbf{k}'\sigma'}\rangle =  0~,~\gamma^{p\dagger}_{\mathbf{k}\sigma}|0_{\mathbf{k}'\sigma'}\rangle =  |\psi_{\mathbf{k}'\sigma', p}\rangle~,~|\psi_{\mathbf{k}'\sigma', p}\rangle = \sqrt{p}|1_{\mathbf{k}'\sigma'}\rangle +\sqrt{1-p}|0_{\mathbf{k}'\sigma'}\rangle~,~~~~ \nonumber\\
&&\nu^{p\dagger}_{\mathbf{k}'\sigma'}|1_{\mathbf{k}'\sigma'}\rangle = 0 ~,~\nu^{p\dagger}_{\mathbf{k}'\sigma'}|0_{\mathbf{k}'\sigma'}\rangle = |\psi^{\perp}_{\mathbf{k}'\sigma', p}\rangle~,~|\psi^{\perp}_{\mathbf{k}'\sigma', p}\rangle =  \sqrt{1-p}|1_{\mathbf{k}'\sigma'}\rangle -\sqrt{p}|0_{\mathbf{k}'\sigma'}\rangle~,~~~~\label{state space action}
\end{eqnarray}
with $\langle\psi_{\mathbf{k}'\sigma', p}|\psi^{\perp}_{\mathbf{k}'\sigma', p}\rangle =0$. Given eq.\eqref{completeness_rel}, the rotated states of eq.\eqref{state space action} also fulfil the completeness relation given above. In the tensor product Hilbert space of $\mathbf{k}\sigma$ and $\mathbf{k}'\sigma'$, we then define the basis states
\begin{eqnarray}
|1_{\mathbf{k}\sigma}\psi_{\mathbf{k}'\sigma', p}\rangle &=& \sqrt{p}|1_{\mathbf{k}\sigma}1_{\mathbf{k}'\sigma'}\rangle +\sqrt{1-p}|1_{\mathbf{k}\sigma}0_{\mathbf{k}'\sigma'}\rangle~, \nonumber\\
|1_{\mathbf{k}\sigma}\psi^{\perp}_{\mathbf{k}'\sigma', p}\rangle &=&  \sqrt{1-p}|1_{\mathbf{k}\sigma}1_{\mathbf{k}'\sigma'}\rangle -\sqrt{p}|1_{\mathbf{k}\sigma}0_{\mathbf{k}'\sigma'}\rangle~,~\label{mixed_valence_configuration1}
\end{eqnarray}
as well as the states $|0_{\mathbf{k}\sigma}\psi_{\mathbf{k}'\sigma', p}\rangle$, $|0_{\mathbf{k}\sigma}\psi^{\perp}_{\mathbf{k}'\sigma', p}\rangle$ similarly to those given above.
\pin
For the states with quantum numbers $(\mathbf{k}\sigma)$ and $(\mathbf{k}'\sigma')$, such that their kinetic energies are ordered as $\epsilon_{\mathbf{k}}>\epsilon_{\mathbf{k}'}>E_F$ (the Fermi energy), the channel $|1_{\mathbf{k}\sigma}\psi_{\mathbf{k}'\sigma', p}\rangle$ is composed of net positive energy states. Bound state formation occurs for such states via the lowering of the total energy below $E_{F}$. This motivates a study of the changes induced in the two-particle ($\Delta H_{[\mathbf{k}\sigma,\mathbf{k}'\sigma']}^{1}$) and three- particle ($\Delta H^{2}_{\mathbf{k}\sigma,\mathbf{k}'\sigma']}$) scattering terms  for the intermediate configuration $|1_{\mathbf{k}\sigma}\psi_{\mathbf{k}'\sigma', p}\rangle$
\begin{eqnarray}
\Delta H^{1}_{[\mathbf{k}\sigma,\mathbf{k}'\sigma']}(\omega) &=& \sum_{\mathbf{q}\mathbf{q}'}\frac{(1-2p)V_{\mathbf{k}\mathbf{k}'\mathbf{q}}V_{\mathbf{k}\mathbf{k}'\mathbf{q}'}}{\omega - E_{p,\mathbf{k},\mathbf{k}'} - \frac{(1-2p)}{4}V_{\mathbf{k}\mathbf{k}'}}~,\nonumber\\
&\times &c^{\dagger}_{\mathbf{k}+\mathbf{q}'\sigma}c^{\dagger}_{\mathbf{k}'-\mathbf{q}'\sigma'}c_{\mathbf{k}'-\mathbf{q}\sigma'}c^{\dagger}_{\mathbf{k}+\mathbf{q}\sigma},\nonumber\\
\Delta H^{2}_{[\mathbf{k}\sigma,\mathbf{k}'\sigma']}(\omega)&=&\sum_{\mathbf{k}''\mathbf{q}'\mathbf{q}}\frac{\sqrt{p(1-p)}V^{\sigma\sigma'}_{\mathbf{k}\mathbf{k}'\mathbf{q}}V^{\sigma\sigma''}_{\mathbf{k}\mathbf{k}''\mathbf{q}'}}{\omega - \frac{1}{2}\epsilon_{\mathbf{k}}-\frac{1}{2}\sum_{\mathbf{k}''}V^{\sigma\sigma'}_{\mathbf{k}\mathbf{k}''}\Theta(\epsilon_{\mathbf{k}''}-E_{F})}~,~~~~~
\end{eqnarray}
where $E_{p,\mathbf{k},\mathbf{k}'} = \frac{p}{2}(\epsilon_{\mathbf{k}}+\epsilon_{\mathbf{k}'})+\frac{(1-p)}{2}(\epsilon_{\mathbf{k}}-\epsilon_{\mathbf{k}'})$. In the regime given by $\omega > E_{p,\mathbf{k},\mathbf{k}'},\frac{1}{2}\epsilon_{\mathbf{k}}>V^{\sigma\sigma'}_{\mathbf{k}\mathbf{k}'}$, we find the optimal spin-charge mixing parameter $p^{*}$ subject to the condition
\begin{eqnarray}
\smash{\displaystyle\max_{p*}}(\omega -E_{p,\mathbf{k},\mathbf{k}'})^{-1} &\equiv &(\omega -E_{p*,\mathbf{k},\mathbf{k}'})^{-1} > (\omega -\frac{1}{2}\epsilon_{\mathbf{k}})^{-1}~.
\end{eqnarray} 
This condition automatically fulfils our goal for the
magnitude of the change in the two-particle vertices in $\Delta H^{1}_{[\mathbf{k}\sigma,\mathbf{k}'\sigma']}(\omega)$ to be of greater than the magnitude of the three-particle vertices in $\Delta H^{2}_{[\mathbf{k}\sigma,\mathbf{k}'\sigma']}(\omega)$.
\section{Unitary matrix as a Scattering matrix and the generalized optical theorem}\label{unitary_scattering}
The unitary matrix described in eq.\eqref{Unitary_op} can be written in terms of the transfer matrix $T$, and also in terms of the exponential of a phase operator $\theta$
\begin{eqnarray}
U_{N} = \exp i\theta_{N} = 1+iT_{N}~, \label{unitary_matrix_T_Matrix}
\end{eqnarray}
such that (using eq.\eqref{eta-operator-rel})
\begin{eqnarray}
\theta_{N}&=& \arctan i\left(\eta_{N}-\eta^{\dagger}_{N}\right)~,\nonumber\\ 
T_{N}  &=& i\left(1-\frac{1}{\sqrt{2}}\right) -\frac{i}{\sqrt{2}}\left(\eta_{N}-\eta^{\dagger}_{N}\right)~. \label{T_matrix}
\end{eqnarray}
The requirement of unitarity $UU^{\dagger}=U^{\dagger}U =I$, together with the form of eq.\eqref{unitary_matrix_T_Matrix}, imposes a constraint on the $T$-matrix. The $T$-matrix form given by eq.\eqref{T_matrix} fulfils the optical theorem
\begin{eqnarray}
i(T_{N} -T^{\dagger}_{N}) =T_{N}T^{\dagger}_{N}~.\label{optical_theorem}
\end{eqnarray} 
\pin\\
\textbf{\Large{References}}
\bibliographystyle{elsarticle-num}
\bibliography{netbib}
\end{document}